\documentclass{ws-ijbc}
\usepackage{ws-rotating}     
\usepackage[english]{babel}
\usepackage{amsmath}
\usepackage{graphicx}
\usepackage{amssymb}

\begin{document}
\bibliographystyle{ws-ijbc}

\catchline{}{}{}{}{} 

\markboth{L.A. DARRIBA}{Comparative study of variational chaos indicators and ODEs' numerical integrators}

\title{Comparative study of variational chaos indicators and ODEs' numerical integrators}

\author{L.A. DARRIBA}
\address{Grupo de Caos en Sistemas Hamiltonianos\\ Facultad de Ciencias
Astron\'omicas y Geof\'isicas, Instituto de Astrof\'isica de La Plata,
Universidad Nacional de La Plata, CONICET-CCT La Plata, \\
Paseo del Bosque s/n, La Plata, B1900FWA, Buenos Aires, Argentina\\
ldarriba@fcaglp.unlp.edu.ar}

\author{N.P. MAFFIONE}
\address{Grupo de Caos en Sistemas Hamiltonianos\\ Facultad de Ciencias
Astron\'omicas y Geof\'isicas, Instituto de Astrof\'isica de La Plata,
Universidad Nacional de La Plata, CONICET-CCT La Plata, \\
Paseo del Bosque s/n, La Plata, B1900FWA, Buenos Aires, Argentina\\
nmaffione@fcaglp.unlp.edu.ar}

\author{P.M. CINCOTTA}
\address{Grupo de Caos en Sistemas Hamiltonianos\\ Facultad de Ciencias
Astron\'omicas y Geof\'isicas, Instituto de Astrof\'isica de La Plata,
Universidad Nacional de La Plata, CONICET-CCT La Plata, \\
Paseo del Bosque s/n, La Plata, B1900FWA, Buenos Aires, Argentina\\
pmc@fcaglp.unlp.edu.ar}

\author{C.M. GIORDANO}
\address{Grupo de Caos en Sistemas Hamiltonianos\\ Facultad de Ciencias
Astron\'omicas y Geof\'isicas, Instituto de Astrof\'isica de La Plata,
Universidad Nacional de La Plata, CONICET-CCT La Plata, \\
Paseo del Bosque s/n, La Plata, B1900FWA, Buenos Aires, Argentina\\
giordano@fcaglp.unlp.edu.ar}

\maketitle

\begin{history}
\received{(to be inserted by publisher)}
\end{history}

\begin{abstract}
The reader can find in the literature a lot of different techniques to study the dynamics of a given system and also, many suitable numerical integrators to compute them. Notwithstanding the recent work of \cite{Maffione11a} for mappings, a detailed comparison among the widespread indicators of chaos in a general system is still lacking. Such a comparison could lead to select the most efficient algorithms given a certain dynamical problem. Furthermore, in order to choose the appropriate numerical integrators to compute them, more comparative studies among numerical integrators are also needed. 

This work deals with both problems. We first extend the work of
 \cite{Maffione11a} for mappings to the 2D \cite{Henon64}
 potential, and compare several variational indicators of chaos: the
 Lyapunov Indicator (LI); the Mean Exponential Growth Factor of Nearby
 Orbits (MEGNO); the Smaller Alignment Index (SALI) and its generalized
 version, the Generalized Alignment Index (GALI); the Fast Lyapunov
 Indicator (FLI) and its variant, the Orthogonal Fast Lyapunov Indicator
 (OFLI); the Spectral Distance (\textit{D}) and the Dynamical Spectras
 of Stretching Numbers (SSNs). We also include in the record the
 Relative Lyapunov Indicator (RLI), which is not a variational indicator
 as the others. Then, we test a numerical technique to integrate
 Ordinary Differential Equations (ODEs) based on the Taylor method
 implemented by \cite{JZ05} (called \texttt{taylor}), and we compare its performance with other
 two well--known efficient integrators: the \cite{PD81} implementation of
 a Runge--Kutta of order 7--8 (DOPRI8) and a Bulirsch--St\"oer
 implementation. These tests are run under two very different systems
 from the complexity of their equations point of view: a triaxial galactic potential model and a perturbed 3D quartic oscillator.

We first show that a combination of the FLI/OFLI, the MEGNO and the
 GALI$_{2N}$ succeeds in describing in detail most of the dynamical
 characteristics of a general Hamiltonian system. In the second part we
 show that the precision of \texttt{taylor} is better than that of
 the other integrators tested, but it is not well suited to integrate
 systems of equations which include the variational ones, like in the
 computing of almost all the preceeding indicators of chaos. The result
 which induces us to draw this conclusion is that the computing times
 spent by \texttt{taylor} are far greater than the times insumed by the DOPRI8 and the Bulirsch--St\"oer integrators in such cases. On the other hand, the package is very efficient when we only need to integrate the equations of motion (both in precision and speed), for instance to study the chaotic diffusion. We also notice that \texttt{taylor} attains a greater precision on the coordinates than either the DOPRI8 or the Bulirsch--St\"oer.
\end{abstract}

\keywords{Chaos Indicators -- Numerical Integrators -- Variational Equations -- ODEs -- Hamiltonian Systems}

\section{Introduction}\label{section_introduction}
In order to study a Hamiltonian system, it is essential to count on both fast and easy--to--compute techniques to determine the chaotic or regular nature of its orbits.

For two degrees of freedom (hereafter d.o.f.), a technique based on a graphical treatment, like the Poincar\'e Surfaces of Section, should suffice to understand such a low dimensional problem. Though generalized to three dimensional systems (\cite{Froeschle70}, \cite{Froeschle72}), the graphical treatment shows fundamental constraints. Therefore, non graphical methods seem to be the logical alternative to deal with systems of higher dimensionality. 

The concept of exponential divergence is thus introduced, and it is one of the two widespread ideas in chaos detection. The introduction of the Lyapunov Characteristic Exponents (LCEs; see e.g. \cite{Rei92}) as well as its numerical implementation (\cite{Benettin80}) were a breakthrough contribution to the field. Notice that the time of integration is limited, so we are able to reach just truncated approximations of the theoretical LCEs, i.e. Lyapunov Indicators, hereafter LI (\cite{Benettin76}, \cite{Froeschle84} or \cite{Tancredi01} and \cite{Skokos10} for reviews on the subject). Therefore, a drawback in the computation of the truncated values of the LCEs is their very slow convergence. Nevertheless, a large number of variational techniques improve the LI's performance, such as: the Mean Exponential Growth factor of Nearby Orbits (MEGNO; \cite{Cincotta00}\footnote{See also \cite{Cincotta03}, \cite{GiCi04}, \cite{GKW05}, \cite{GB08}, \cite{LDV09}, \cite{HCAG10}, \cite{Maffione11}, \cite{Compere11}.}); the Smaller Alignment Index (SALI; \cite{Skokos01}\footnote{See also \cite{Skokos04}, \cite{Szell04}, \cite{BS06}, \cite{C08}, \cite{AVB10}.}) and its generalized version, the Generalized Alignment Index (GALI; \cite{Skokos07}\footnote{See also \cite{Skokos08}, \cite{MA11}.}); the Fast Lyapunov Indicator (FLI; \cite{Froeschle97}, \cite{Froeschle97a}\footnote{See also \cite{Froeschle98}, \cite{Froeschle00}, \cite{Lega01}, \cite{Guzzo02}, \cite{Froeschle06}, \cite{PFL08}, \cite{TLF08}, \cite{LGF10}.}) and its variant, the Orthogonal Fast Lyapunov Indicator (OFLI; \cite{Fouchard02}); the Spectral Distance (\textit{D}; \cite{Voglis99}) and the Dynamical Spectras of Stretching Numbers (SSNs; \cite{Voglis94}\footnote{See also \cite{Contopoulos96}, \cite{Contopoulos97}, \cite{Contopoulos97a}, \cite{Voglis98}.}). Finally, we include the Relative Lyapunov Indicator (RLI; \cite{Sandor00}\footnote{See also \cite{Szell04}, \cite{Sandor04}, \cite{SSEPD07}.}), which is not based on the evolution of the solution of the first variational equations as the others, but on the evolution of two different but very close orbits. 

These are just a sample of all the variational Chaos Indicators (CIs) widely used nowadays. They are the ones selected to be used in the forthcoming investigation. However, the record could indeed continue to include the Average Power Law Exponent (APLE; \cite{Lukes08}); the Orthogonal Fast Lyapunov Indicator of order two (OFLI$_{2}^{TT}$; \cite{Barrio05}, \cite{Barrio09}, \cite{Barrio10}) which is a second order variational method; the Dynamical Spectras of Helicity or Twist Angles (\cite{Contopoulos96}, \cite{Contopoulos97}, \cite{Froeschle98}, \cite{Lega98}). Moreover, evolutionary algorithms (\cite{Petalas09}) are also interesting alternatives. 

The other widespread idea in the field of CIs is the analysis of some particular quantities (e.g. the frequency) of a single orbit. In this group we find the Frequency Map Analysis (FMA; \cite{Laskar90}\footnote{See also \cite{Laskar92}, \cite{Papaphilippou96}, \cite{Papaphilippou98}.}) and its variant, the Frequency Modified Fourier Transform (FMFT; \cite{SN96}) or other alternatives using Fast Fourier Transform techniques (\cite{MF95}, \cite{FMBC05}, \cite{MLCF10}), among others. Nevertheless, we leave such family of indicators for a future comparison.

In order to compute the above variational CIs efficiently, it is fundamental to appropriately recognize numerical techniques for the integration of Ordinary Differential Equations (ODEs hereafter), which is the goal of the second part of this review. That is, in most of the problems that arise from the investigation of dynamical systems, as well as in every realistic model, the corresponding differential equations have to be solved by means of numerical techniques. Its computational implementation should be as fast and precise as possible, in order to minimize both the numerical error and the computing time.

Among all numerical integrators, there are those which use recurrent power series in order to compute the solutions with a considerably high precision, and if we take advantage of the recurrence, the computing time is reduced (\cite{Sit79} \& \cite{Sit89}). One of the most used power series integrators is the Taylor method (\cite{Mon91}, \cite{Mon92}, \cite{Gof92}, \cite{GM95}, \cite{Sim06}, \cite{Sim08} \& \cite{GS10}), whose usefulness is at all means clear; a particular implementation of which will be tested in this work.

In \cite{JZ05} the authors  presented an integration package (coded in C) based on the Taylor method for numerical integration of ODEs. They showed that this package turned out to be a very efficient integrator (both in speed and precision) when applied to the Lorentz system, the forced pendulum model and the Restricted Three--Body Problem.

The reason for selecting this particular package for our numerical experiments is that it implements several features which makes it a very versatile and nearly automatic package. This is due to the fact that the software encompasses the necessary subroutines to obtain a complete integrator, such as the inclusion of the so--called \emph{automatic differentiation} (\cite{BKSF59}, \cite{Wen64}, \cite{Moo66}, \cite{Ral81}, \cite{GC91}, \cite{BCCG92}, \cite{BBCG96} and \cite{Gri00}), an algorithm to choose the order of the power series $p$ and the step size $h$, the possibility to incorporate customized mathematical routines that use higher precisions than the standard \emph{double precision}, and even a FORTRAN wrapper to call the integrator from a main program coded in FORTRAN 77. 

Taking into account those advantageous features of Jorba \& Zou package, we have decided to test their implementation of Taylor method by analizing its performance when applied to three different dynamical problems for which we have compared its efficiency --in both computing time and precision-- against that of other two widely used integrators.

This work comprises two parts: in the first part of the paper, we make a comparative study of variational CIs including the LI, the RLI, the MEGNO, the \textit{D} and the SSNs, the FLI and the OFLI, the SALI and the GALIs. In Section \ref{section_introductionCIs}, we briefly introduce the above--mentioned CIs and in Section \ref{section_Flows}, we perform some experiments on the simple model of \cite{Henon64} (hereafter HH). There, we deal with the CI's final values and time evolution curves. In Section \ref{final values}, we study the reliability of the CI's thresholds in order to separate chaotic from regular motion, and also the level of dynamical information provided by the CI's final values. Then, in Section \ref{Qualitative study-representative groups}, we examine the capability of identifying periodicity and instabilities by means of the CI's time evolution, and the sensitivity of the CIs on initial deviation vectors (hereafter, i.d.v.). Finally, we discuss the computing times and the advantage of the saturation times (see Section \ref{cpu-times}). Our purpose in this first part of the reported research is to extend the results shown in \cite{Maffione11a} to a Hamiltonian flow for mappings. 

Then, in the second part of the paper, the comparison among numerical techniques for the resolution of ODEs in order to compute the CIs just mentioned is presented. Some of the features of Jorba \& Zou package will be briefly described in Section \ref{section_taylor}. We introduce the results of testing the latter package over three different dynamical problems as follows: the first two problems (Section \ref{section_pot_tri} and Section \ref{section_quartic}) concern the computation of two dynamical indicators, i.e. the MEGNO (Section \ref{section_MEGNO_pot_tri} and Section \ref{section_MEGNO_quartic}) and the LI (Section \ref{section_LCN_pot_tri} and Section \ref{section_LCN_quartic}). Both  problems require the rather accurate integration of the equations of motion along with their first variationals and, in the first case,  some additional differential equations must be included. The third problem under consideration is the solely integration of the equations of motion (see Section \ref{section_solomov_pot_tri} and Section \ref{section_solomov_quartic}). Though this is a rather simple problem, it has many useful applications, such as the study of chaotic diffusion in phase space, where we need to follow accurately the coordinates on phase space over rather long time intervals. For the above--mentioned problems, we have chosen two particular Hamiltonian systems: the potential corresponding to an auto--consistent model of a triaxial elliptical galaxy \cite{MCW05}, which actually involves complicated expressions for the ODEs, and a rather simpler one, a perturbed three dimensional quartic oscillator. In Section \ref{section_trans_error} we only compute the equations of motion for the three integrators to test the precision of the coordinates of phase space, and compare which one preserves better the position on the orbit. Finally, we discuss the results in Section \ref{section_discussion}.

\section{Selected CIs revisited}\label{section_introductionCIs}
 The aim of the present section is to introduce the CIs that we will use in the
 forthcoming investigation. However, for the sake of simplicity, we describe only the main properties defined for flows.

\subsection{The LI and the MEGNO}\label{MEGNO}
Consider a continuous dynamical system defined on a differentiable manifold $S$, where $\mathbf{\Phi}^t(\mathbf{x})$ characterizes the state of the system at time $t$ and being the state of the system at time $t=0$, $\mathbf{x}(0)$. Therefore, the state of the system after two consecutive time steps $t$ and $t'$ will be given by the composition law: $\mathbf{\Phi}^{t+t'}=\mathbf{\Phi}^t\circ\mathbf{\Phi}^{t'}$.

Moreover, the tangent space of $\mathbf{x}$ maps onto the tangent space of $\mathbf{\Phi}^t(\mathbf{x})$ according to the operator $d_{\mathbf{x}}\mathbf{\Phi}^t$ and following the rule $\mathbf{w}(t)=d_{\mathbf{x}}\mathbf{\Phi}^t(\mathbf{w}(0))$ where $\mathbf{w}(0)$ is an i.d.v. The action of such operator at consecutive time intervals satisfies the equation: 

$$d_{\mathbf{x}}\mathbf{\Phi}^{t+t'}=d_{\mathbf{\Phi}^{t'}(\mathbf{x})}\mathbf{\Phi}^t\circ d_{\mathbf{x}}\mathbf{\Phi}^{t'}.$$

If we suppose that our manifold $S$ has some norm denoted by $\|\cdot\|$, we can define the useful quantity:

$$\lambda_t(\mathbf{x})=\frac{\|d_{\mathbf{x}}\mathbf{\Phi}^t\mathbf{w}\|}{\|\mathbf{w}\|}$$
called ``growth factor'' in the direction of $\mathbf{w}$. 

Let us now say that $H({\mathbf{p}},{\mathbf{q}})$ with ${\mathbf{p}},\,{\mathbf{q}}\in \mathbb{R}^N$ be an $N$--dimensional Hamiltonian, that we suppose autonomous just for the sake of simplicity. Introducing the following notation:

$${\mathbf{x}}=({\mathbf{p}},{\mathbf{q}})\in\mathbb{R}^{2N},\, \mathbf{f}(\mathbf{x})=(-\partial H/\partial{\mathbf{q}},\ \partial H/\partial{\mathbf{p}})\in\mathbb{R}^{2N},$$

the equations of motion can be written in a simple way like

\begin{equation}
\dot{\mathbf{x}}={\mathbf{f}}({\mathbf{x}}).
\label{megno1}
\end{equation}

Let $\gamma(\mathbf{x_{0}};t)$ be an arc of an orbit of the flow (\ref{megno1}) over a compact energy surface: $M_h\subset\mathbb{R}^{2N}$, $M_h=\{{\mathbf{x}}: H({\mathbf{p}},{\mathbf{q}})=h\}$ with $h=\,$constant, then

$$\gamma(\mathbf{x_{0}};t)=\{{\mathbf{x}}(t';{\mathbf{x}}_0):{\mathbf{x}}_0\in M_h,\ 0\le t'< t\}.$$

We can gain fundamental information about the Hamiltonian flow in the neighborhood of any orbit $\gamma$ through the largest LCE (lLCE, \cite{Skokos10}) defined as:

\begin{equation}
\chi[\gamma(\mathbf{x_{0}};t)]=\lim_{t\to\infty}\frac{1}{t}\ln\lambda_t[\gamma(\mathbf{x_{0}};t)]
\label{megno2}
\end{equation}
with 

$$\lambda_t[\gamma(\mathbf{x_{0}};t)]=\frac{\|d_{\gamma}\mathbf{\Phi}^t\mathbf{w}\|}{\|\mathbf{w}\|}$$ 
where $\|d_{\gamma}\mathbf{\Phi}^t\mathbf{w}\|$ is an ``infinitesimal displacement'' from $\gamma$ at time $t$. The fact that the lLCE (and its truncated value, the so--called LI$=\lim_{t\to T}\frac{1}{t}\ln\lambda_t[\gamma(\mathbf{x_{0}};t)]$ for T, finite) measures the mean exponential rate of divergence of nearby orbits is clearly understood when written Eq. (\ref{megno2}) in an integral fashion:

\begin{equation}
\chi[\gamma(\mathbf{x_{0}};t)]=\lim_{t\to\infty}{1\over t}\int_0^t{\|\dot{d}_{\gamma}\mathbf{\Phi}^{t'}\mathbf{w}\|
\over\|d_{\gamma}\mathbf{\Phi}^{t'}\mathbf{w}\|}{\rm d}t'= \overline{\left(\|\dot{d}_{\gamma}\mathbf{\Phi}^{t'}\mathbf{w}\|\over\|d_{\gamma}\mathbf{\Phi}^t\mathbf{w}\|\right)},
\label{megno3}
\end{equation}
where the bar denotes time average. 

Now we are in condition to introduce the MEGNO (\cite{Cincotta00}), $Y[\gamma(\mathbf{x_0};t)]$, through the expression:

\begin{equation}
  Y[\gamma(\mathbf{x_0};t)]={2\over t}\int_0^t{\|\dot{d}_{\gamma}\mathbf{\Phi}^{t'}\mathbf{w}\|\over\|d_{\gamma}\mathbf{\Phi}^{t'}\mathbf{w}\|}t'{\rm d}t',
  \label{MEGNO_eq}
\end{equation}

which is related to the integral in Eq. (\ref{megno3}); i.e., in case of an exponential increase of $\|d_{\gamma}\mathbf{\Phi}^{t}\mathbf{w}\|$, $\|d_{\gamma}\mathbf{\Phi}^{t}\mathbf{w}\|=\|\mathbf{w}\|\cdot\exp(\chi t)$, the quantity $Y[\gamma(\mathbf{x_0};t)]$ can be considered as a weighted variant of the integral in Eq. (\ref{megno3}). Instead of using the instantaneous rate of increase, $\chi$, we average the logarithm of the growth factor, $\ln(\lambda_t)=\chi t$.

Introducing the time average:

\begin{equation}
  \overline{Y}[\gamma_q(\mathbf{x_0};t)]\equiv{1\over t}\int_0^tY[\gamma_q(\mathbf{x_0};t')]{\rm d}t',
  \label{MEGNO_avg_eq}
\end{equation}

we notice that the time evolution of the MEGNO can be briefly described in a suitable and unique expression for all kinds of motion. Indeed, its asymptotic behavior can be summarized in the following way:  $\overline{Y}[\gamma(\mathbf{x_0};t)]\approx a_{\gamma}t+b_{\gamma}$, where $a_{\gamma}=\chi_{\gamma}/2$ and $b_{\gamma}\approx 0$ for irregular, stochastic motion, while $a_{\gamma}=0$ and $b_{\gamma}\approx 2$ for quasi--periodic motion. Deviations from the value $b_{\gamma}\approx 2$ indicate that $\gamma$ is close to some particular objects in phase space, being $b_{\gamma}\lesssim 2$ or $b_{\gamma}\gtrsim 2$ for stable periodic orbits (resonant elliptic tori), or unstable periodic orbits (hyperbolic tori), respectively (see \cite{Cincotta00}).

\subsection{The FLI and the OFLI}\label{The FLI}
The FLI (see \cite{Froeschle97}, \cite{Froeschle97a},
\cite{Froeschle06}) is a quantity intimately related to the lLCE (and
the MEGNO, see \cite{Mestre11}), which is able to distinguish between
chaotic (weakly chaotic) and regular motion and also, between resonant and non resonant motion (\cite{Froeschle00}, \cite{Lega01}) using only the first part of the computation of the lLCE.

On an $N$--dimensional flow, we follow the time evolution of $N$ unit i.d.v. Therefore, the FLI at time $t$ is defined by the highest norm among the unit i.d.v. of the basis, as follows: 

$$FLI(t)=sup_t\left[\|\mathbf{w}(t)_1\|,\|\mathbf{w}(t)_2\|,\ldots,\|\mathbf{w}(t)_N\|\right].$$ 
For further details, refer to \cite{Froeschle97a}.

For both kinds of motion, the FLI tends to infinity as the time increases, with completely different rates, though (it behaves linearly for regular motion and grows exponentially fast for chaotic motion).

In the case of the OFLI (see \cite{Fouchard02}), we take the component orthogonal to the flow for each unit i.d.v. of the basis at every time step: 

$$OFLI(t)=sup_t\left[w(t)^{\perp}_1,w(t)^{\perp}_2,\ldots,w(t)^{\perp}_N\right].$$ 

This modification makes the OFLI a CI that can easily distinguish periodicity among the regular component. The OFLI for periodic orbits oscillates around a constant value, while for quasiperiodic motion and chaotic motion has the same behavior as the FLI. 

\subsection{The SSNs and the \textit{D}}\label{The SSN and the SD}
The stretching number $s_i$ is defined as (see \cite{Contopoulos96}, \cite{Contopoulos97a}, \cite{Contopoulos97}, \cite{Voglis98}):

$$s_i=\frac{1}{\delta_t}\ln\frac{|d_{\mathbf{x}}\mathbf{\Phi}^{t+i\cdot\delta_t}(\mathbf{w}(0))|}{|d_{\mathbf{x}}\mathbf{\Phi}^{t+(i-1)\cdot\delta_t}(\mathbf{w}(0))|}$$
where $d_{\mathbf{x}}\mathbf{\Phi}^{t+i\cdot\delta_t}(\mathbf{w}(0))=\mathbf{w}(t+i\cdot\delta_t)$, i.e. the tangent vector at time $t+i\cdot\delta_t$, with $i=1,2,\ldots,p$ and $p$ is the number of steps of integration $\delta_t$ (in case of mappings, the $\delta_t \equiv 1$) in which the interval of integration is divided. Therefore, the serie of $s_i$ with $i\rightarrow\infty$ converges to the lLCN. 

Another way to obtain dynamical information is analyzing the spectra of the $s_i$, i.e. the SSNs. However, when the sample to study is very large, the single analysis of the orbits by means of the SSNs is no longer reliable. Thus, in \cite{Voglis99} the authors introduce the \textit{D}. 

The SSNs are defined by the probability density of the values $s$ the $s_i$ can take:

$$S(s)=\frac{dZ(s,s+ds)}{Z\cdot ds},$$
with $Z$ the total number of $s_i$ and $dZ(s,s+ds)$ the number of $s_i$ in the interval $(s,s+ds)$.  

There are certain properties of the SSNs histograms that are worth mentioning:

\begin{enumerate}
\item The dynamical spectra $S(s)$ is invariant regarding the initial condition (hereafter, i.c.) along the same orbit (invariant with respect to time). 
\item The dynamical spectra $S(s)$ is invariant regarding the i.c. in a connected chaotic domain (invariant with respect to space).
\end{enumerate}

Finally, we define the \textit{D} like:

$$D^2=\sum_{s}[S_1(s)-S_2(s)]^2\cdot ds,$$
where $S_j(s)$ is the normalized number of $s_i$ corresponding to the unit i.d.v. $\mathbf{w}_j(0)$ which have values in the interval $(s,s+ds)$ (it is necessary to follow the evolution of two unit i.d.v. for each i.c. to compute the \textit{D}). 

Therefore, based on the properties of the SSNs, it is straighforward to see that if the orbit is regular, the \textit{D} tends to a constant non--zero value. If the orbit is chaotic, the \textit{D} decreases approaching to zero. 

\subsection{The SALI and the GALI}\label{The SALI}
In \cite{Skokos01} the author introduces the SALI defining the alignment indices. The parallel alignment index:

$$d_-=\|\mathbf{w}_1-\mathbf{w}_2\|$$
and the anti--parallel alignment index:

$$d_+=\|\mathbf{w}_1+\mathbf{w}_2\|$$
where the $\|\cdot\|$ is the usual euclidean norm. If the deviation vectors are normalized, then in case of chaotic motion: $d_-\rightarrow 0$ and $d_+\rightarrow 2$ (the deviation vectors eventually become aligned with the same direction) or $d_-\rightarrow 2$ and $d_+\rightarrow 0$ (with opposite directions). If the motion is regular, $d_-$ and $d_+$ oscillate within the interval (0,2). Therefore, the SALI is defined as the smaller of those indices:  

$$SALI(t)=min{(d_+,d_-)}.$$
The behavior of the SALI for chaotic orbits follows an exponential law:

$$SALI(t)\propto e^{-(\chi_1-\chi_2)\cdot t}$$
shown in \cite{Skokos04}, where $\chi_i$ with $i=1,2$ are the two largest LCEs of the orbit. If the orbit is regular, the SALI oscillates within the interval (0,2).

In \cite{Skokos07} the authors generalize the SALI introducing an alternative way to compute it. They evaluate the quantity:

$$P(t)=d_+\cdot d_-,$$
for each time step, where 

$$\frac{P(t)}{2}=\|\mathbf{w}_1\wedge\mathbf{w}_2\|$$
with $\mathbf{w}_1\wedge\mathbf{w}_2$ the wedge product of two deviation
vectors. This represents the area of the parallelogram formed by the two
deviation vectors $\mathbf{w}_1$ and $\mathbf{w}_2$. In fact, the wedge product can be generalized to represent the volume of a parallelepiped formed by the deviation vectors $\mathbf{w}_1,\mathbf{w}_2,\ldots,\mathbf{w}_k$ with $2< k\le 2N$, of an $N$--d.o.f. Hamiltonian system, or a $2N$--dimensional symplectic mapping. Therefore, the GALI$_k$ is defined as the volume of the $k$--parallelepiped formed by the $k$ initially linearly independent unit deviation vectors $\hat{w}_i(t)=\frac{\mathbf{w}_i(t)}{\|\mathbf{w}_i(t)\|}$, $i=1,2,\ldots,k$ and can be computed through the wedge product as follows:

\begin{equation}
GALI_k(t)=\|\hat{w}_1(t)\wedge\hat{w}_2(t)\wedge\ldots\wedge\hat{w}_k(t)\|.
\label{gali1}
\end{equation}
From Eq. \eqref{gali1} it is easy to see that if GALI$_k$(t) tends to zero, this implies that the volume of the parallelepiped having the unit i.d.v. as edges also shrinks to zero, as at least one of the deviation vectors becomes linearly dependent on the remaining ones. Thus, the behavior of the GALI$_k$(t) follows the exponential law:

$$GALI_k(t)\propto e^{-[(\chi_1-\chi_2)+(\chi_1-\chi_3)+\ldots+(\chi_1-\chi_k)]\cdot t},$$
where $\chi_1,\chi_2,\ldots,\chi_k$ are the $k$ largest LCEs of the orbit.  

On the other hand, if GALI$_k$(t) is different from zero as $t$ increases, the linear independence of the unit i.d.v. involved becomes clear. This second case corresponds to regular motion. Regular orbits move on tori of dimension $M\le N$. In general, the dimension of the tori is $N$ because there are no resonances involved. The number of resonances diminishes the dimensionality of the manifold and, for periodic motion, the orbit moves on an invariant curve, i.e. a manifold of dimension one: $M=1$. The behavior of the GALI$_k$ is given by (\cite{CB06}, \cite{Skokos07}, \cite{Skokos08}, \cite{MSA11}):

$$GALI_k(t)\propto\left\{\begin{array}{cl}constant&if\,2\le k\le M\\\frac{1}{t^{(k-M)}}&if\,M< k\le 2N-M\\\frac{1}{t^{2(k-N)}}&if\,2N-M< k\le 2N.\end{array}\right.$$

\subsection{The RLI}\label{The RLI}
The definition of the RLI is straightforward (see
\cite{Sandor04}). Consider the LI for a given i.c. $\mathbf{x}_0$ and
after $j$ steps of integration: $LI(\mathbf{x}_0;j)$. The RLI is thus
defined according to the LI difference for the ``base'' orbit and its ``shadow'': 

$$\Delta LI(\mathbf{x}_0;j)=\|LI(\mathbf{x}_0+\mathbf{\Delta x};j)-LI(\mathbf{x}_0;j)\|$$,
where $\mathbf{x}_0$ and $\mathbf{x}_0+\mathbf{\Delta x}$ are the i.c. for the base orbit and its shadow, respectively. 

The initial separation between the base orbit and its shadow, i.e. $\|\mathbf{\Delta x}\|$, is a free parameter. 

It is advisable to smooth the evolution curve of the LI for the base
orbit to eliminate fast fluctuations. Thus, we define the RLI as the
total average each step of the integration:

$$RLI(t)=\langle\Delta LI(\mathbf{x}_0)\rangle_t=\frac{1}{t}\sum^{t/\delta_t}_{i=1}\Delta LI(\mathbf{x}_0,i\cdot\delta_t),$$
$\delta_t$ being the step of integration. 

RLI values for chaotic motion are several orders of magnitude higher
than those associated with regular motion.


\section{A CIs' function for a Hamiltonian flow}\label{section_Flows}
There are important characteristics that make a CI suitable for a given study. In order to select the adequate CIs, it is useful to compare their performances according to these main aspects. Therefore, it is essential to study the resolving power, i.e. the CI's capability to distinguish chaotic orbits from regular orbits, to describe the levels of chaoticity and the levels of regularity and to identify periodicity and stickiness. Moreover, we need to analyze the integration time that the CI requires to identify clearly chaotic and regular orbits, i.e. the so--called speed of convergence. Finally, it is also important to study the CPU or computing times of the CIs, as we will see in Section \ref{cpu-times}. 

The resolving power and the speed of convergence tell us about the possibility of gathering detailed dynamical information in short integration times. The CPU times tell us about the limitations of the numerical implementation of the algorithms to compute the CIs. 

There are two ways of collecting this dynamical information. The first one is using the time evolution of the CI. The second one is by means of the CI's final value or its value at the end of the integration process. The information provided by the CI's time evolution is much more detailed than only a record of the final value. However, the study of large samples of CI's time evolution curves is not as efficient as the study of the corresponding final values despite the difference in the amount of information provided by each method. Thus, the use of the CI's time evolution is recommended to analyze small samples of orbits or to study particular interesting cases. On the other hand, we use the CI's final values to study extended regions of the phase space with large samples of orbits. 

The aim of the work done by \cite{Maffione11a} was finding a ``CIs' function'' (hereafter, CIsF) for mappings, which means a small package of CIs that represents the most efficient way to gather dynamical information. There, the authors dealt with many CIs' characteristics previously mentioned, such as the identification of chaotic, regular and sticky orbits, the sensitivity to hyperbolicity and stability levels, the speed of convergence and the reliability on the thresholds. However, they did not consider their capability to identify periodic orbits, their performances' dependence on i.d.v. and their computing times. In this section, we do not only extend their work considering a Hamiltonian flow, the $2$--d.o.f. HH potential, but we also increase the number of variational CIs in the comparison, including the OFLI and the GALI$_k$ with $2\le k\le 4$. 

In Section \ref{Thresholds study} we consider the robustness and reliability of the CIs' thresholds in distinguishing chaotic from regular motion. As we are interested in studying extended regions of the phase space, following the time evolution of the CIs for every single orbit proves unsuitable. Therefore, it is very important to count on thresholds that make a confident distinction between chaotic and regular motion. Final values are required for big samples of orbits whereas the time evolution of the CIs is the most efficient way to study small samples of orbits or individual orbits. In Section \ref{Qualitative study-statistical sample} we deal with the CIs' final values and in Section \ref{Qualitative study-representative groups} we deal with the CIs' time evolution. We analyze the capability of the CIs to identify periodicity (Section \ref{periodicity}) and the sensitivity of the methods on dynamic instabilities  (Section \ref{instabilities}) using the CIs' time evolution, and the CIs' dependence on the i.d.v. (Section \ref{initial deviation vectors}). Finally, in Section \ref{cpu-times} we briefly discuss the computing times of the different CIs.  

All the computations in the investigation were done using the following configuration. Hardware: CPU, 2 x Dual XEON 5450, Dual Core 3.00GHz; M.B., Intel S5000VSA; RAM, 4GB(4x1GB), Kingston DDR--2, 667MHz, Dual Channel. Software: gfortran 4.2.3. 

The computation of the CIs was accomplished by means of a single code we are developing, the so--called \textit{La Plata Variational Indicators Code} (LP--VIcode). The code uses the DOPRI8 routine. 

We use the following configuration for the rest of Section \ref{section_Flows} for the computation of the experiments unless stated otherwise: we record the data every $10^3$ units of time (hereafter u.t.) until a final integration time of $10^4$ u.t. The initial separation taken for the calculation of the RLI is $10^{-10}$ (\cite{Sandor04}). The \textit{D} is computed at intervals of $10^2$ time steps, and the number of cells considered for the generation of the histograms for the SSNs is $10^3$. The basis of unit i.d.v. (or simply i.d.v.) is the canonical basis. Let us recall that it is important to keep the same i.d.v. for the whole sample along the single experiments because some dependency of the CIs on the i.d.v. may be found (\cite{Froeschle00}). In Section \ref{Qualitative study-representative groups} we deal with such dependence. 

\subsection{The CIs' final values}\label{final values}
We start considering the reliability on the thresholds because they are essential to study the resolving power and the speed of convergence by means of the CIs' final values in Section \ref{Qualitative study-statistical sample}.
 
\subsubsection{The reliability on the thresholds}\label{Thresholds study}
The following study is undertaken on the 2D HH potential on an energy surface defined by $h=0.118$. We adopt a sample of $1.25751\times 10^5$ i.c. taken in the region defined by $x=0$, $y\in[-0.1:0.1]$ and $p_y\in[-0.05:0.05]$.

The well--known 2D HH system is described by the Hamiltonian:

$$H=\frac{1}{2}\left(p_x^2+p_y^2\right)+\frac{1}{2}\left(x^2+y^2\right)+x^2\cdot y-\frac{1}{3}y^3$$
where $x,y,p_x,p_y$ are the usual phase space variables.

We applied the time--dependent threshold of the LI (Table \ref{tablethreshold}) with $t$ as the time. It is known that an empirical adjustment of the LI's theoretical threshold is strongly advisable, in particular when studying large samples of orbits. Yet, our choice here is to avoid this fine tunning and to consider the raw theoretical estimation for the sake of a fair comparison. The critical value used for the RLI (Table \ref{tablethreshold}) was computed following \cite{Sandor04} and the remarks discussed in \cite{Maffione11a}. The way to determine the time--dependent threshold for the \textit{D} is not known yet, thus it is not considered in this section.

 In the case of the MEGNO, the threshold is a fixed value (Table
 \ref{tablethreshold}) which also needs an empirical
 adjustment. However, as we did for the LI, we use the theoretical
 fixed value.

 For the SALI there are two thresholds commonly used, namely, $10^{-12}$
 and $10^{-4}$ (see e.g. \cite{Skokos04}). In between, the orbits are
 called ``sticky chaotic''. Nevertheless, we consider them also as
 chaotic orbits. Then, the threshold to be analyzed is $10^{-4}$, which
 separates regular orbits from chaotic and sticky chaotic ones. Once
 again, this is done to avoid any advantages in taking more than one
 critical value.

 The threshold associated with the FLI or the OFLI (sometimes used with
 two thresholds also, see \cite{PFL08}) is time--dependent and follows
 the formulae presented in Table \ref{tablethreshold}. 

Finally, we define the thresholds for the GALIs. The HH potential is a
 2--d.o.f. Hamiltonian system and we can compute three GALIs: the GALI$_2$,
 the GALI$_3$, and the GALI$_4$. In order to establish their thresholds for
 large samples of orbits, we considered the formulaes for chaotic motion
 first: GALI$_2\propto e^{-\chi_1t}$, GALI$_3\propto e^{-2\chi_1t}$, and
 GALI$_4\propto e^{-4\chi_1t}$, where $\chi_1$ is the lLCE of the orbit
 (see \cite{CB06}, \cite{Skokos07} and \cite{MSA11} or Section \ref{The
 SALI} of this paper for further details). The lLCE was approximated by
 the LI, which has the threshold $ln(t)/t$ (see Table
 \ref{tablethreshold}). Then, the thresholds for the GALIs to study
 large samples of orbits are given in Table \ref{tablethreshold}. We are
 aware that the behaviors of the GALIs for regular motions change with
 the i.d.v. initially tangent to the torus. However, the selected time--dependent thresholds remain a good approximation to separate regular from chaotic motion. 

\begin{table}[!ht]\centering
\begin{tabular}{cc}
\hline\hline  \vspace*{-2ex} \\ 
\emph{} CI  & Threshold \vspace*{1ex} \\ 
\hline 
LI & $ln(t)/t$ \vspace*{1ex} \\
\hline 
RLI & $10^{-10}$ \vspace*{1ex} \\
\hline 
MEGNO & $2$ \vspace*{1ex} \\
\hline 
SALI & $10^{-4}$ \vspace*{1ex} \\
\hline 
FLI/OFLI & $t$ \vspace*{1ex} \\
\hline 
GALI$_2$ & $t^{-1}$ \vspace*{1ex} \\
\hline 
GALI$_3$ & $t^{-2}$ \vspace*{1ex} \\
\hline 
GALI$_4$ & $t^{-4}$ \vspace*{1ex} \\
\hline\hline  \vspace*{-4ex} 
\end{tabular}
\vspace{3mm}
\caption{Thresholds for the LI, the RLI, the MEGNO, the SALI, the FLI and the OFLI and the GALIs.}
\label{tablethreshold}
\end{table}

In \cite{Maffione11a} the authors tested the reliability of the thresholds of the CIs by evaluating the percentage variation of the chaotic component according to a change in their thresholds by $\pm1\%$. This change emulates a fine tunning of the critical value. Here, we repeat the experiment in the HH potential with a slight modification and we find similar results to those showed in the above mentioned work for mappings. 

First, we recorded the time evolution of the difference between the percentages of chaotic orbits according to a change in the threshold by $\pm1\%$ for the LI, and used it to measure the robustness of the other CIs' thresholds. That is, we computed the ratios between the CIs and the LI's percentage variation of the chaotic component according to the change above--mentioned. If the ratios are above one, it means that the robustness of the CIs' thresholds are worse than the LI's. On the other hand, ratios below one mean that the CIs' thresholds work better in the experiment than the ones associated with the LI. The variational CIs were introduced to improve the performances of the LI. Therefore, we expect ratios below one, and this is the case for many of the CIs. 

\begin{figure}[ht!]
\begin{center}
\begin{tabular}{cc}
\hspace{-5mm}\resizebox{63mm}{!}{\includegraphics{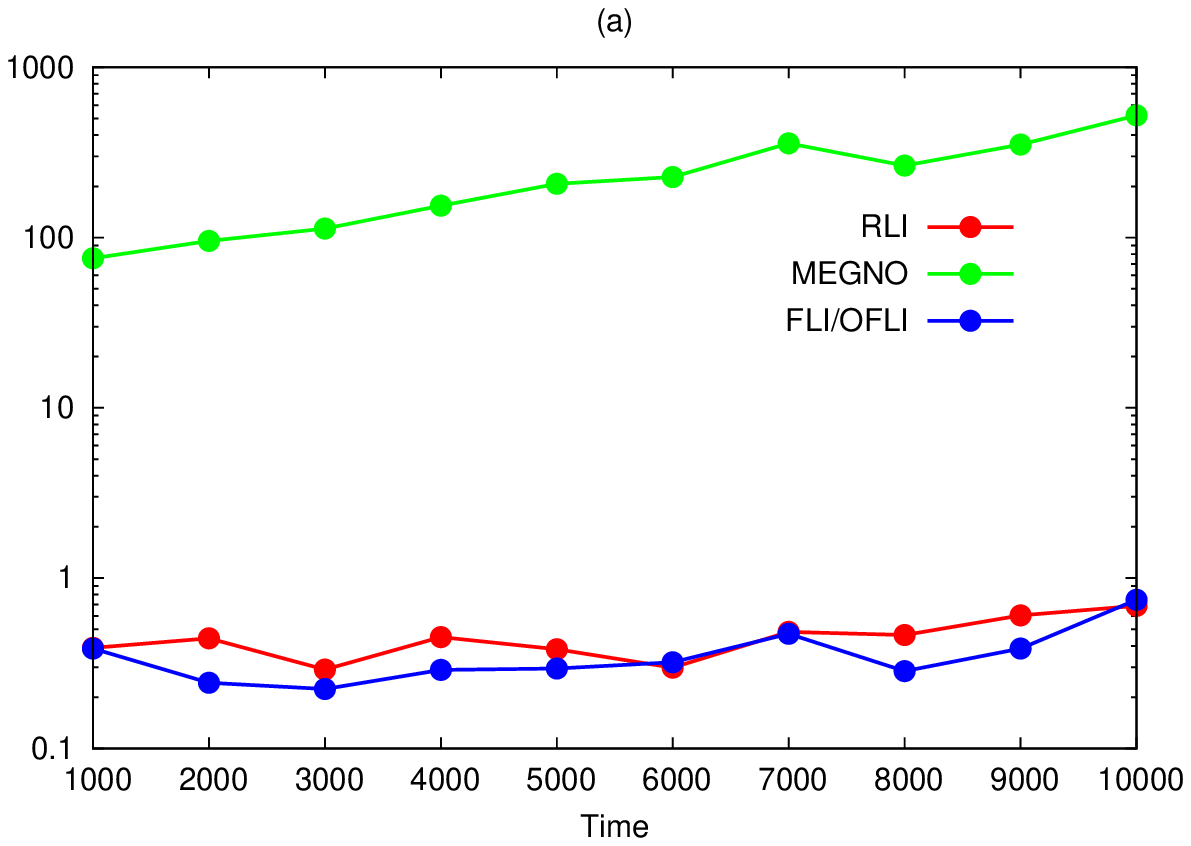}}& 
\hspace{-5mm}\resizebox{63mm}{!}{\includegraphics{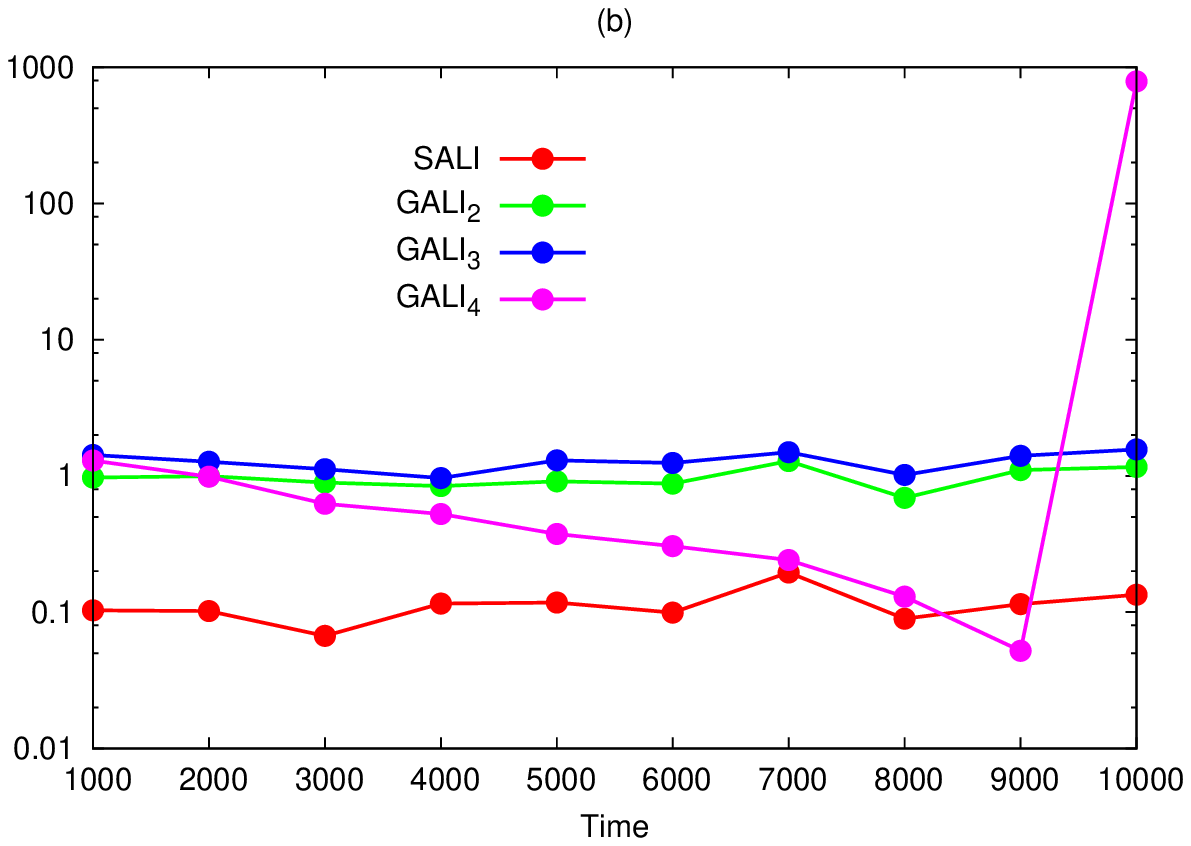}}\\
\hspace{-5mm}\resizebox{63mm}{!}{\includegraphics{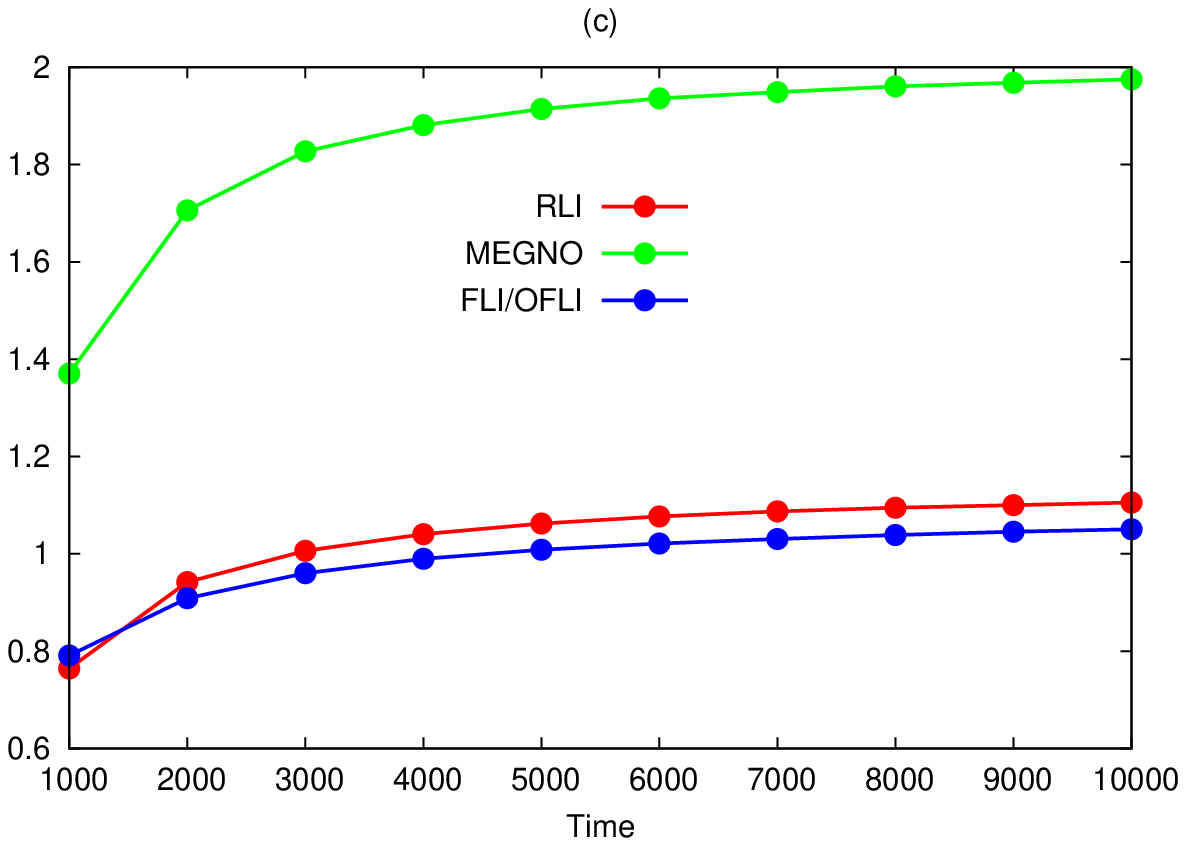}}& 
\hspace{-5mm}\resizebox{63mm}{!}{\includegraphics{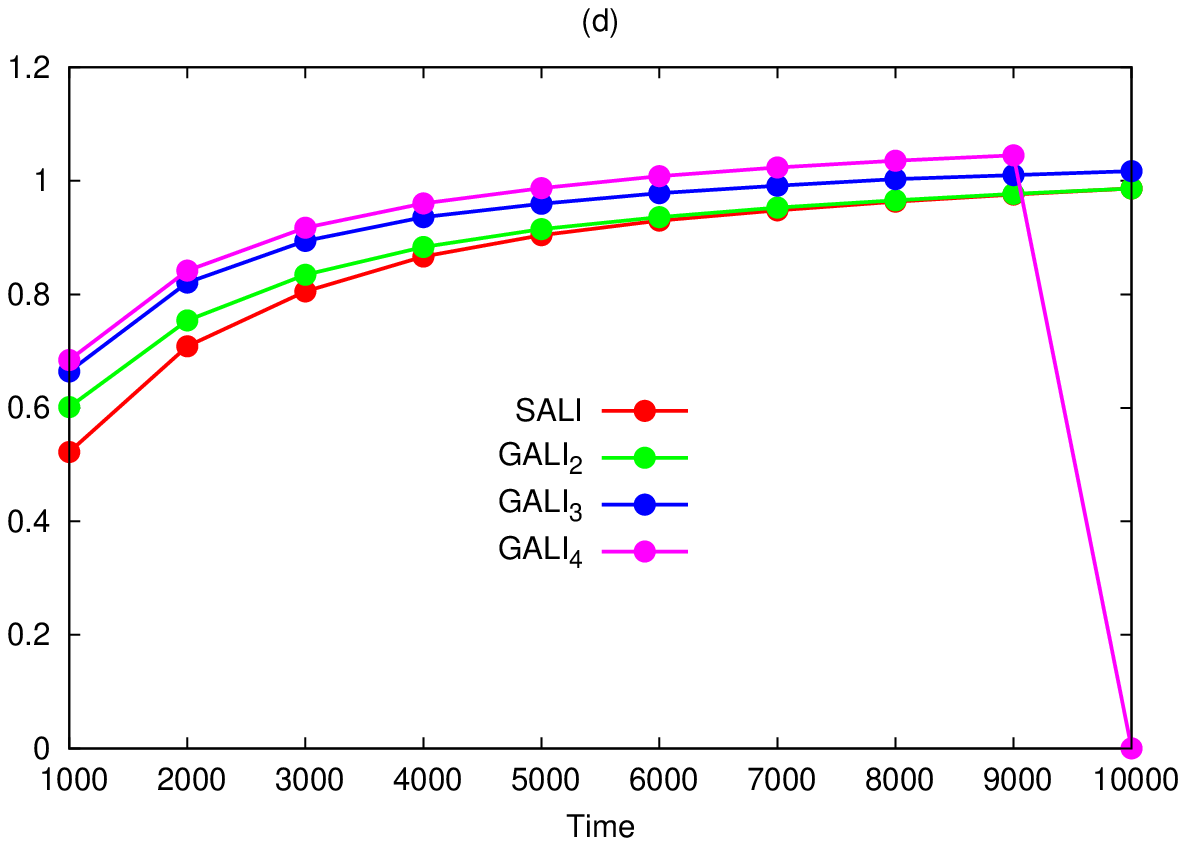}}
\end{tabular}
\caption{The normalized time evolution of the difference between the percentages of chaotic orbits according to a change in the thresholds of Table \ref{tablethreshold} by $\pm1\%$ (a) for the RLI, the MEGNO and the FLI/OFLI, and (b) for the SALI, the GALI$_2$, the GALI$_3$ and the GALI$_4$. The time evolution of the percentage of chaotic orbits according to the thresholds of Table \ref{tablethreshold} normalized by $\sim 39.92$\% (c) for the RLI, the MEGNO and the FLI/OFLI and (d) for the SALI, the GALI$_2$, the GALI$_3$ and the GALI$_4$. See text for further details.}
\label{thresholds}
\end{center}
\end{figure}

The normalized time evolution of the difference between the percentages of chaotic orbits according to a change in the thresholds by $\pm1\%$ is shown on the top panels of Fig. \ref{thresholds}. The weakness of the theoretical fixed threshold for the MEGNO becomes evident (green linepoints on the top left panel of Fig. \ref{thresholds}). This weakness is a logical consequence of the asymptotic nature of the threshold. On the other hand, we find that the RLI and the SALI have very reliable thresholds (i.e. ratios below 1), despite their empirical nature (red linepoints on the top left and top right panels of Fig. \ref{thresholds} for the RLI and the SALI, respectively), and the FLI/OFLI and the GALI$_4$, as well (blue linepoints on the top left panel of Fig. \ref{thresholds} for the FLI/OFLI, and green, blue and violet linepoints on the top right panel of Fig. \ref{thresholds} for the GALI$_2$, the GALI$_3$ and the GALI$_4$, respectively). The GALI$_4$ starts with a value higher than one, but after a short initial transient, its rate decreases below one.

Now we consider the percentages of chaotic orbits. The value of a final percentage of chaotic orbits is not as important as the approximation rate to that value. The former can be fixed by an empirical adjustment. The latter is a combination of the threshold's reliability and the efficiency of the indicator. Therefore, it is not easy to adjust in order to improve the results. Notwithstanding the relative importance of the CIs' final percentages of chaotic orbits, we would like to test the reliability of the thresholds given in Table \ref{tablethreshold}. We consider the percentage of chaotic orbits given by the LI by $10^4$ u.t. (i.e. $\sim 39.92$\%) the ``true percentage'' of chaotic orbits. We support this idea because the LI is the most tested CI in the literature and $10^4$ u.t. is a confident convergent time for this indicator.

To take these considerations into account we are going to normalize the time evolution of the percentage of chaotic orbits according to every method with the ``true percentage'' given by the LI. Thus, values higher than 1 show percentages of chaotic orbits higher than the ``true percentage'' given by a confident value of the LI.
 
The time evolution of the percentage of chaotic orbits according to the thresholds of Table \ref{tablethreshold} normalized by $\sim 39.92$\% is shown on the bottom panels of Fig. \ref{thresholds}. 
The disagreement with the LI's final percentage of chaotic orbits (see the bottom left panel of Fig. \ref{thresholds}) forces an immediate empirical adjustment of the MEGNO's threshold. Actually, on considering the recommendations of \cite{Maffione11}, a threshold value close to $3$ seems to be more appropriate. The SALI has the robustest threshold according to the top right panel of Fig. \ref{thresholds} (red linepoints), as we have just seen. However, the convergency of both SALI and GALI$_4$'s to the LI's final percentage of chaotic orbits (see the red and violet linepoints on the bottom right panel of Fig. \ref{thresholds} for the SALI and the GALI$_4$, respectively) is slower than the convergency of the RLI and the FLI/OFLI (see the red and blue linepoints on the bottom left panel of Fig. \ref{thresholds} for the RLI and the FLI/OFLI, respectively). The final percentages of the RLI and the FLI/OFLI are higher than that of the LI, though. Nevertheless, the final percentage of chaotic orbits can be fixed with a small adjustment of their thresholds, as previously mentioned.

Thus, among the above--mentioned CIs, the RLI and the FLI/OFLI show the most reliable thresholds in the experiment and, on comparison, the FLI/OFLI's threshold seems to work better.

By $10^4$ u.t. the threshold taken for the GALI$_4$ (i.e. $t^{-4}$) reaches the computer's precision of the computer ($10^{-16}$) and thus, every chaotic orbit lies beyond such precision. Therefore, the last point on the right panels of Fig. \ref{thresholds} (violet linepoints) falls apart from the tendency.

\subsubsection{Qualitative study of a sample of orbits through the CIs' final values}\label{Qualitative study-statistical sample}
In this section we only use the information provided by the CIs' final values and times of saturation to test their speed of convergence and resolving power. 

The FLI and the OFLI, the SALI and the GALIs increase or decrease exponentially fast for chaotic motion (see Section \ref{The FLI} for the FLI/OFLI and Section \ref{The SALI} for the SALI/GALIs, and references therein). If the chaotic nature of an orbit is well characterized when the CI reaches a certain value, it is worthless to continue the calculation, and the integration process should be stopped. Such value is a saturation value and an immediate consequence of that saturation value are the times of saturation. The times of saturation are the time by which the CIs' final values reach the corresponding saturation value. The times of saturation recover the hyperbolicity levels of the chaotic component (see \cite{Skokos07} and \cite{Maffione11a}).
   
The parameters used for the computation of the CIs and the thresholds (see Table \ref{thresholds}) are the same as those applied in Section \ref{Thresholds study}. We present the performances of the CIs on the sample of $1.25751\times 10^5$ i.c. on the energy surface defined by $h=0.118$ constraining the values of $x$, $y$ and $p_y$ to $x=0$, $y\in[-0.1:0.1]$ and $p_y\in[-0.05:0.05]$ (the same used in Section \ref{Thresholds study}) with an integration time of $10^4$ u.t. 

\begin{figure}[ht!]
\begin{center}
\begin{tabular}{cc}
\hspace{-5mm}\resizebox{63mm}{!}{\includegraphics{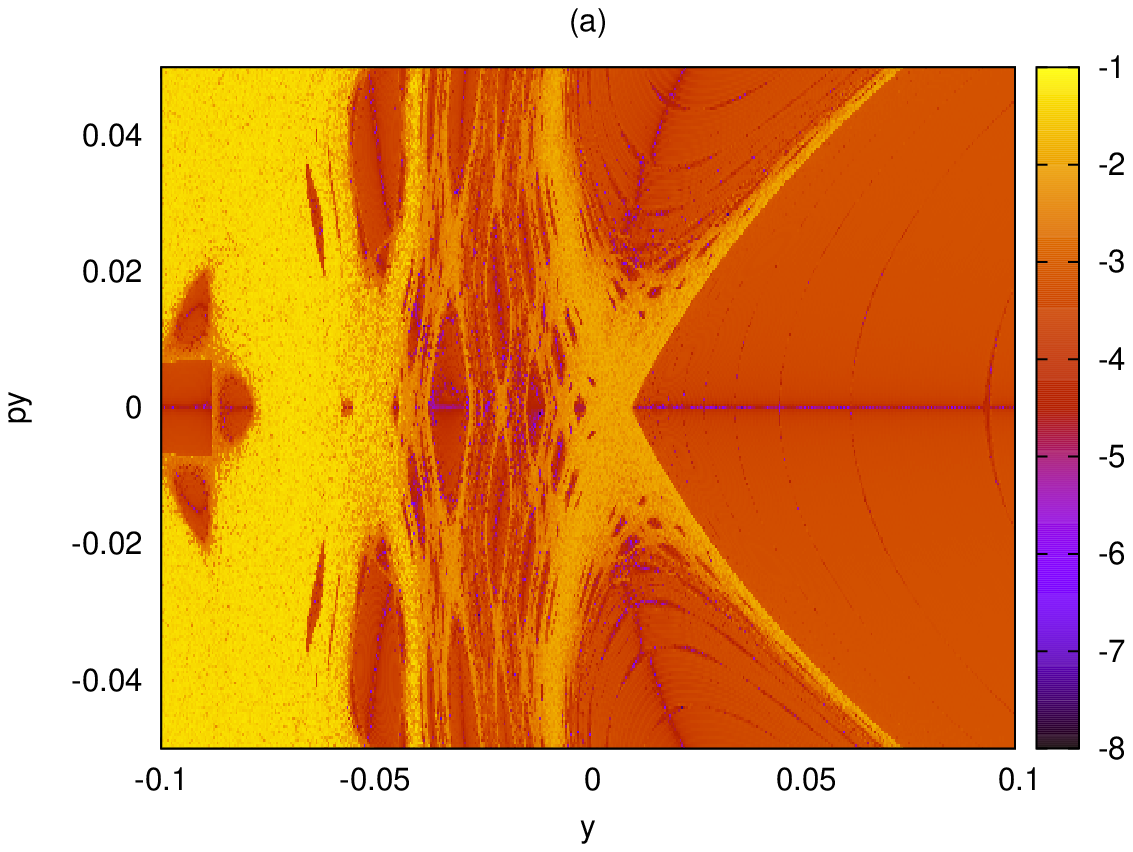}}&
\hspace{-5mm}\resizebox{63mm}{!}{\includegraphics{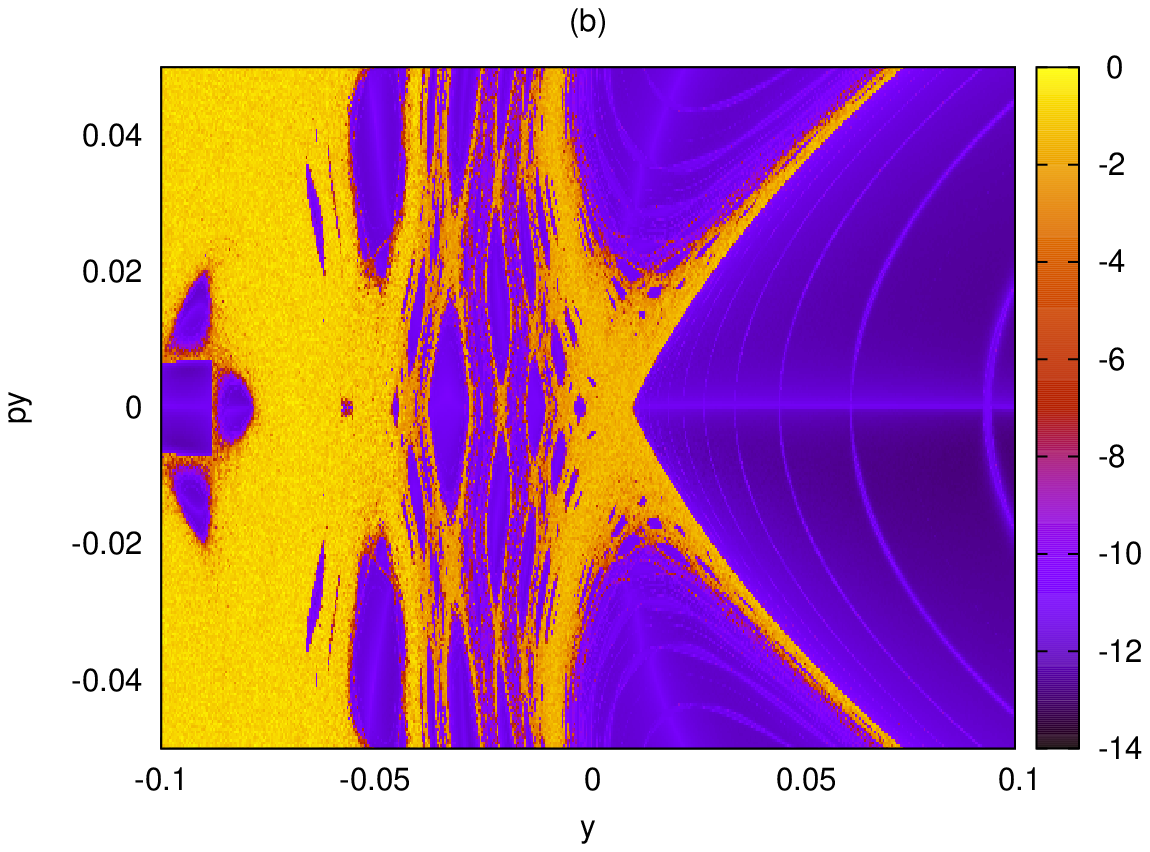}}\\
\hspace{-5mm}\resizebox{63mm}{!}{\includegraphics{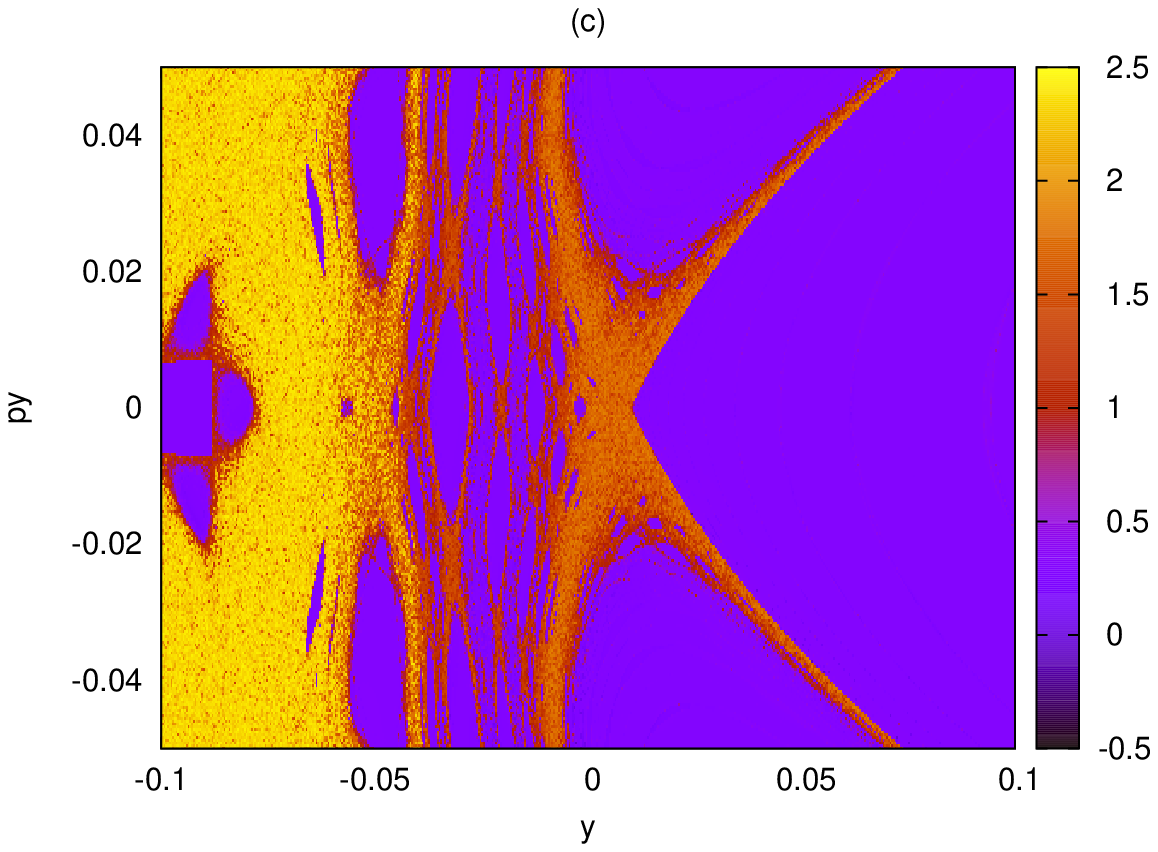}}&
\hspace{-5mm}\resizebox{63mm}{!}{\includegraphics{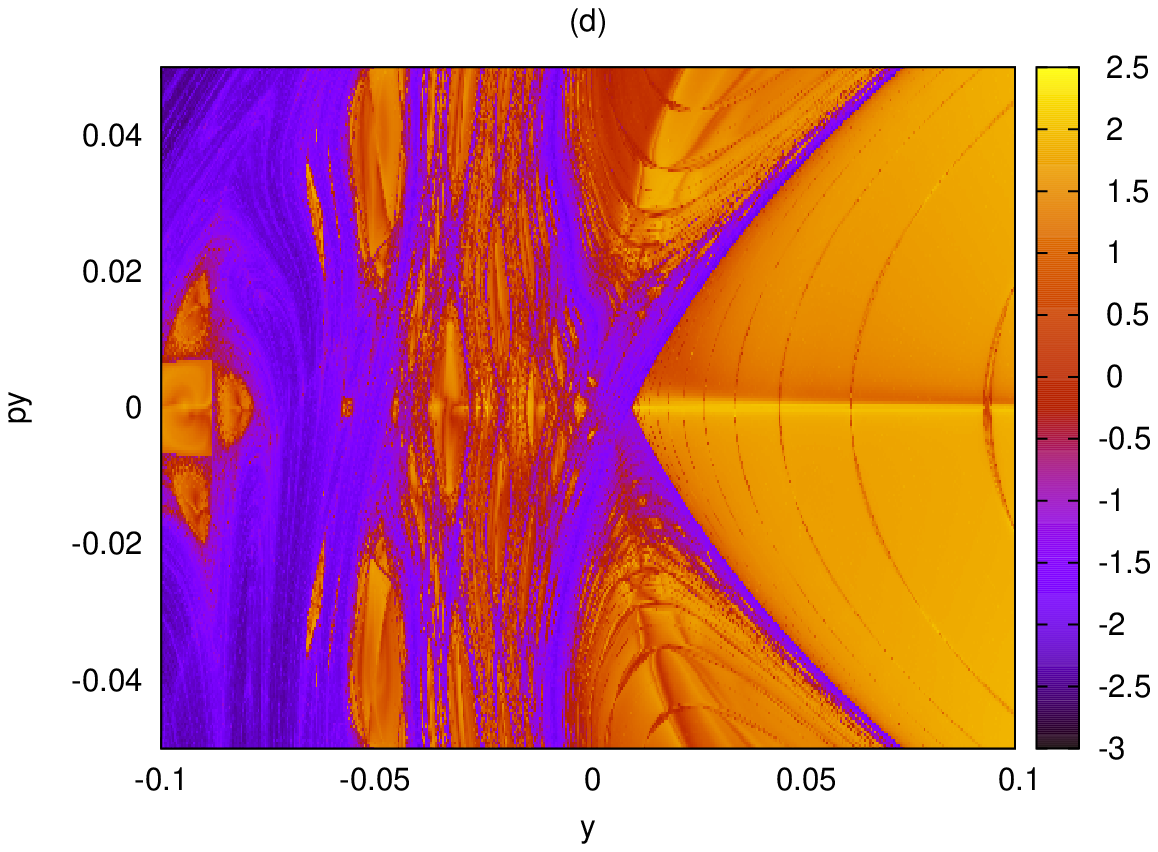}}\\
\hspace{-5mm}\resizebox{63mm}{!}{\includegraphics{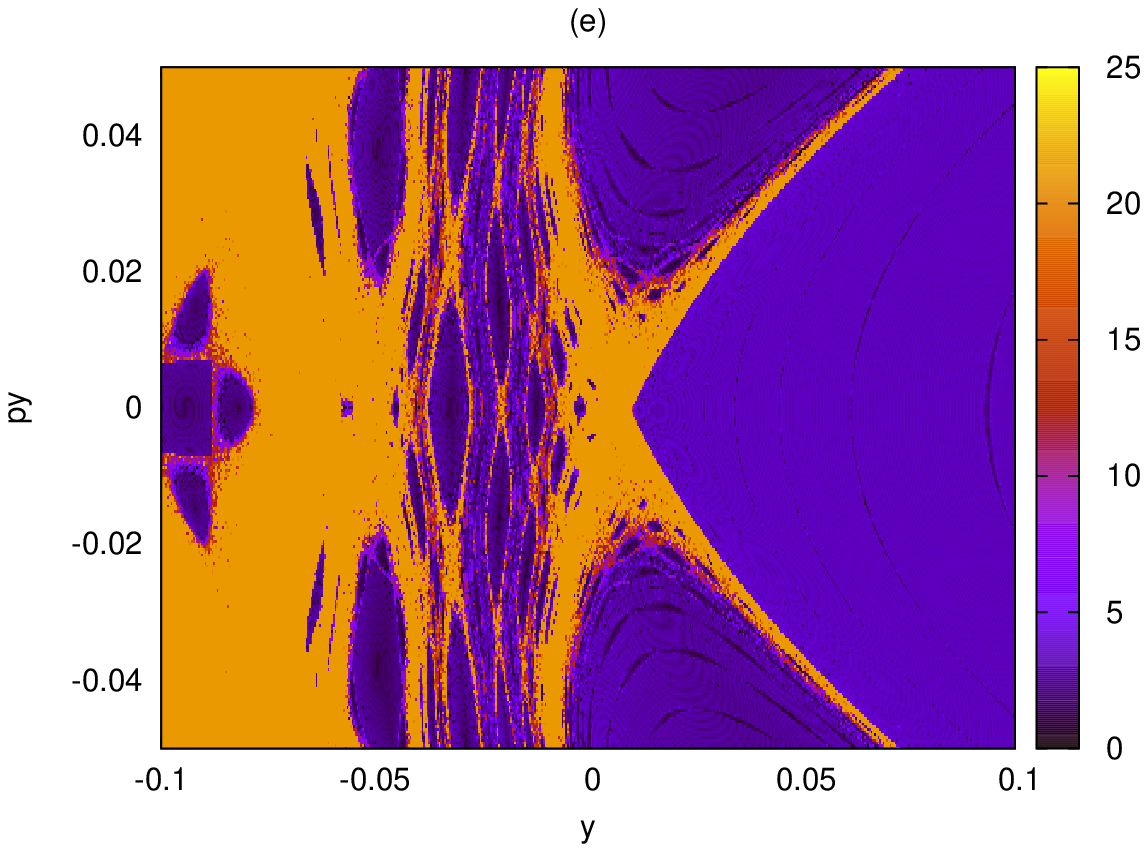}}&
\hspace{-5mm}\resizebox{63mm}{!}{\includegraphics{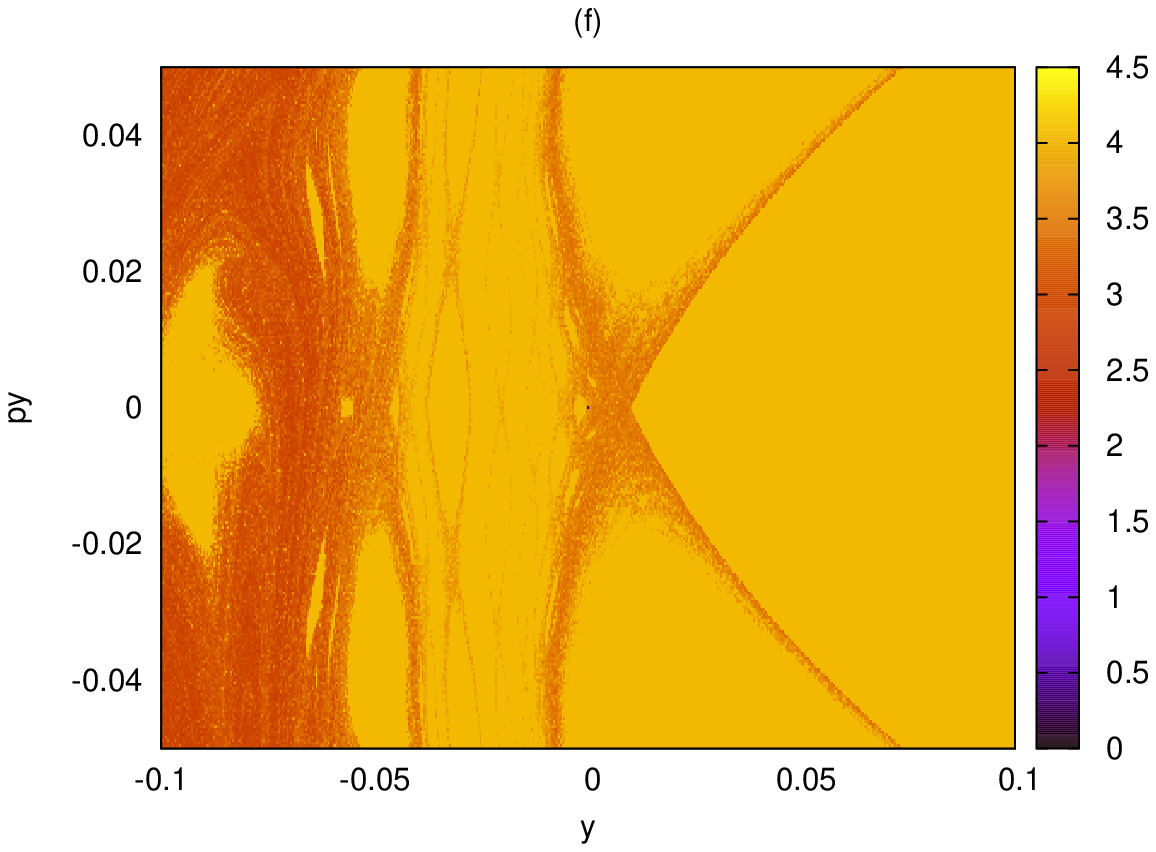}}
\end{tabular}
\caption{Phase space portraits with $1.25751\times 10^5$ i.c. in the region defined by $x=0$, $y\in[-0.1:0.1]$ and $p_y\in[-0.05:0.05]$ in the HH potential on an energy surface $h=0.118$. The phase space portraits are computed by $10^4$ u.t., and the color scale represents the final value of each indicator, (a) for the LI, (b) the RLI, (c) the MEGNO, (d) the \textit{D} and the OFLI. In case of the OFLI we include (e) the final values and (f) the corresponding times of saturation by a final time of integration of $10^4$ u.t. in logartithmic scale.}
\label{li-rli-mengo-sd-ofli-maps}
\end{center}
\end{figure}

\begin{figure}[ht!]
\begin{center}
\begin{tabular}{cc}
\hspace{-5mm}\resizebox{63mm}{!}{\includegraphics{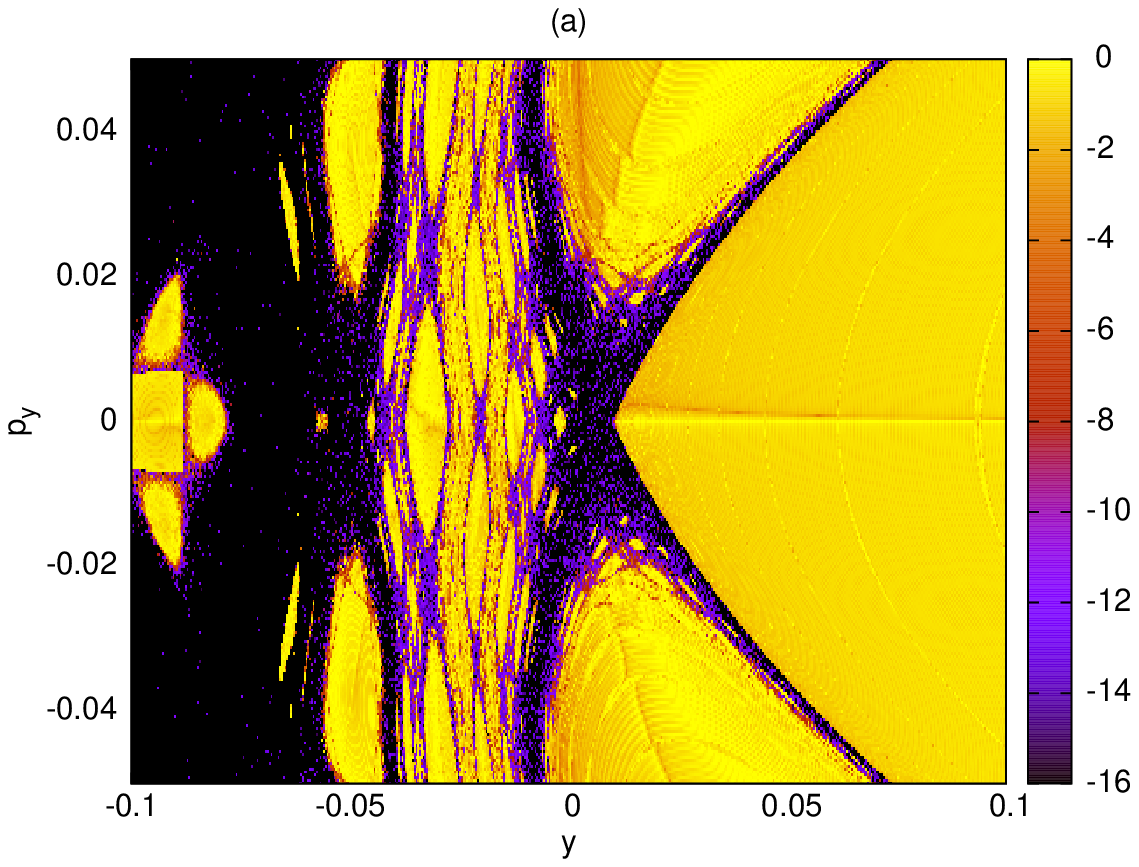}}&
\hspace{-5mm}\resizebox{63mm}{!}{\includegraphics{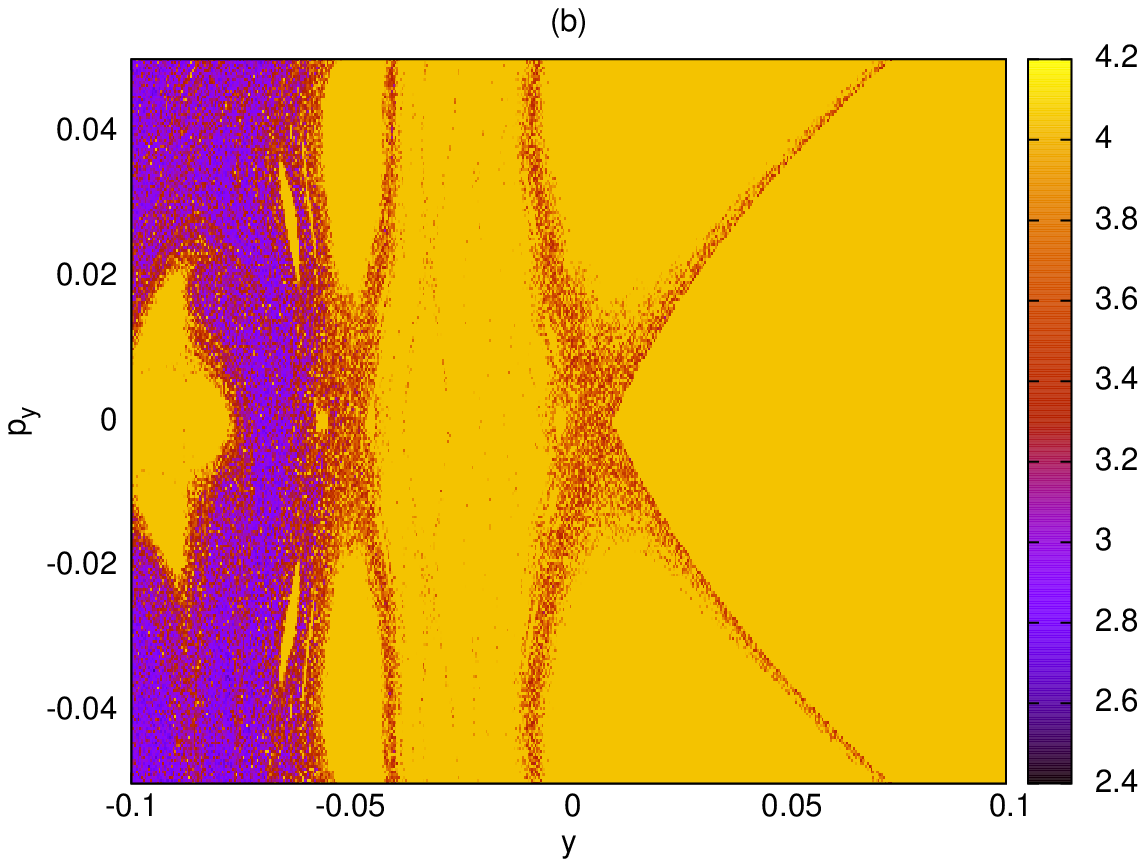}}\\
\hspace{-5mm}\resizebox{63mm}{!}{\includegraphics{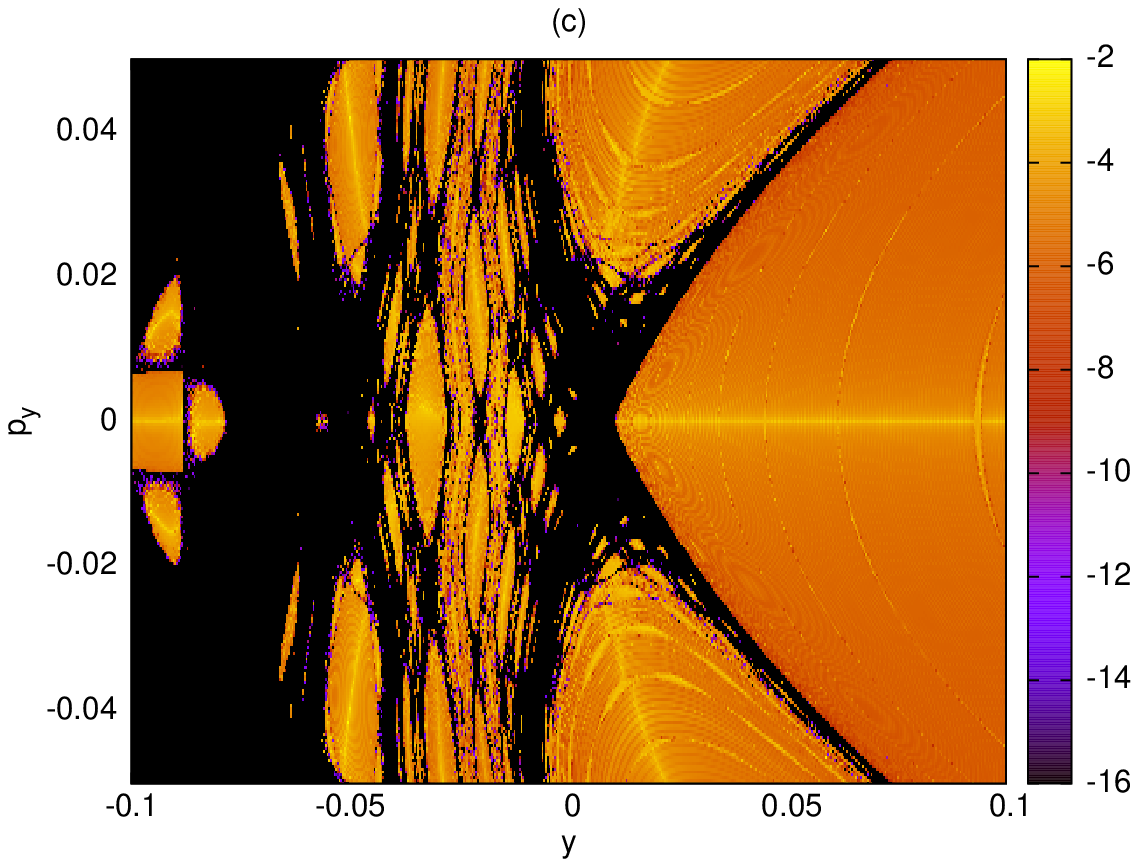}}&
\hspace{-5mm}\resizebox{63mm}{!}{\includegraphics{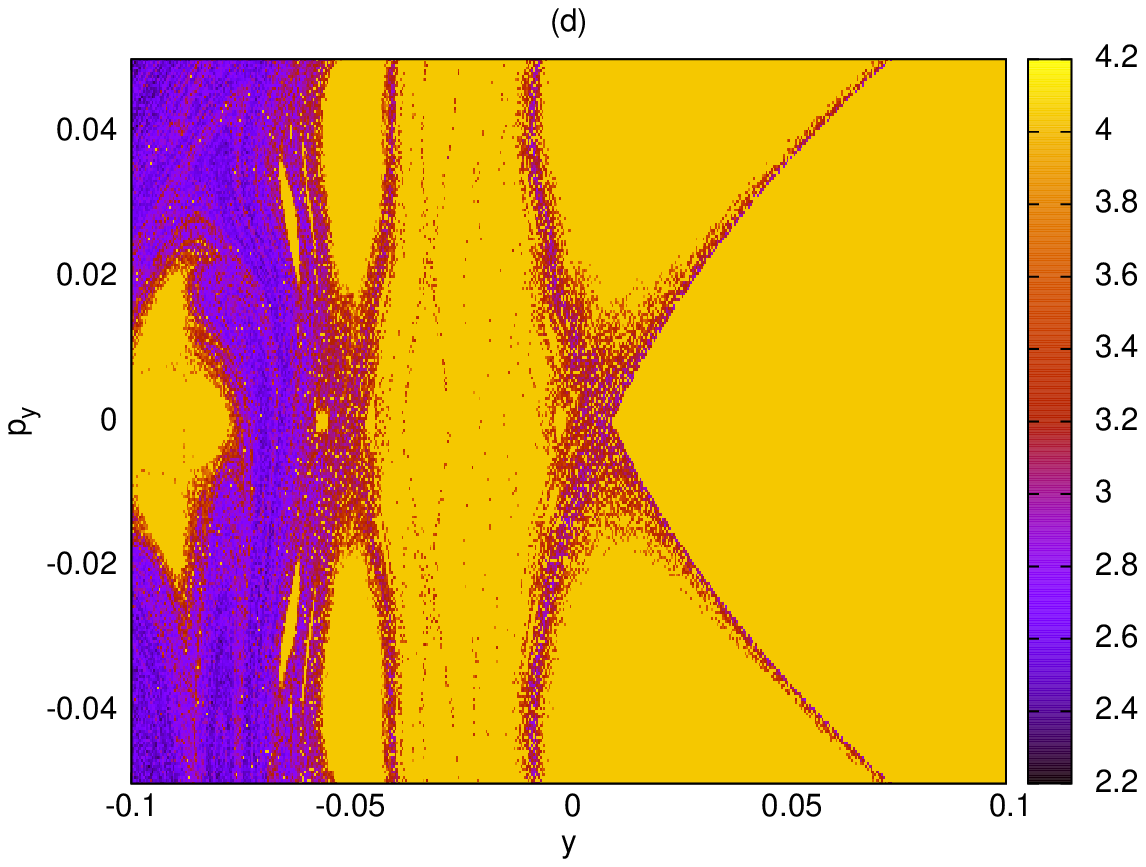}}\\
\hspace{-5mm}\resizebox{63mm}{!}{\includegraphics{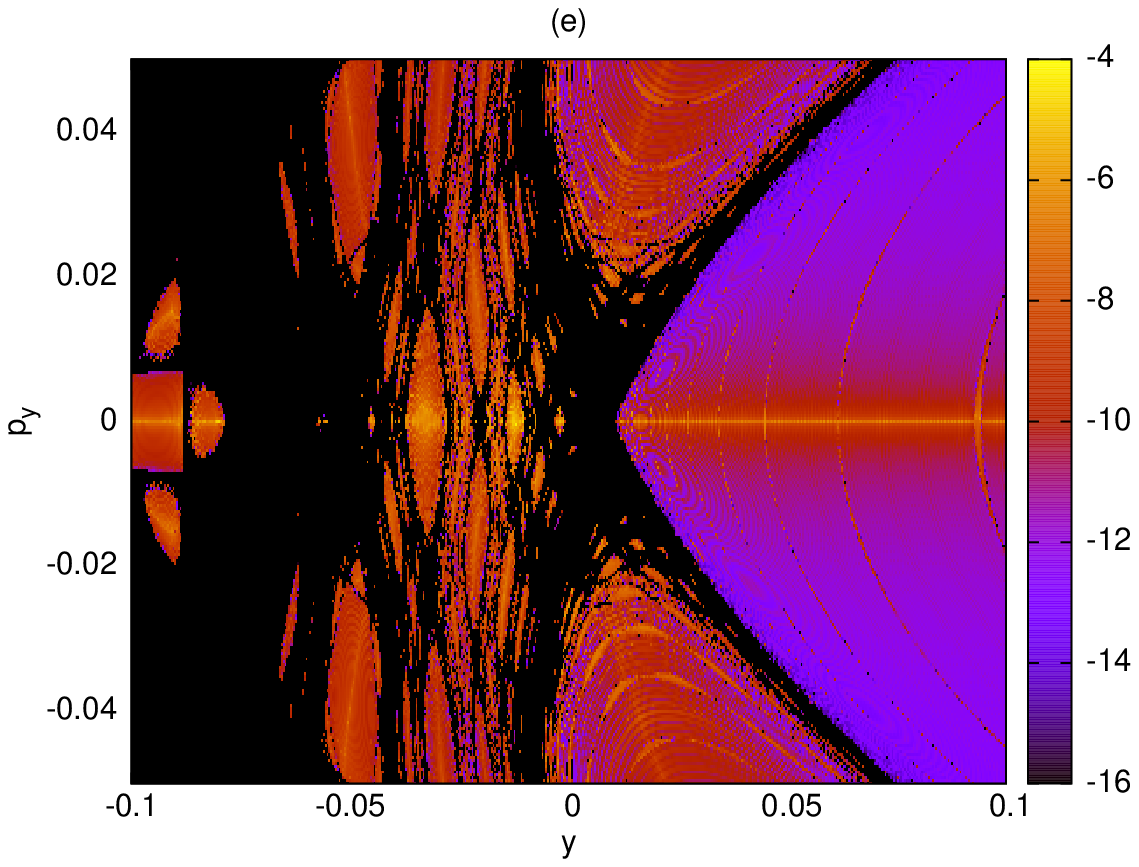}}&
\hspace{-5mm}\resizebox{63mm}{!}{\includegraphics{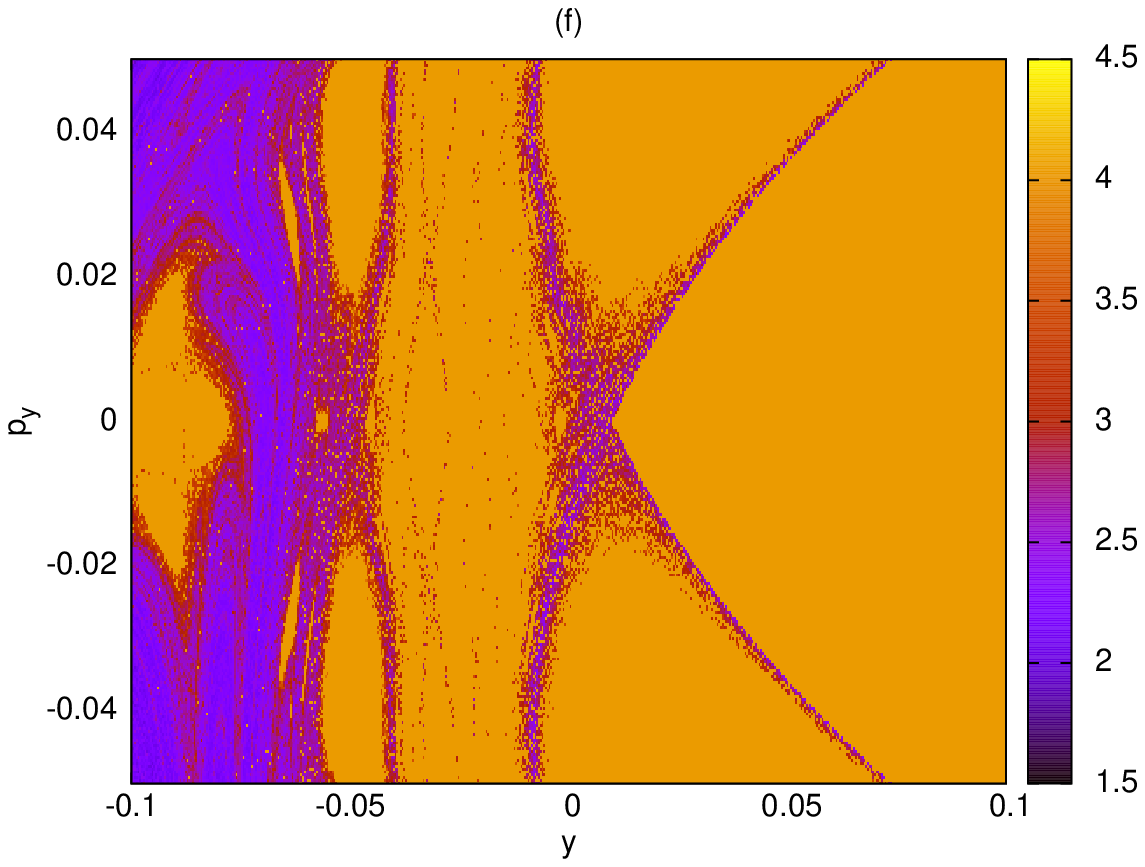}}
\end{tabular}
\caption{Left panels: phase space portraits of the GALI method with $1.25751\times 10^5$ i.c. in the region defined by $x=0$, $y\in[-0.1:0.1]$ and $p_y\in[-0.05:0.05]$ in the HH potential on an energy surface $h=0.118$. (a), (c), (e) for the phase space portraits by $10^4$ u.t. of the GALI$_2$, the GALI$_3$ and the GALI$_4$, respectively. (b), (d), (f) for the corresponding times of saturation by a final time of integration of $10^4$ u.t. in logarithmic scale. The color scale has the same meaning as in Fig. \ref{li-rli-mengo-sd-ofli-maps}}
\label{ofli-gali-maps}
\end{center}
\end{figure}

Figure \ref{li-rli-mengo-sd-ofli-maps} shows the performances of the LI and the RLI (top left and top right panels, respectively), the MEGNO and the \textit{D} (middle left and middle right panels, respectively) and the OFLI's final values and times of saturation (bottom left and bottom right panels, respectively). The left panels of Fig. \ref{ofli-gali-maps} show the phase space portraits of the GALI method for the above--mentioned region. The right panels of Fig. \ref{ofli-gali-maps} show the corresponding times of saturation (from top to bottom panels, the GALI$_2$, the GALI$_3$ and the GALI$_4$, respectively).

The FLI and the SALI have not been included because they have similar performances to those of the OFLI and the GALI$_2$, respectively.

First, we deal with the identification of two domains in the chaotic component, namely the chaotic sea for $y\lesssim -0.05$ and the stochastic layers surrounding the main stability islands. We show that the CIs can be divided into two groups according to the different ways of describing such a component. In the first group we include the LI, the MEGNO and the \textit{D}; in the second group, the RLI, the OFLI and the GALIs. 

Both the LI (top left panel of Fig. \ref{li-rli-mengo-sd-ofli-maps}) and the MEGNO (middle left panel of Fig. \ref{li-rli-mengo-sd-ofli-maps}) identify the two domains with their final values. However, the MEGNO gives a clear distinction between them. The \textit{D} (middle right panel of Fig. \ref{li-rli-mengo-sd-ofli-maps}) does not show such a clear distinction between both chaotic domains, but it shows a structure in the chaotic sea which is not reflected in the portraits of the LI or the MEGNO.

In the second group, the RLI (top right panel of Fig. \ref{li-rli-mengo-sd-ofli-maps}) shows a slight difference between the colours used to identify the two domains due to its high speed of convergence for chaotic motion. The OFLI (bottom left panel of Fig. \ref{li-rli-mengo-sd-ofli-maps}) and the GALIs (left panels of Fig. \ref{ofli-gali-maps}) make no distinction at all, because of the same high speed of convergence for chaotic orbits. However, the OFLI and the GALIs count with the times of saturation to recover the hyperbolicity levels (bottom right panel of Fig. \ref{li-rli-mengo-sd-ofli-maps} for the OFLI and right panels of Fig. \ref{ofli-gali-maps} for the GALIs). Thus, the distinction between both chaotic domains is perfectly seen. Moreover, the structure seen by the \textit{D} in the chaotic sea is also recovered.

The OFLI and the GALIs show the best way to describe the chaotic component thanks to the combination of a high speed of convergence and their times of saturation. Nevertheless, a similar quantity to the times of saturation could be defined for the RLI.
  
Second, we deal with the description of the regular component and show that the CIs can be divided into two groups again. The division comes from the appearance of spurious structures, which might be due to different reasons: numerical artifacts, short final integration times and a poor selection of initial conditions of the variational equations (see \cite{Barrio09} for a thorough discussion). 

The LI, the RLI and the \textit{D} show spurious structures in the regular component (Fig. \ref{li-rli-mengo-sd-ofli-maps}). Those structures are also seen in the GALIs' phase space portraits (Fig. \ref{ofli-gali-maps}). These structures are seen as lines cutting the islands of stability in half. 

On the other hand, neither the MEGNO (because of its universal value for quasiperiodic motion, $\sim 2$, it does not show any structure for the regular component) nor the OFLI presented spurious structures in the experiment. 

Finally, we deal with the separation of both chaotic and regular components and the speed of convergence. The distinction between domains of chaotic and regular orbits is successfully drawn by every single method of the package. However, the OFLI and the GALIs seem to make the distinction faster. 

In the case of the GALI$_k$, as we increase the index $k$ of the method (from top to bottom panels of Fig. \ref{ofli-gali-maps}), we find more detailed descriptions of the regular and chaotic components by means of the GALIs' final values and the corresponding times of saturation, respectively. Furthermore, the GALI$_4$ has the highest speed of convergence in the chaotic orbits. 

Therefore, a combination of the FLI/OFLI and the GALI$_4$ seems to carry enough information to successfully describe the strongly divided phase space of the studied region in the HH potential by means of their final values and times of saturation. 

\subsection{Study of some representative orbits through the CIs' time evolution}\label{Qualitative study-representative groups}
In \cite{Maffione11a}, Section 3, the authors used the CIs' time evolution to study the speed of convergence and the sensitivity to distinguish different kinds of motions. The results found by \cite{Maffione11a} for mappings remain the same for the HH potential. Therefore, we deal with some CIs' features not discussed in their work. We start studying the ability of the CIs to distinguish periodic orbits and their CIs' dependency on the i.d.v. Then, we deal with the sensitivity of the GALIs to identify instabilities.

We study a small sample of six orbits in the HH potential: 

\begin{itemize}
 \item Orbit \textbf{(a)}: the 5--periodic orbit with i.c.: $(x^{0(a)},y^{0(a)},p_x^{0(a)},p_y^{0(a)})\sim(0,0.35207,0,0.14979)$, and two close orbits on the same energy surface, $h=0.125$. The close orbits are defined by $\Delta y=0,00793$ and $\Delta y=0.02793$, $\Delta y$ being the difference in the y--component of the i.c. of orbit \textbf{(a)} and the y--component of the i.c. of the close orbits. They were taken from \cite{MSA11}. 
 \item Orbit \textbf{(b)}: a quasiperiodic orbit with i.c.: $(x^{0(b)},y^{0(b)},p_x^{0(b)},p_y^{0(b)})\sim(0,0,0.5,0)$, which is associated with the 1--periodic orbit in the HH potential on the energy surface $h=0.125$. The orbit was taken from \cite{Skokos07}.  
 \item Orbit \textbf{(c)}: a regular orbit close to the separatrix with i.c.: $(x^{0(c)},y^{0(c)},p_x^{0(c)},p_y^{0(c)})\sim(0,0.5085,0.25512,0)$. 
 \item Orbit \textbf{(d)}: a chaotic orbit inside the chaotic sea with i.c.: $(x^{0(d)},y^{0(d)},p_x^{0(d)},p_y^{0(d)})\sim(0,0.6,0.14142,0)$. 
\end{itemize}

The last two orbits belong to an energy surface of $h=0.118$ and were taken from \cite{Cincotta99}.

The final integration time is $2.4\times 10^4$ u.t. (unless stated otherwise) and ensures the convergency of the LI for both energy surfaces to properly determine the regular or chaotic nature of the orbits. 

\subsubsection{Identification of periodic orbits}\label{periodicity}
We start with the identification of periodic orbits. Thus, we evaluate the CIs in the orbit \textbf{(a)} and the quasiperiodic orbits close to \textbf{(a)}. The behaviors of the LI, the SSNs, the MEGNO, the FLI, the GALI$_3$ and the GALI$_4$ are nearly the same for the three orbits.

\begin{figure}[ht!]
\begin{center}
\begin{tabular}{cc}
\hspace{-5mm}\resizebox{63mm}{!}{\includegraphics{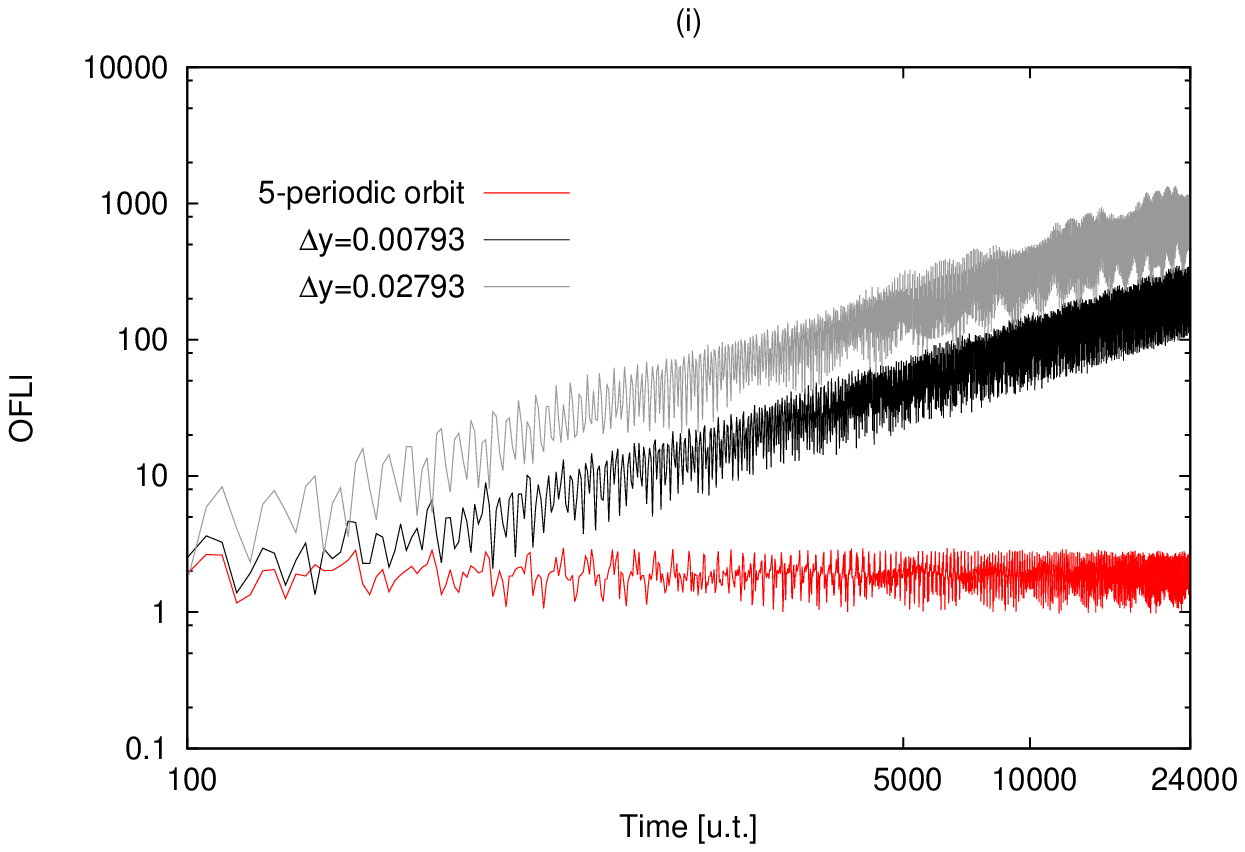}}& 
\hspace{-5mm}\resizebox{63mm}{!}{\includegraphics{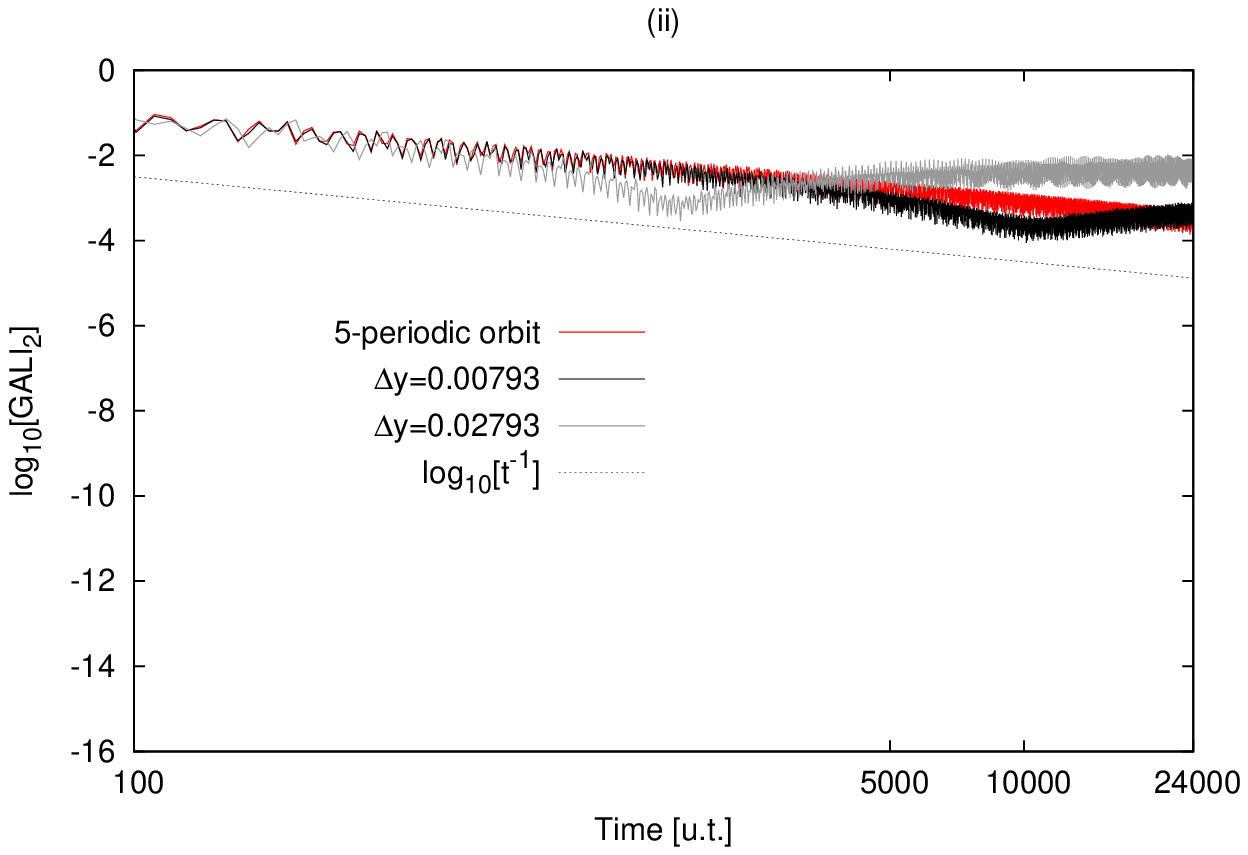}}\\
\hspace{-5mm}\resizebox{63mm}{!}{\includegraphics{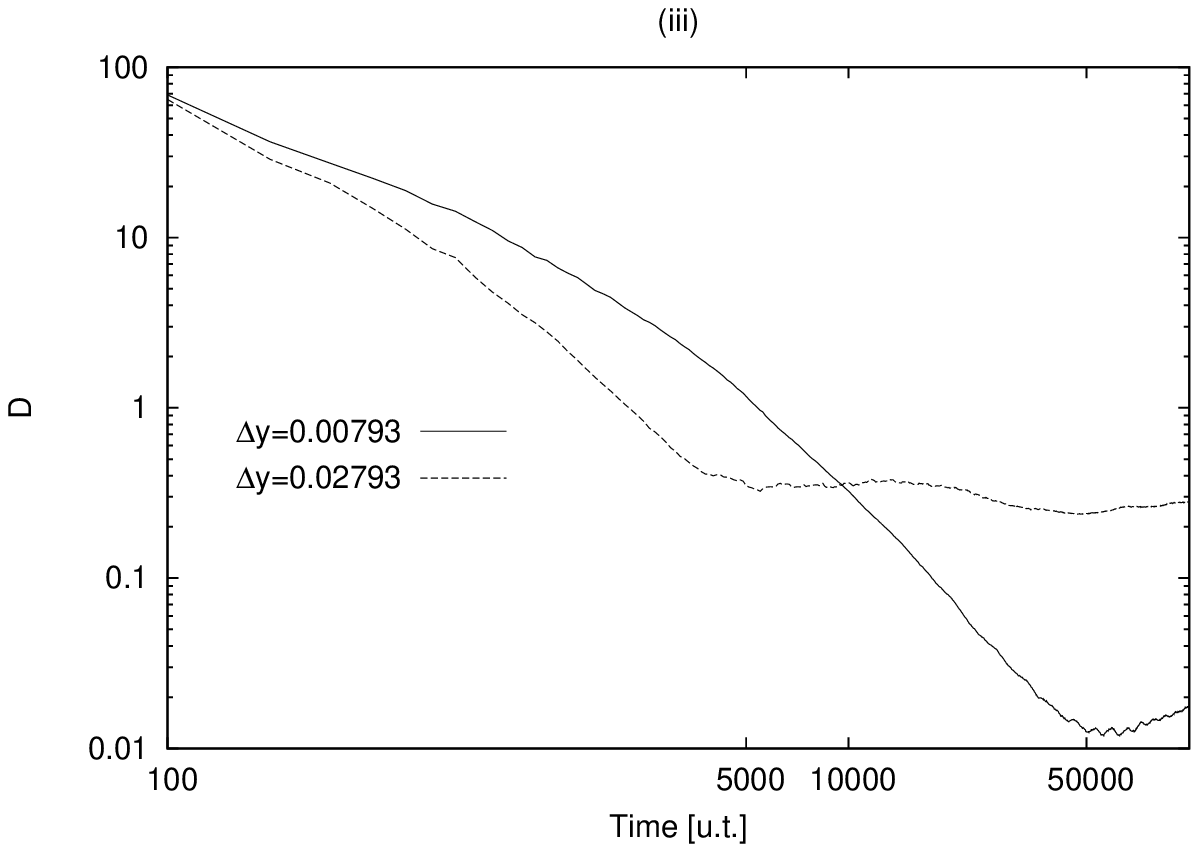}}& 
\hspace{-5mm}\resizebox{63mm}{!}{\includegraphics{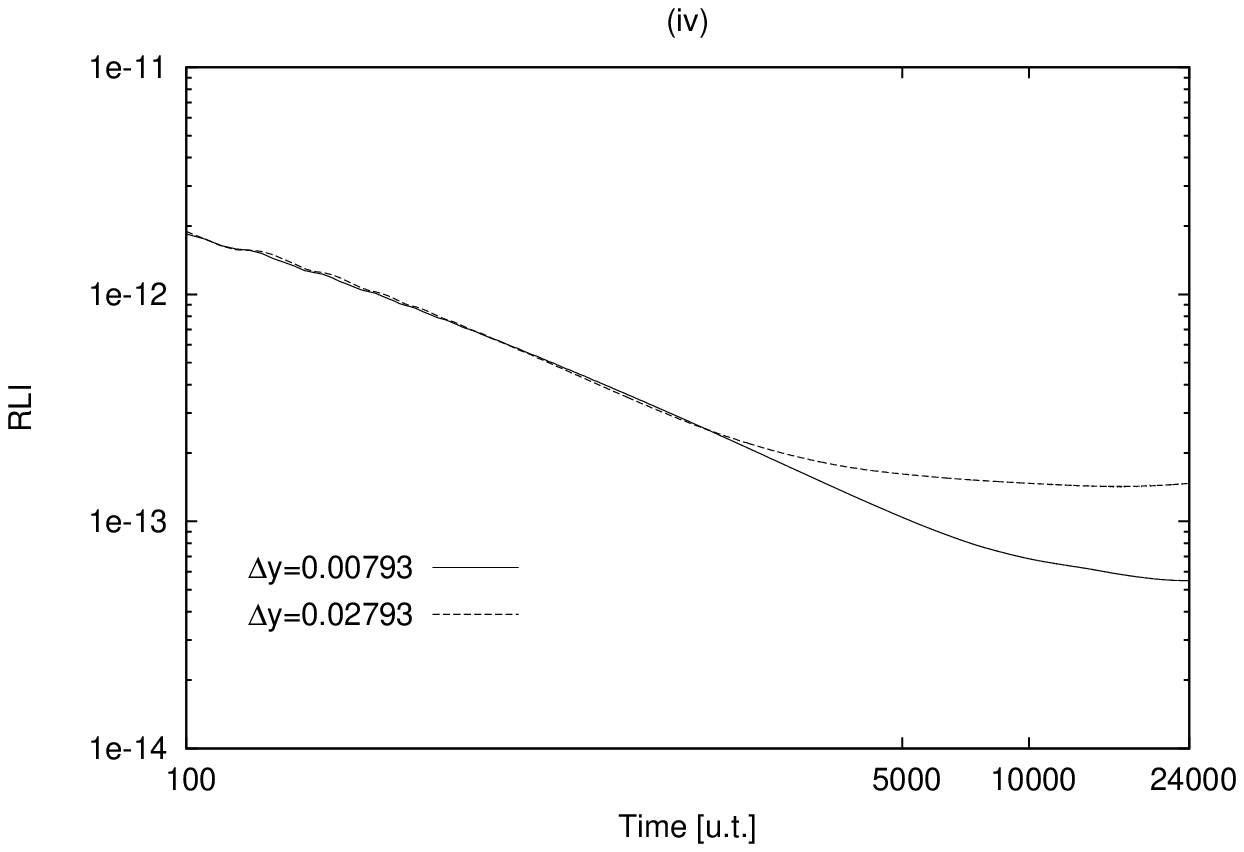}}
\end{tabular}
\caption{Performances of the CIs for the orbit \textbf{(a)} and for the quasiperiodic orbits close to \textbf{(a)}, (\textit{i}) for the OFLI and (\textit{ii}) the GALI$_2$ (the predicted behavior for the periodic orbit is also included in the left chart). Performaces of the CIs for the quasiperiodic orbits close to the orbit \textbf{(a)}, (\textit{iii}) for \textit{D} and (\textit{iv}) the RLI.} 
\label{experiment-bandc}
\end{center}
\end{figure}

The OFLI is precisely defined to reveal periodicity (\cite{Fouchard02}). The top left panel of Fig. \ref{experiment-bandc} shows the performances of the OFLI for the orbit \textbf{(a)} and for the two quasiperiodic orbits close to \textbf{(a)}. Although, the OFLI shows an oscillatory regime around a constant value for the periodic orbit, an increment following a linear rate is seen for quasiperiodic motion, as in the case of the FLI.

The top right panel of Fig. \ref{experiment-bandc} shows the performances of the GALI$_2$ (as the SALI has similar behavior, it is not included) on the periodic and quasiperiodic orbits of the sample. The GALI$_2$ is the only GALI capable of clearly distinguishing periodic motion in the experiment.

Let us analyze the behavior of the GALIs (and the SALI) for the special case of a 2--d.o.f. Hamiltonian system as the HH potential. In a 2--d.o.f. system, the quasiperiodic orbits lie on $2$--dimensional tori. All deviation vectors tend to fall on the $2$--dimensional manifold tangent to the torus where the motion takes place. If we start the computation with any two linearly independent deviation vectors (in order to compute the GALI$_2$ or the SALI), they will remain linearly independent on the $2$--dimensional tangent space of the torus. Then, the GALI$_2$ (or the SALI) is constant and different from zero. On the other hand, the GALI$_k$ (with $k=3,4$) tends to zero, since  some i.d.v. become linearly dependent on the rest (\cite{MSA11}). 

In the case of the periodic orbits, the motion takes place on $1$--dimensional tori (an invariant curve), the tangent space being also $1$--dimensional. Thus, the behavior of the GALI$_2$ (or the SALI) is $\propto t^{-1}$. The following asymptotic formulae summarizes the general behaviors in the HH potential of the GALI$_k$ for $m$ initially tangent i.d.v. and regular orbits lying on $M$--dimensional tori (with  $M=1,2$):

$$GALI_k(t)\propto\left\{\begin{array}{cc} constant & if\, k=2, M=2, m=0,1,2\\&\\t^{-1} & \left\{\begin{array}{c} if\, k=2, M=1, m=0,1\\if\, k=3, M=2, m=1,2\end{array}\right.\\&\\t^{-2} & \left\{\begin{array}{c} if\, k=3, M=2, m=0\\if\, k=4, M=2, m=2\end{array}\right.\\&\\t^{-4}\, or\, t^{-3} & if\, k=4, M=2, m=0,1\end{array}\right.$$

For the general formulae see \cite{CB06} or Section \ref{The SALI} of this paper.

On the top right panel of Fig. \ref{experiment-bandc}, we see that the GALI$_2$ shows the same initial behavior for the three orbits (i.e. following a power law of $t^{-1}$, which is also included in the figure). After an initial transient of similar behaviors, the GALI$_2$ oscillates around a constant value for the quasiperiodic orbits.

In a $2$--d.o.f. Hamiltonian system, the \textit{D} is not able to distinguish between periodic and chaotic motion (a decay is observed for both motions). Nevertheless, the ambiguity disappears when dealing with Hamiltonian systems of more d.o.f. (see \cite{Skokos01} for further information). The bottom left panel of Fig. \ref{experiment-bandc}, shows the behavior of the \textit{D} for the two regular orbits close to the orbit \textbf{(a)} and for an extended integration time of $10^5$ u.t. The closer the orbit to the periodic orbit, the longer it takes the \textit{D} to converge to a constant value, characterizing the regularity of the orbit. 

On the bottom right panel of Fig. \ref{experiment-bandc} we show the performances of the RLI. The RLI converges to lower values for the orbit \textbf{(a)} than for the quasiperiodic orbits. The closer the quasiperiodic orbit to the orbit \textbf{(a)}, the lower the final value of the RLI. Nevertheless, neither is there a significant change in the behavior of the CI nor a reference value is defined to separate the periodic orbits from the associated quasiperiodic orbits.  

According to the previous results, the OFLI and the GALI$_2$ (and the SALI) are the only CIs capable of clearly identifying the periodic orbit. The identification of the periodic orbits may fail due to an inappropriate choice of the i.d.v. Though the dependency of some fast CIs on the i.d.v. has already been pointed out by \cite{Froeschle00} and \cite{Barrio05}, it is worth reviewing this dependency.

\subsubsection{Dependency of various CIs on the i.d.v.}\label{initial deviation vectors}
In Section \ref{final values} we used the canonical basis of i.d.v. for the whole sample of orbits along the experiments. However, herein we analyze only the orbit \textbf{(b)} and we use three different bases of i.d.v. to test the dependency of the CIs on the i.d.v. These bases are characterized by the parameter $m$ (the number of i.d.v. initially tangent to the torus). The three bases are selected to have $m=0,1$ and $2$ ($2$ is the upper limit in a $2$--d.o.f. Hamiltonian system like the HH potential). From \cite{Skokos07} we know that the unit vectors (1,0,0,0) and (0,0,0,1) are initially tangent to the torus where the orbit \textbf{(b)} moves. Then, using a Gram-Schmidt process we build three bases of i.d.v. with $m=0,1,2$. For example, the basis with $m=2$ has four orthonormal vectors, two of them being the above--mentioned i.d.v. 

Although the LI and the RLI show some dependency on the parameter $m$, the differences are meaningless. The MEGNO, the FLI and the OFLI do not show any dependency at all. 

\begin{figure}[ht!]
\begin{center}
\begin{tabular}{cc}
\hspace{-5mm}\resizebox{63mm}{!}{\includegraphics{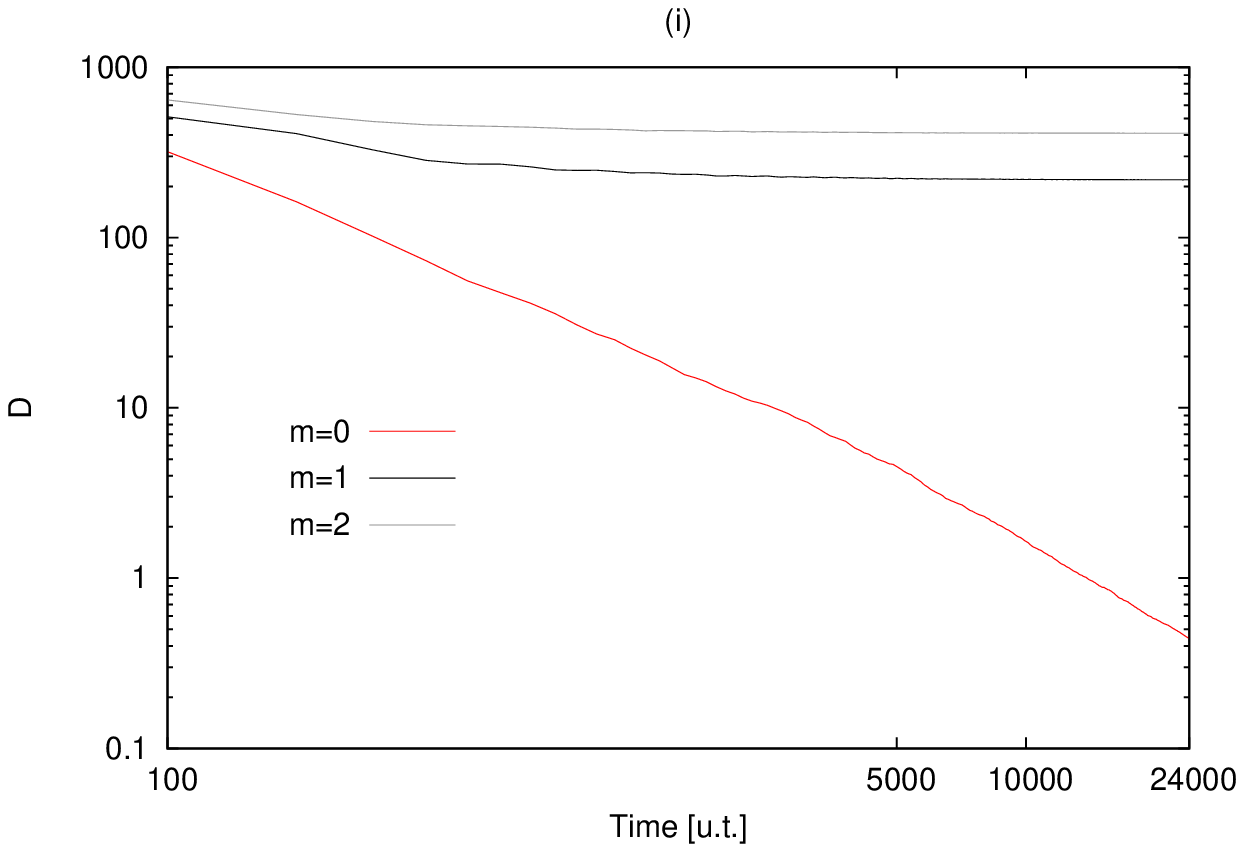}}& 
\hspace{-5mm}\resizebox{63mm}{!}{\includegraphics{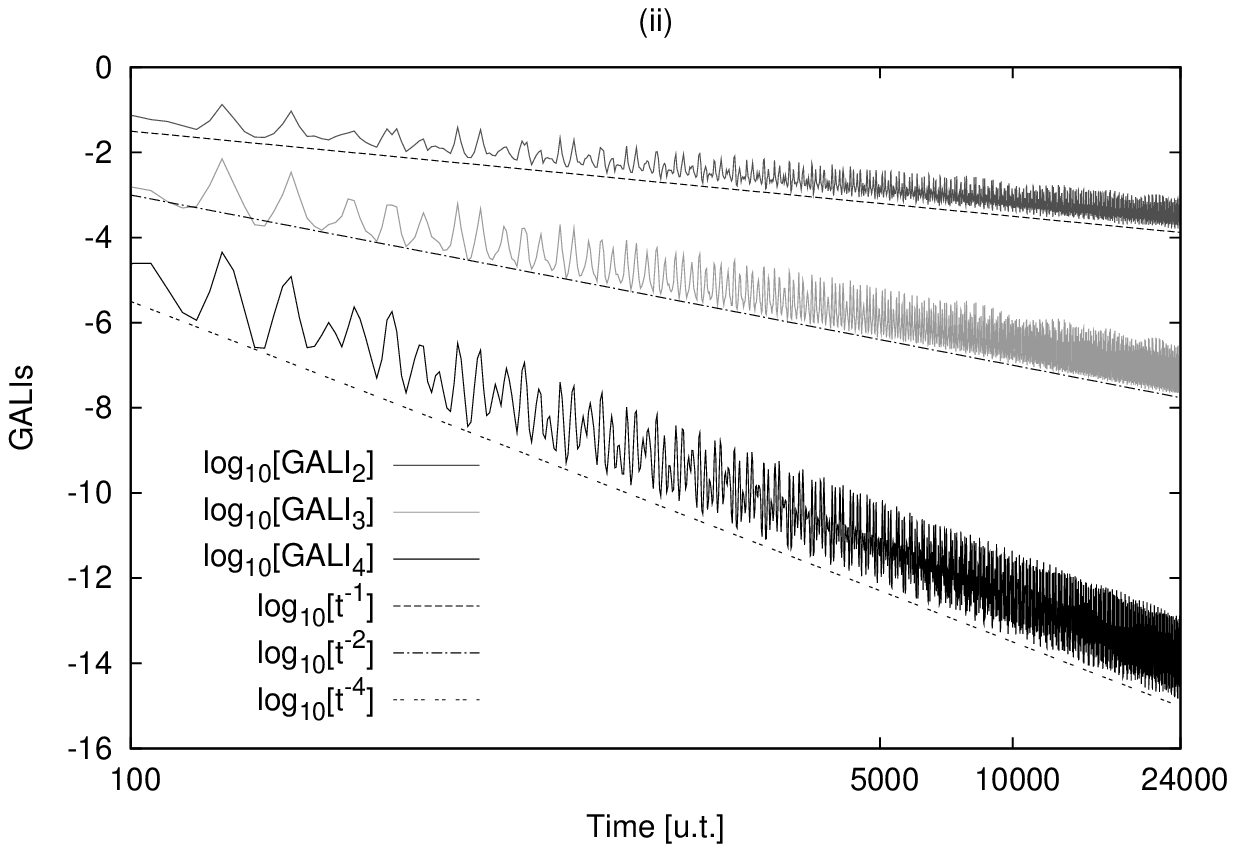}}\\
\hspace{-5mm}\resizebox{63mm}{!}{\includegraphics{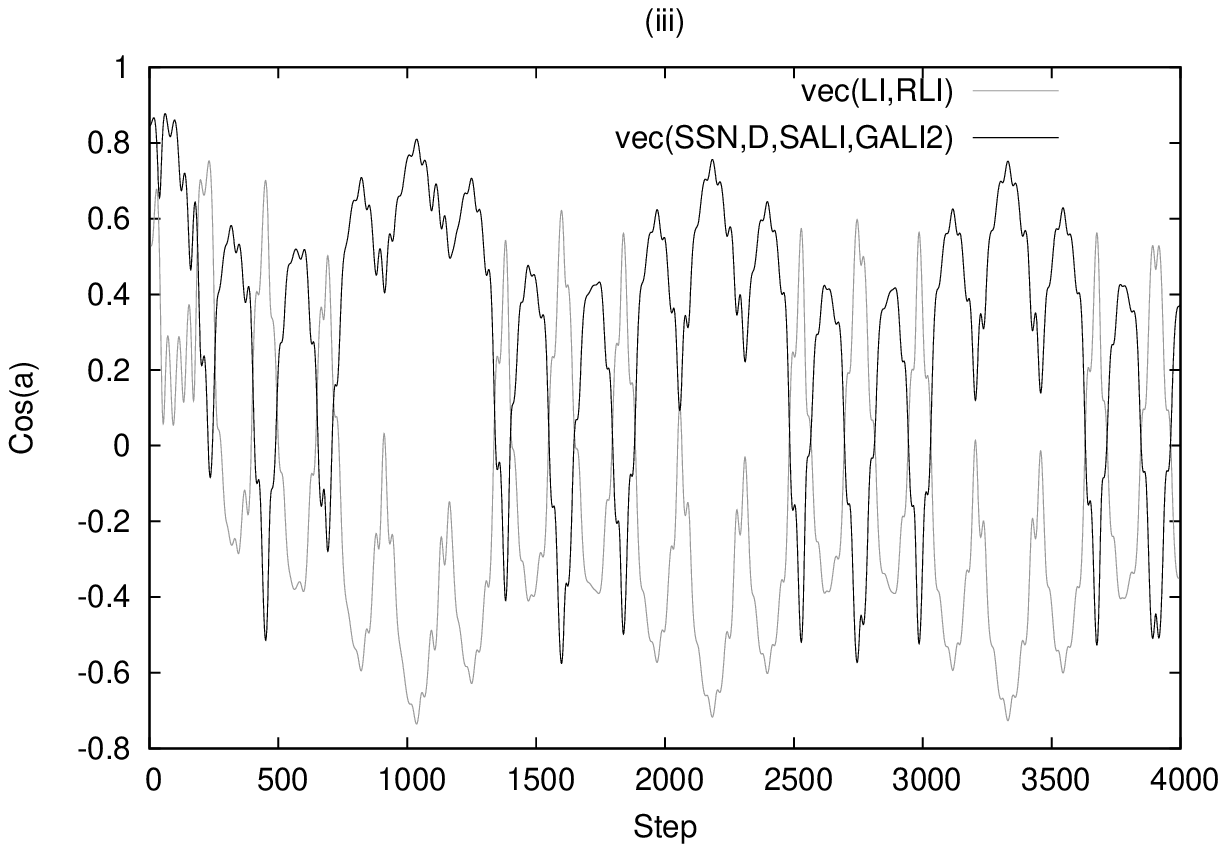}}& 
\hspace{-5mm}\resizebox{63mm}{!}{\includegraphics{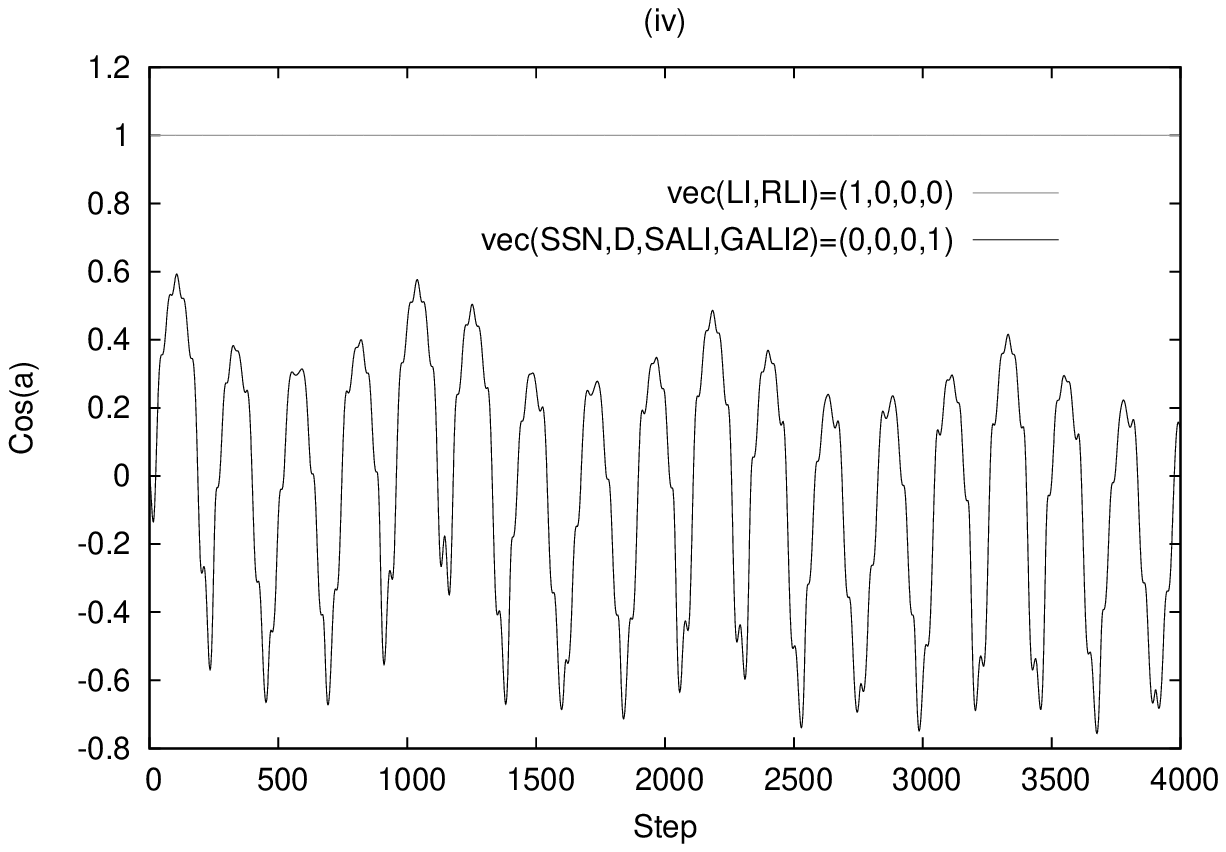}}
\end{tabular}
\caption{(\textit{i}) Performaces of the \textit{D} for the regular orbit \textbf{(b)} and $m=0,1,2$. (\textit{ii}) GALIs' performances for the orbit \textbf{(b)} with $m=0$. The power laws associated with each GALI are also included (note the logarithmic scale). Behavior of the quantity $\cos(a)$ for both the i.d.v. considered in the computation of the GALI$_2$ (\textit{iii}) with $m=0$ and (\textit{iv}) with $m=2$.}
\label{experiment-a}
\end{center}
\end{figure}

The top left panel of Fig. \ref{experiment-a} shows the behavior of the \textit{D} for the orbit \textbf{(b)} and for the three bases of i.d.v. We identify the regular nature of the orbit \textbf{(b)} using the bases with $m=1,2$. The \textit{D} reaches a constant value with both bases, with different convergent times, though. On the other hand, with $m=0$, the \textit{D} decreases like for chaotic motion. Despite extending the integration time to $10^6$ u.t., the result has not changed. 

To study the behaviors of the GALI$_k$ with $k=2,3,4$, we reproduced the experiment shown in Fig. 4 of \cite{Skokos07}. They analyzed the orbit \textbf{(b)} in the HH potential with $m=0,1,2$. Although the results agree with $m=1,2$, we see a discrepancy in the behavior of the GALI$_2$ with $m=0$ (see the dark gray line on the top right panel of Fig. \ref{experiment-a} and their top left panel of Fig. 4). On the top left panel of Fig. 4 in \cite{Skokos07}, the authors showed an almost constant value for the SALI and the GALI$_2$ (we do not include the SALI in our figure because the results are similar to the GALI$_2$). Although we select zero initially tangent i.d.v. as they did, our two i.d.v. used for the computation of the GALI$_2$ align with each other. On the bottom left panel of Fig. \ref{experiment-a} we show the opposite behaviors of the quantity $\cos(a)$ ($a$ being the angle between the direction of the i.d.v. and the direction of the flow) for both i.d.v. Therefore, they are linearly dependent and we see a $t^{-1}$ power law for the GALI$_2$ (top right panel of Fig.  \ref{experiment-a}), like for a periodic orbit. This linear dependency on both i.d.v. also explains the behavior of the \textit{D} for the orbit \textbf{(b)} and with $m=0$. 

On the bottom right panel of Fig. \ref{experiment-a}, we show the behaviors of the quantity $\cos(a)$ for the i.d.v. (1,0,0,0) and (0,0,0,1). They correspond to the orthonormal basis of i.d.v. with $m=2$. It is clear that the first i.d.v. aligns parallel to the flow while the second i.d.v. oscillates around one of the orthogonal directions to the flow. The last case with $m=2$ is the general case, and the behaviors of the GALIs are well predicted by the formulae given by \cite{CB06} or \cite{Skokos07}. Therefore, there is no need to check the choice of the i.d.v. to analyze large samples of orbits. It is convenient to track the behavior of the i.d.v. for a short period of time to analyze small sample of orbits, though.

\subsubsection{Sensitivity of the GALIs on dynamic instabilities}\label{instabilities}
We are also interested in studying the CIs' sensitivity to instabilities. In \cite{Maffione11a} the authors dealt with the sensitivity to stability and chaoticity levels of many of the CIs of the package, i.e. the LI, the RLI, the MEGNO, the SSNs and the D, the FLI and the SALI. Although their study was done on a 4D mapping, the results do not change in the HH potential. Furthemore, the FLI and the OFLI show no difference in their performances. Therefore, here we deal with the sensitivity to the instabilities of the GALIs alone and, for this purpose, we take the orbits \textbf{(c)} and \textbf{(d)}. 

\begin{figure}[ht!]
\begin{center}
\begin{tabular}{cc}
\hspace{-5mm}\resizebox{63mm}{!}{\includegraphics{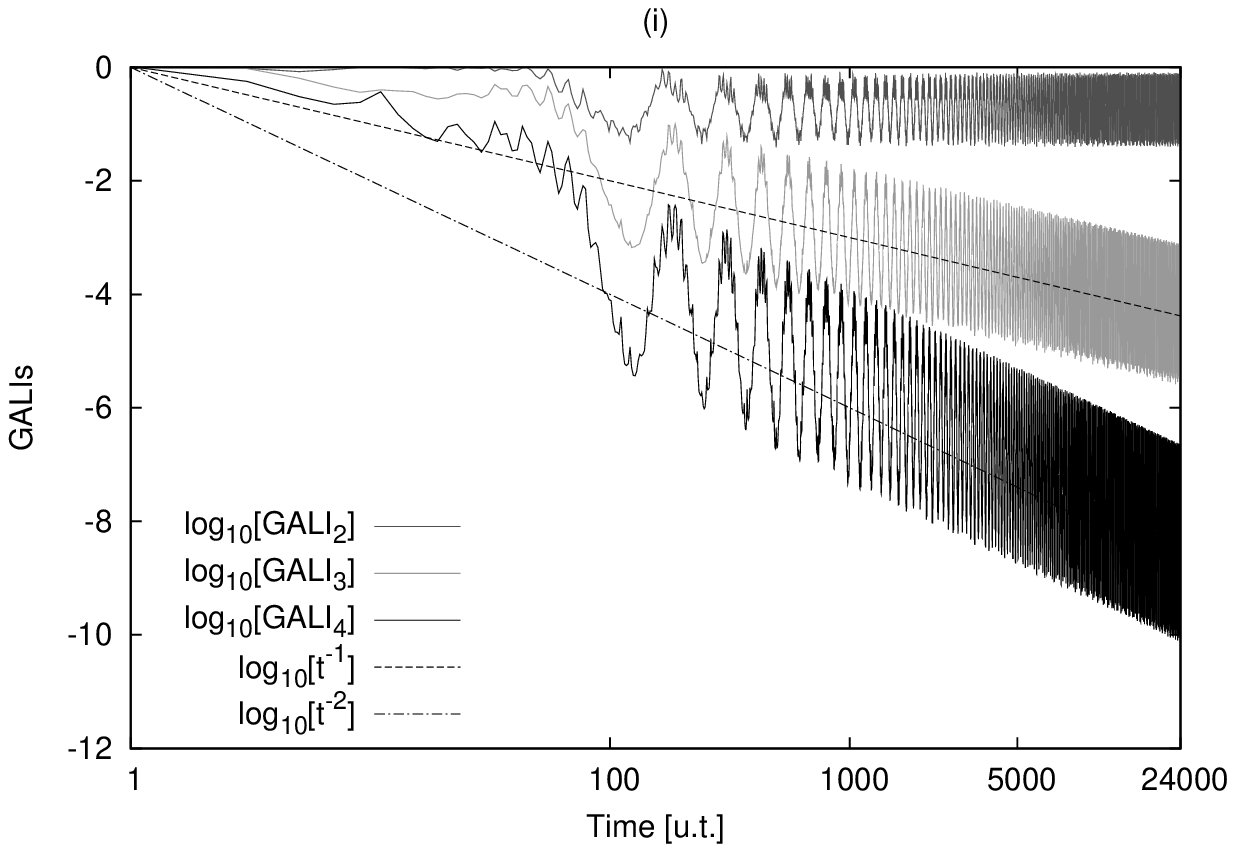}}&
\hspace{-5mm}\resizebox{63mm}{!}{\includegraphics{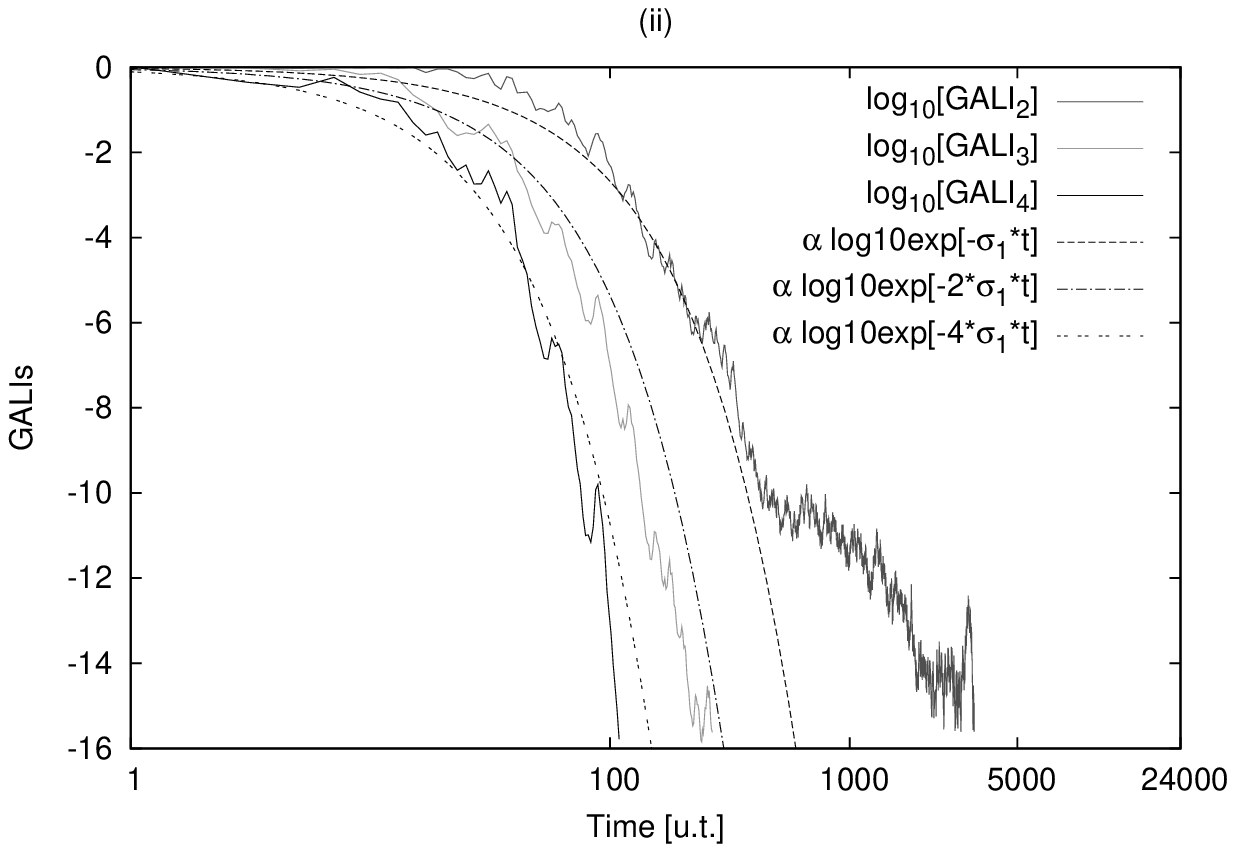}}
\end{tabular}
\caption{(\textit{i}) Performances of the GALIs for the orbit \textbf{(c)}. (\textit{ii}) Performances of the GALIs for the orbit \textbf{(d)}. The predicted behavior is also included, in logarithmic scale, in both charts.}
\label{grouppartI}
\end{center}
\end{figure}

On Fig. \ref{grouppartI} we show the behaviors of the GALI$_2$, the GALI$_3$ and the GALI$_4$ for the orbits \textbf{(c)} (left panel) and \textbf{(d)} (right panel). The GALIs for the orbit \textbf{(c)} show an unstable behaviour for a regular orbit because of the large amplitude of the oscillations around the predicted behavior for regular motion (i.e. GALI$_2\propto$ constant; GALI$_3\propto t^{-1}$; GALI$_4\propto t^{-2}$, with $m=2$). On the other hand, the GALIs for the orbit \textbf{(d)} show a chaotic orbit because their behavior follows exponential laws (i.e. GALI$_2\propto e^{-\chi_1t}$; GALI$_3\propto e^{-2\chi_1t}$; GALI$_4\propto e^{-4\chi_1t}$, with $\chi_1\sim 0.4118490\times 10^{-1}$, $\chi_1$ being the LI value of the orbit). The GALI$_4$ has the highest speed of convergence for the chaotic orbits among all the indicators tested so far. However, it is not economical in terms of computational time, see \cite{Skokos07} and Section \ref{cpu-times} of this paper for further details. 

Finally, we show that the OFLI and the GALI$_2$ (and the SALI) are the only CIs that clearly identify the periodic orbit \textbf{(a)} due to an evident change in their behaviours (Section \ref{periodicity}). However, some dependency on the i.d.v. it is shown in the experiment, which may lead to an incorrect identification of periodic orbits (Section \ref{initial deviation vectors}). Thus, it is convenient to track their behaviors for a short period of time in order to select the best i.d.v. The dependency of first order variational indicators on the i.d.v. is further discussed in \cite{Barrio05}.

The GALI$_2$, like the other GALIs, shows a dependency on the parameter $m$ which gives information of the dimensionality of the tori where the regular motion takes place (see \cite{MSA11}). Moreover, in the case of the chaotic orbits, the GALI$_4$ has the highest speed of convergence among all the indicators tested so far (Section \ref{instabilities}). 

The MEGNO showed the same good performances for single studies of orbits in the HH potential (not shown) as for mappings (see \cite{Maffione11a}), in particular because it is simple to characterize the stability levels, see Section \ref{MEGNO}. Thus, for studies of this kind, a combination of the OFLI and the MEGNO seems to be the best choice if we can use the time evolution of the CIs. However, the GALI$_2$ (or the SALI) is also appropriate for the task.

\subsection{The computing times}\label{cpu-times}
Before going to the second part of the work, we finish the comparative study of variational CIs studying their computing times. 

\begin{figure}[ht!]
\begin{center}
\begin{tabular}{cc}
\hspace{-5mm}\resizebox{63mm}{!}{\includegraphics{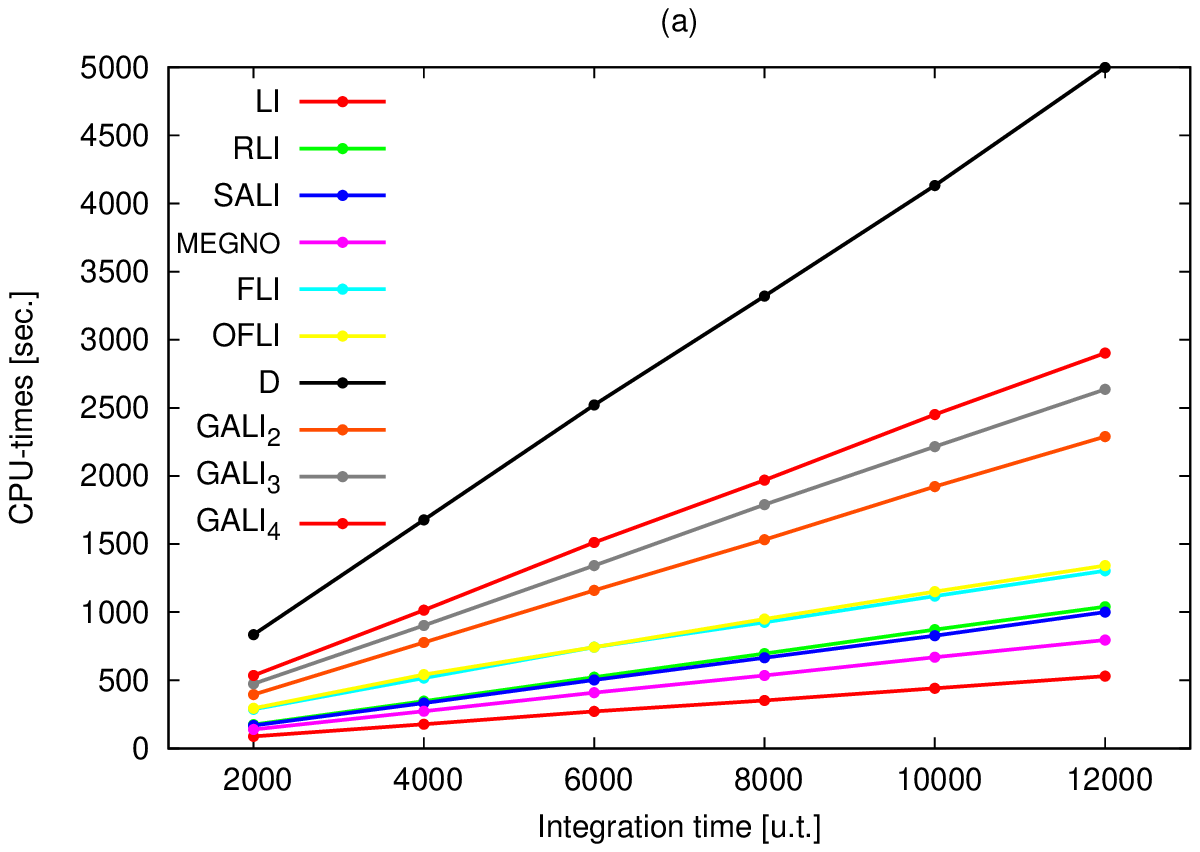}}& 
\hspace{-5mm}\resizebox{63mm}{!}{\includegraphics{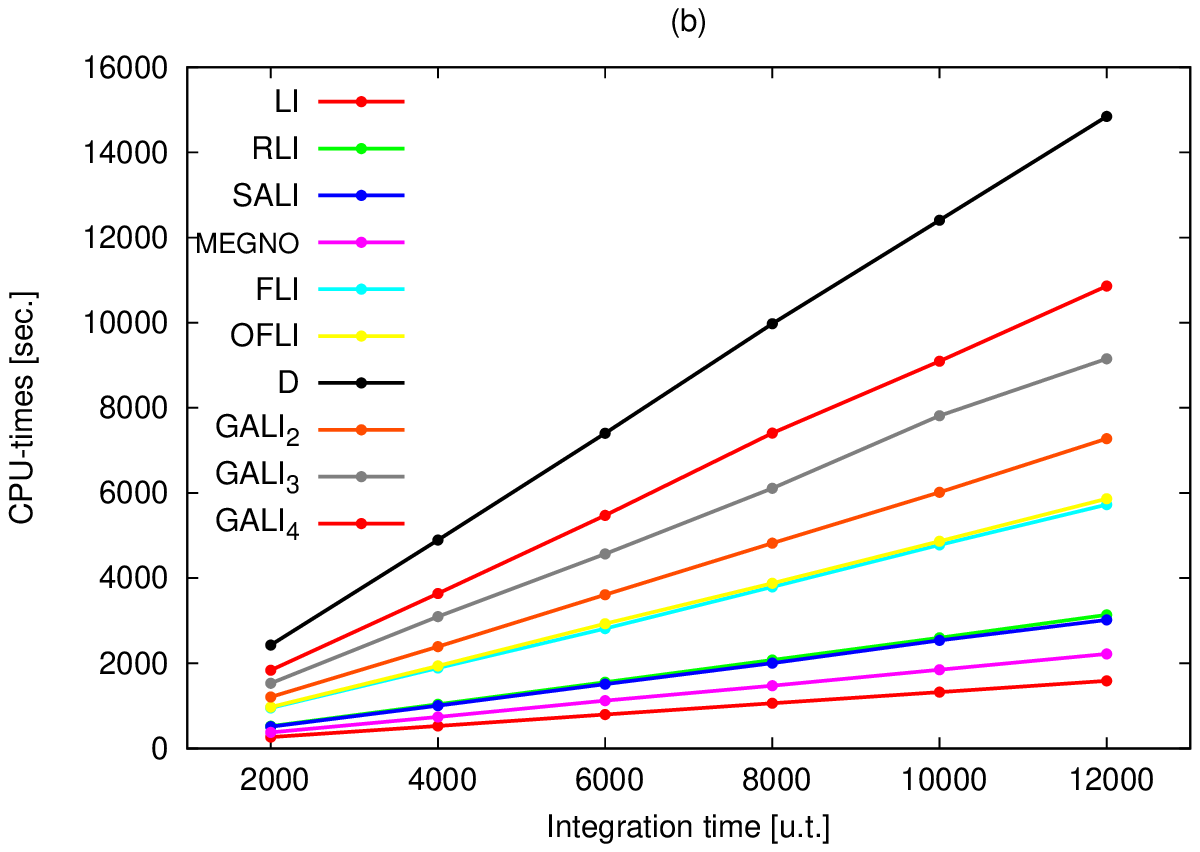}}
\end{tabular}
\caption{CPU--times for all the CIs under study (a) for region A and (b) for region B.}
\label{cputimes-1}
\end{center}
\end{figure}

On Fig. \ref{cputimes-1}, we show the CPU-times for all the CIs currently under study. On the left, we compute the CIs for a sample of $10^3$ i.c. located in the interval defined by $y\in[0.55:0.6]$, $x=p_y=0$ and $h=0.118$ (hereafter region A). Region A has a divided phase space, and a high percentage of orbits that reach the saturation values (i.e. $10^{20}$ for the FLI/OFLI and $10^{-16}$, the precision of the computer, for the SALI or the GALIs) within the interval of integration ($1.2\times 10^4$ u.t.). On the right, we computed the CIs for a sample of $3\times 10^3$ i.c. located inside the stability island associated with a periodic orbit of period one. We call this region ``B''; it is defined by $y\in[0.3:0.506]$, $x=p_y=0$ and $h=0.118$. In this region there are no orbits that reach any of the saturation values by $1.2\times 10^4$ u.t.

From Fig. \ref{cputimes-1} it is clear that the CPU--time increases with a growing percentage of regular orbits for each CI (compare both panels). Also, the CPU--time increases with the complexity of the CI's algorithm. Therefore, the LI is the least time consuming method while the \textit{D} is the most CPU--time demanding CI. The order would be: the LI, the MEGNO, the SALI and the RLI, the FLI and the OFLI, the GALI$_2$, the GALI$_3$, the GALI$_4$ and the \textit{D}. The MEGNO uses two more equations than the LI. The SALI and the RLI have very similar computing times because the SALI uses the computation of two deviation vectors while the RLI, the computation of two orbits. The FLI and the OFLI use the computation of four deviation vectors (the HH potential has a 4D phase space). The GALI$_k$ uses $k$ deviation vectors, but their numerical implementation is time consuming (despite the fact that we used the ``Singular Value Decomposition'' or SVD routine, see \cite{Skokos08}). Finally, the \textit{D} uses a statistical approach which is not numerically efficient.

\begin{figure}[ht!]
\begin{center}
\begin{tabular}{c}
\hspace{-5mm}\resizebox{90mm}{!}{\includegraphics{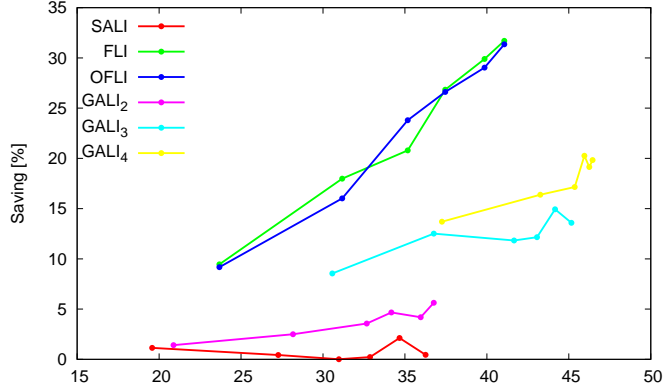}}
\end{tabular}
\caption{Percentage of orbits that reach the saturation values against the reduction in the CPU--times for all the CIs that have a saturation value.}
\label{cputimes-sattimes}
\end{center}
\end{figure}

Using a saturation value for some methods has several advantages which were studied in this work (Section \ref{Qualitative study-statistical sample}), like recovering the chaoticity levels with the times of saturation. Furthermore, there is also a numerical advantage with such kind of methods. If the orbit reaches the saturation value of the CI, the computation can be stopped, thus reducing the CPU--time (see \cite{Skokos07} and \cite{Maffione11a}). On Fig. \ref{cputimes-sattimes}, we tried to measure such relationship. Therefore, we computed the CPU--time gained due to the percentages of orbits that reach the corresponding saturation value. This relation gives an idea of how much efficient an indicator is. For example, with a $35\%$ of orbits that reach the saturation value $10^{-16}$, the SALI gains less than $5\%$ of the CPU--time, the GALI$_2$ and the GALI$_3$ gain around $5\%$ and $12\%$, respectively and the FLI and the OFLI gaining between $20\%$ and $25\%$. Although the GALI$_4$ shows higher percentages of orbits that reach the associated saturation value ($10^{-16}$), the CPU--time gained is not as high as that for the FLI or the OFLI. 

Finally, this experiment shows that the FLI/OFLI is the most efficient method because it gains most computing time by means of a succesful implementation of the saturation values.

\section{The software package using Taylor method}
\label{section_taylor}
Here we summarize the main features of the software package that
integrates ODEs using the Taylor method (\texttt{taylor}\footnote{The package \texttt{taylor} is released under the GNU Public License, and it can be retrieved from \texttt{http://www.ma.utexas.edu/$\sim$mzou/taylor} (US)
} hereafter) that is described in detail in \cite{JZ05}. 

As mentioned in Section \ref{section_introduction}, one of the main
distinguishable features of this software is the use of the so-called
\emph{automatic differentiation}, which computes the derivatives of the
functions involved up to an arbitrarily high order. This particular
implementation of automatic differentiation generates its code from a
file with the differential equations system, and makes the
differentiation in an automatic fashion, without losing accuracy on the
derivatives. There are many general purposes computer programs which
build its code from a program (\cite{BKSF59}, \cite{Gib60} \& \cite{CC94}), while others  are built up by the user for a specific
problem (see for instance, \cite{Bro71} for the $N-$body
problem). Further alternatives to apply the Taylor method can be found
in \cite{SV87} and in \cite{IS90}. In fact, the key point of the Taylor method is the precise computation of the derivatives of $x(t_m)$, $x^{(j)}(t_m)$, which the automatic differentiation plays a crucial role.

The way in which \texttt{taylor} operates can be summarized as follows: the package reads the differential equations system from a plain text file, and uses it to build the integrator (for a full syntax and parameters list see ``Taylor user's manual'', which comes with the package \texttt{taylor}). \texttt{taylor} subroutines are written in ANSI C code. It includes the subroutines for the automatic differentiation, the time stepper, as well as step-size and order control. The subroutine which computes the automatic differentiation reads each equation from the file, decomposes the expression into binary or unitary operations and generates recurrences in order to compute the derivatives analytically (for an example of some recurrences see \cite{JZ05}, Section 2, Proposition 2.1). Thus,  the derivatives of the function $f$ up to an arbitrarily high order can be computed without losing accuracy. For a more detailed description of the software implementation see Section 4 of \cite{JZ05}.

For the calculation of the step size $h$ and the order of the method $p$, \texttt{taylor} uses variable-stepsize and variable-order algorithms. The package computes, for each time-step, the order $p_m$ that guarantees a truncation error of the order of a given threshold $\varepsilon_m$ (for simplicity, in the numerical computation sections we call it $\varepsilon$). With this value of $p_m$ the step size $h_m$ is computed. These two values are chosen in such a way that not only the required precision $\varepsilon$ is achieved but  the number of operations needed to advance one time step is minimized. 

An interesting feature of Taylor method, unlike other integrators, is that to increase accuracy, it requires much less work to increase the order $p$ than to reduce the step size $h$. Its explanation can be seen in \cite{JZ05}, on Section 3.3.

Another useful feature of \texttt{taylor} is the possibility to
implement easily the use of extended arithmetic, using high precision
libraries like \emph{GMP} \footnote{For further informatin see \texttt{http://gmplib.org/}}. This is
possible due to the implementation of abstract operations with a
floating point type called \texttt{MY\_FLOAT}. Then, in the header file,
these operations are replaced by the ones corresponding to the mathematical library we are using (the default is \emph{double}).

\section{Application to a model of triaxial elliptical galaxy}
\label{section_pot_tri}

We considered a triaxial potential, corresponding to an auto-consistent model of elliptical galaxy \cite{MCW05}, given by

\begin {equation}
  V(x,y,z) = -f_0(\it{p})-f_x(\it{p})(x^2-y^2)-f_z(\it{p})(z^2-y^2),
  \label{pot_tri_potential}  
\end{equation}

\noindent
where $\it{p}$ is the softened radius
\begin{equation}
\it{p}^2 = r^2 + \epsilon_1^2 \qquad \mbox{for} \ f_0,
\end{equation}
\noindent
and
\begin{equation}
\it{p}^2 = r^2 + 2\epsilon_1^2 \qquad \mbox{for}\ f_x,\ f_z,
\end{equation}
with the softening parameter fixed at $\epsilon_1=0.01$. 

The functions $f_s$ were computed from an $N-$body simulation and fit with equations of the form

\begin{equation}
  f_s=\frac{C_s}{(\it{p}^{k_s}+{q_s}^{k_s})^{l_s/k_s}}, \qquad s=0,x,z.
  \label{pot_tri_eq}
\end{equation}

\noindent
The adopted values for $C_s$, $k_s$, $q_s$, $l_s$ are given in Table 2

\begin{table}[!htbp]\centering
    \tbl{Values of the coefficients of the functions $f_s(\it{p})$ given by Eq. \ref{pot_tri_eq} (taken from Muzzio {\it et al.}, [2005]).}
    {\begin{tabular}{ccccc}\\
      \toprule
      & $C_s$ & $k_s$ & $q_s$ & $l_s$ \\
      \hline
      $s = 0$ & $0.92012657$ & $1.15$ & $0.1340$ & $1.03766579$\\
      \hline
      $s = x$ & $0.08526504$ & $0.97$ & $0.1283$ & $4.61571581$ \\
      \hline 
      $s = z$ & $-0.05871011$ & $1.05$ & $0.1239$ & $4.42030943$ \\
      \botrule
      \end{tabular}}
  \label{table_pot_tri_coefs}
\end{table}

We adopted value $E=-0.5$ for the system's energy (which
corresponds to an $x$-axis central orbit with period $\sim 10$ u.t.). We
considered central orbits (i.e. $x_0=y_0=z_0=0$) and, by defining
$\Delta V = E-V(x_0,y_0,z_0)$, we took initial values for $p_x$, $p_y$ and
$p_z$ which satisfy ${p_x}^2+{p_y}^2+{p_z}^2=2\Delta V$ using a grid
made of points which swept the values of $p_x$ and $p_y$ with a step of
$\Delta p_x= \Delta p_y=0.2$, yielding a total of 2080 orbits. The
integrations were carried out for $t=5\times 10^3$ and $t=5 \times 10^4$
u.t., but only  the results for $t_{final}=5 \times 10^4$ are shown here since the same behavior is observed in both cases. We chose as $t_{initial}=0.1$ in order to avoid a division by zero in the computation of the MEGNO (note that there is a $t$ in the denominators of eqs. \ref{MEGNO_eq} and \ref{MEGNO_avg_eq}), and $t_{initial}=0$ in Subsections \ref{section_LCN_pot_tri} and \ref{section_solomov_pot_tri}. The required tolerances were $\varepsilon=10^{-15}$ (the same for both the absolute and relative tolerances).

For all the computations done in this second part of the work we used variable-stepsize. For \texttt{taylor}, we used the stepsize control method mentioned on \cite{JZ05}, Section 3.2.1, and for DOPRI8 and BS their own stepsize control method, included on both programs. All the results presented here have been computed in a $2\times$Dual XEON 5120, Dual Core $1.86$ GHz.

\subsection{The MEGNO}
\label{section_MEGNO_pot_tri}
\begin{figure}[!htbp]
  \begin{center}
    \epsfig{file=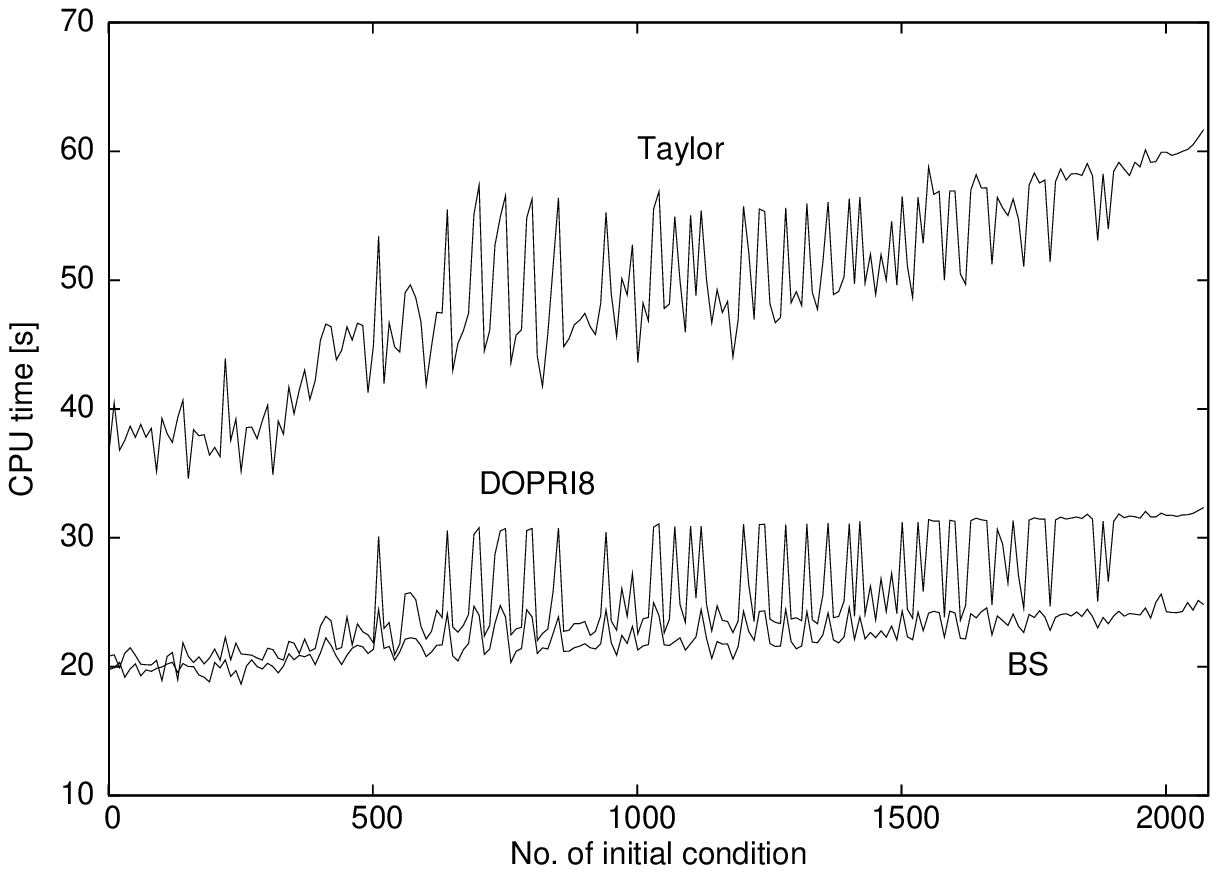, width=8cm}
    \epsfig{file=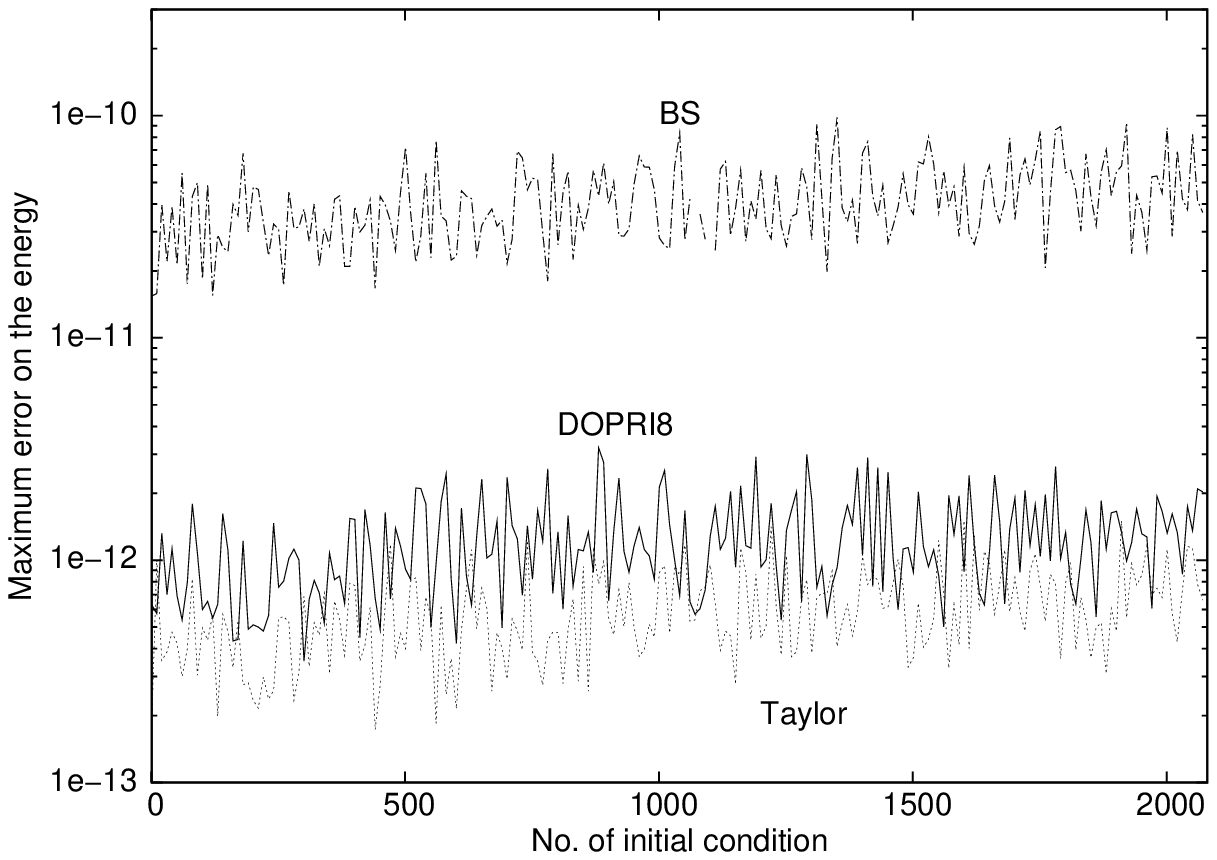, width=8cm}
    \caption{CPU times (Left) and Maximum energy error (Right) for
   several i.c. in the calculation of the MEGNO for the triaxial
   potential, using \texttt{taylor}, DOPRI8 and BS, with $t_{final}=5
   \times 10^4$ u.t.}
    \label{MEGNO_pot_tri_tcpu_maxerr}
  \end{center}
\end{figure}

The skillful computation of the MEGNO definitely requires an efficient integrator. Therefore, it would be interesting to confirm the superiority of \texttt{taylor} claimed out in \cite{JZ05}. To this end, the package was compared with other two well--known integrators: a Runge-Kutta 7/8 method, the so--called DOPRI8 \cite{PD81}, and the Bulirsch-St\"oer method (BS) \cite{NR97}.
The results of the comparison regarding both speed and precision of the integrations performed by means of  \texttt{taylor}, DOPRI8 and BS are presented in Figure \ref{MEGNO_pot_tri_tcpu_maxerr}, where we have plotted the respective CPU times and maximum energy errors for one out of ten i.c..

\noindent
The left side of Figure \ref{MEGNO_pot_tri_tcpu_maxerr} reveals  that
\texttt{taylor} is considerably slower than  DOPRI8 and BS. Its
computation time is $\sim 2$ or $2.2$ times the time insumed by the
other two integrators (i.e. it's slower in a factor of $\sim 2$,
$2.2$). When we look at the maximum energy error displayed on the right side of Figure \ref{MEGNO_pot_tri_tcpu_maxerr}, we observe that \texttt{taylor} is slightly more precise than DOPRI8 (in a factor $\sim 2$), while the errors arising from the integration by BS are about 2 orders of magnitude larger (in a factor $\sim 70$).

Therefore we can assert that although \texttt{taylor} is nearly as precise as DOPRI8, it clearly demands much more computation time than both DOPRI8 and BS. Thus, at least for this particular system of ODEs, the \texttt{taylor} package turns out to be a somewhat  inefficient integrator.

This is a surprising fact, since \texttt{taylor} showed up as a very
competitive integrator  in \cite{JZ05}. However it seems that
there are some particular features in these ODEs that could be
conflictive for the \texttt{taylor} package. In fact, the differential equations for the MEGNO 

\begin{equation}
 \left\{ \begin{array}{lll}
  Z' & = & t\frac{\dot{\delta}\gamma(\bm{x}_0;t)}{\delta\gamma(\bm{x}_0;t)} \\
	  \tilde{Y}' & = & \frac{2}{t}Z\\      
	 \end{array}\right.
 \label{MEGNO_diff_eqs}
\end{equation}

\noindent
encompass the temporal variable $t$ in the expressions for $Z'$ and
$\tilde{Y}'$ in the numerator and denominator, repectively. As a
consequence, we infer that, at the beginning of the integration when $t$
is very small and near its end, when we consider a large period, we
obtain large values of $Z'$ or $\tilde{Y}'$,
respectively. \texttt{taylor} might compensate this effect by taking
very small step sizes, and these two equations might be responsible for
the inefficient performance of the package.

In order to test this conjecture we proceed to the computation of
another dynamical indicator, namely, the LI. This suits perfectly our
needs, since  only the equations of motion and their first variational
ones are to be integrated, and the two equations (\ref{MEGNO_diff_eqs}) that calculate the MEGNO are no longer involved.

\subsection{The LI}
\label{section_LCN_pot_tri}

\begin{figure}[!htbp]
  \begin{center}
    \epsfig{file=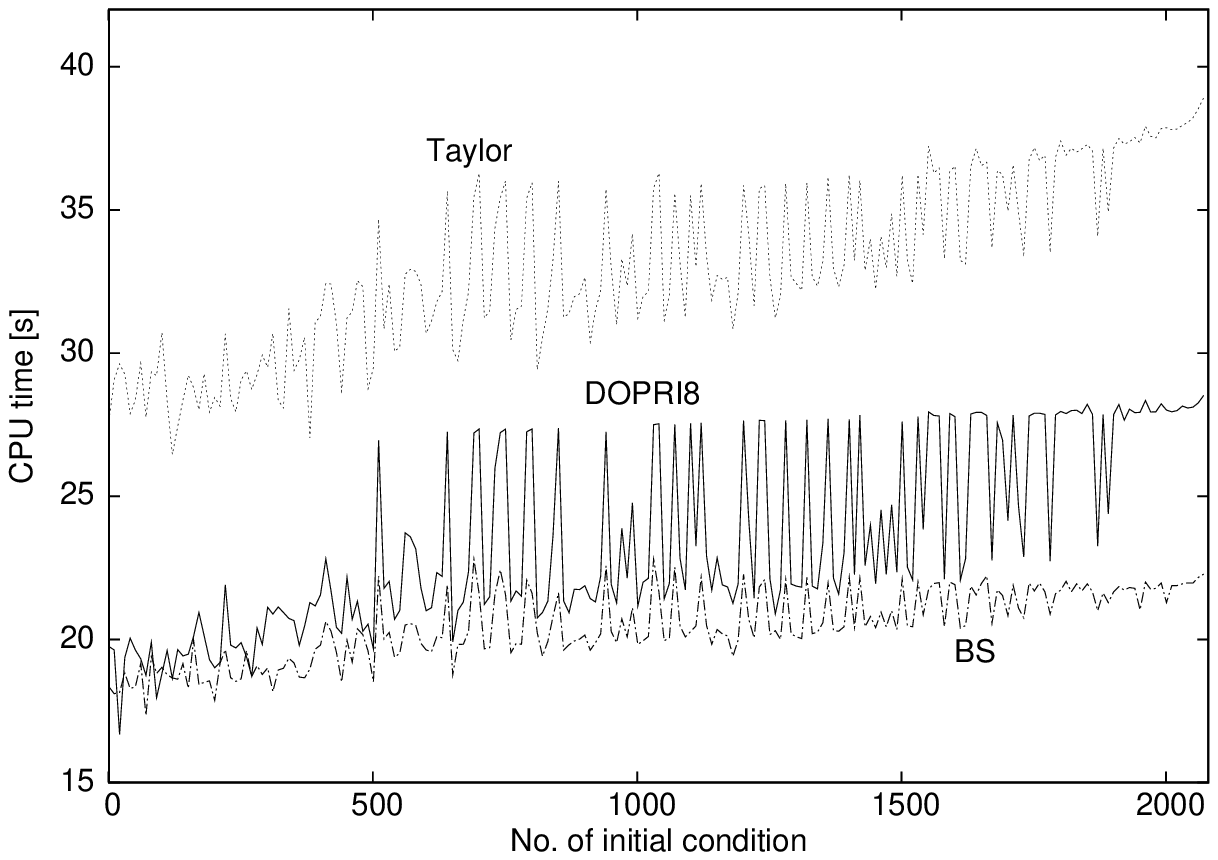, width=8cm}
    \epsfig{file=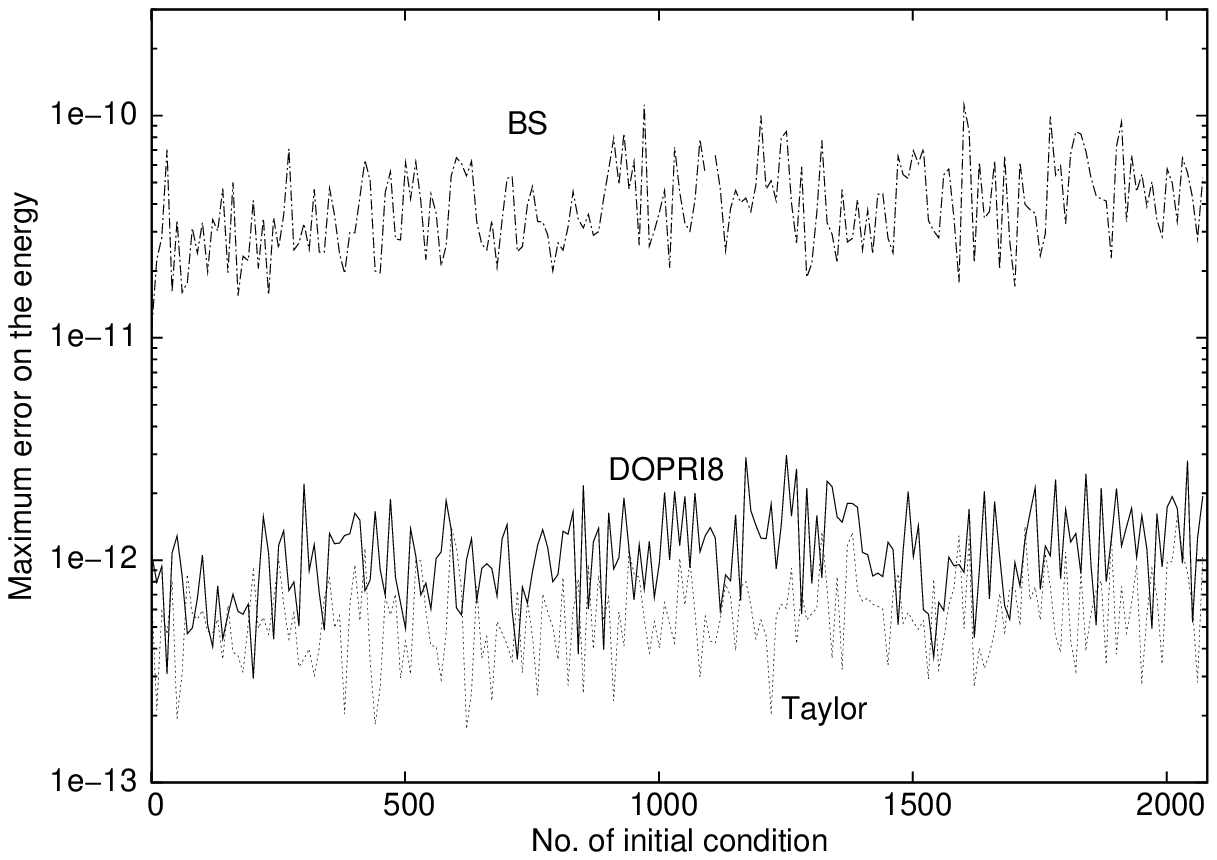, width=8cm}
  \end{center}
  \caption{CPU times (Left) and maximum error on the energy (Right) for
 several i.c. in the calculation of the LI for the triaxial potential,
 using \texttt{taylor}, DOPRI8 and BS, with $t_{final}=5 \times 10^4$ u.t.}
 \label{LCN_pot_tri_tcpu_maxerr}
\end{figure}

The task was carried out using the very same i.c. and parameters for the
galactic model as those of the previous section and the same tolerance
was imposed. The CPU times demanded by the computation of the LI values
and the resulting maximum energy errors are displayed in Fig. \ref{LCN_pot_tri_tcpu_maxerr}.

Regarding to the CPU time, \texttt{taylor} still remains slower than
both DOPRI8 (a $\sim 1.4$ factor) and BS ($\sim 1.65$), but it is
evident that the differences have decreased significantly, i.e. from a
$\sim 2$ to $\sim 1.4$ factor. However, as long as the energy error is concerned, no significant dissimilarity from the results shown in Figure \ref{MEGNO_pot_tri_tcpu_maxerr} is observed.

These results support our assumption that the equations for the MEGNO
decrease \texttt{taylor}'s performance. In fact, the reduction of the system of ODEs by avoiding those required for the MEGNO considerably dimished the difference between the  computational times invested by the compared integrators. Nonetheless, \texttt{taylor} kept on being the least efficient, in discrepancy with the results given in \cite{JZ05}, Table 4. 

The question arises whether for integrating only the system of equations
of motion, \texttt{taylor} succeds in showing its benefit regarding the speed of computation  stated by \cite{JZ05}, Table 4. This issue is addresed in the forthcoming subsection.

\subsection{Integration of the equations of motion}
\label{section_solomov_pot_tri}

\begin{figure}[!htbp]
  \begin{center}
    \epsfig{file=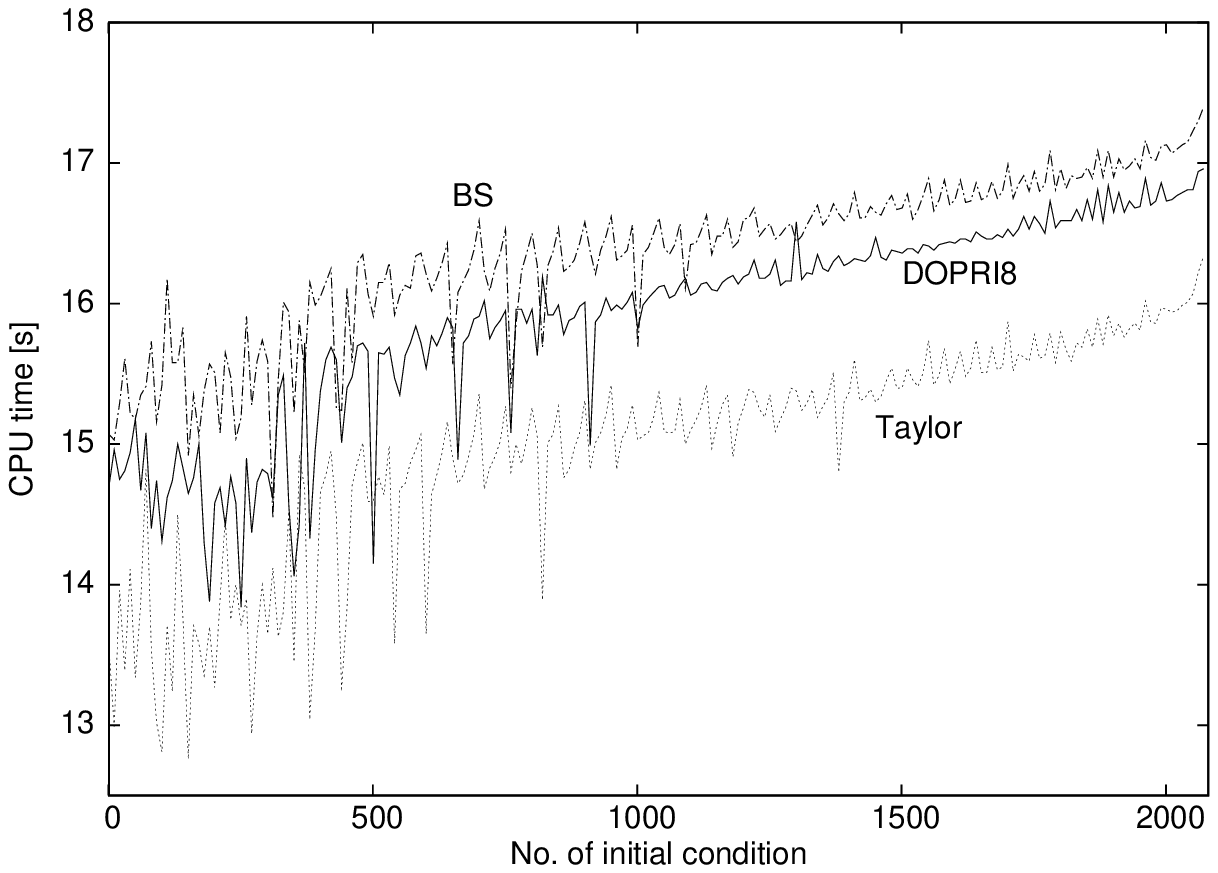, width=8cm}
    \epsfig{file=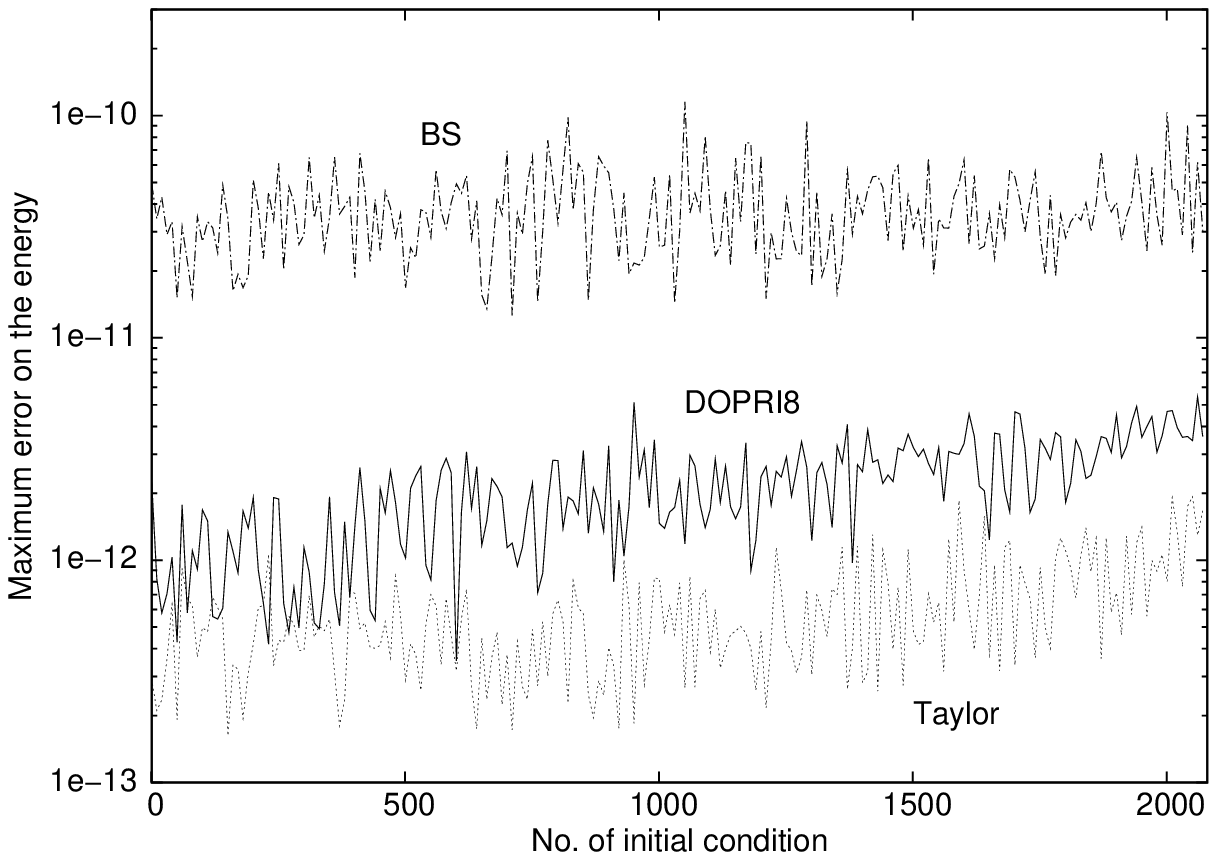, width=8cm}
  \end{center}
  \caption{CPU times (Left) and maximum error on the energy (Right) for
 several i.c. when computing the equations of motion for the triaxial
 potential, using \texttt{taylor}, DOPRI8 and BS, at $t_{final}=5 \times
 10^4$ u.t.}
 \label{solomov_pot_tri_tcpu_maxerr}
\end{figure}

On the left of figure \ref{solomov_pot_tri_tcpu_maxerr} it is displayed
the CPU time insumed by \texttt{taylor}, DOPRI8 and BS for several
i.c. when we integrate the equations of motion for the galactic
model. The plot evinces that for this particular problem,
\texttt{taylor} results slightly faster than both DOPRI8 (a factor of
$\sim 1.06$) and BS ($\sim 1.08$).

Regarding to the maximum error in the energy, a similar behavior is revealed on the right side of Fig. \ref{solomov_pot_tri_tcpu_maxerr}, being \texttt{taylor} slightly more accurate than DOPRI8 and about 2 orders of magnitude more precise than BS. 

Let us remark that when solving the three different sets of ODEs for
this particular system (the MEGNO, the LI and the equations of motion),
\texttt{taylor} showed up to be as precise as DOPRI8 and much more
precise than BS; but as far as the required CPU times are concerned,
only when integrating the equations of motion \texttt{taylor} turned out
to be the best, in a smaller degree, though.

The results achieved so far suggest  that  \texttt{taylor} is not well suited for dealing with systems involving variational equations or ODEs like the ones defining the MEGNO, but it offers some advantage when only the equations of motion are integrated.

In fact, the \texttt{taylor} package consumes a quite significant amount of CPU time to compute the automatic differentiation, build the power series and evaluate them at the requested points. Thus, it seems fair to wonder whether using this dynamical system for the comparison is a fortunate selection, due to the rather cumbersome expressions involved.  

\section{Application to a  simpler problem: the perturbed 3D quartic oscillator}
\label{section_quartic}

In this section we present the results of the reproduction of the same numerical experiments for a somewhat simple dynamical problem, namely, a 3D quartic oscillator subject to a cubic perturbation of the form $ax^2(y+z)$, where $a=5\times 10^{-4}$ is the perturbative parameter. Therefore, the potential can be recasted as

 \begin{equation}
   V(x,y,z)=\frac{1}{4}(x^4+y^4+z^4)+ax^2(y+z).
 \end{equation}

For these experiments we adopted the energy value  $E \approx 0.485$, for which
the central $x$-axis orbit has a period of $2\pi$ u.t., and we
constructed a lattice of i.c. by taking $y=p_x=p_z=0$  and values of $x$ and $z$
in the range $0 \leq x \leq 1.5$ and $-1.5 \leq z \leq 1.5$, with
step-sizes  $\Delta x=\Delta z=0.05$, yielding ${p_y}^2 \geq 0$
(i.e. $p_y \in \mathbb{R}$). The integrations for the resulting 1056
i.c. extended over $10^4$ u.t. and $10^5$ u.t., though only the results
for $t=10^5$ are included herein, for the same reason as in the case of the $t_{final}$ for the triaxial potential. As in Section \ref{section_MEGNO_pot_tri}, the initial time $t_{initial}=0.1$ was adopted to avoid a division by zero in the calculation of the MEGNO, but the value $t_{initial}=0$ was fixed in Sections \ref{section_LCN_quartic} and \ref{section_solomov_quartic}. The tolerance was again set to $\varepsilon=10^{-15}$.

\subsection{The MEGNO}
\label{section_MEGNO_quartic}

\begin{figure}[!htbp]
  \begin{center}
    \epsfig{file=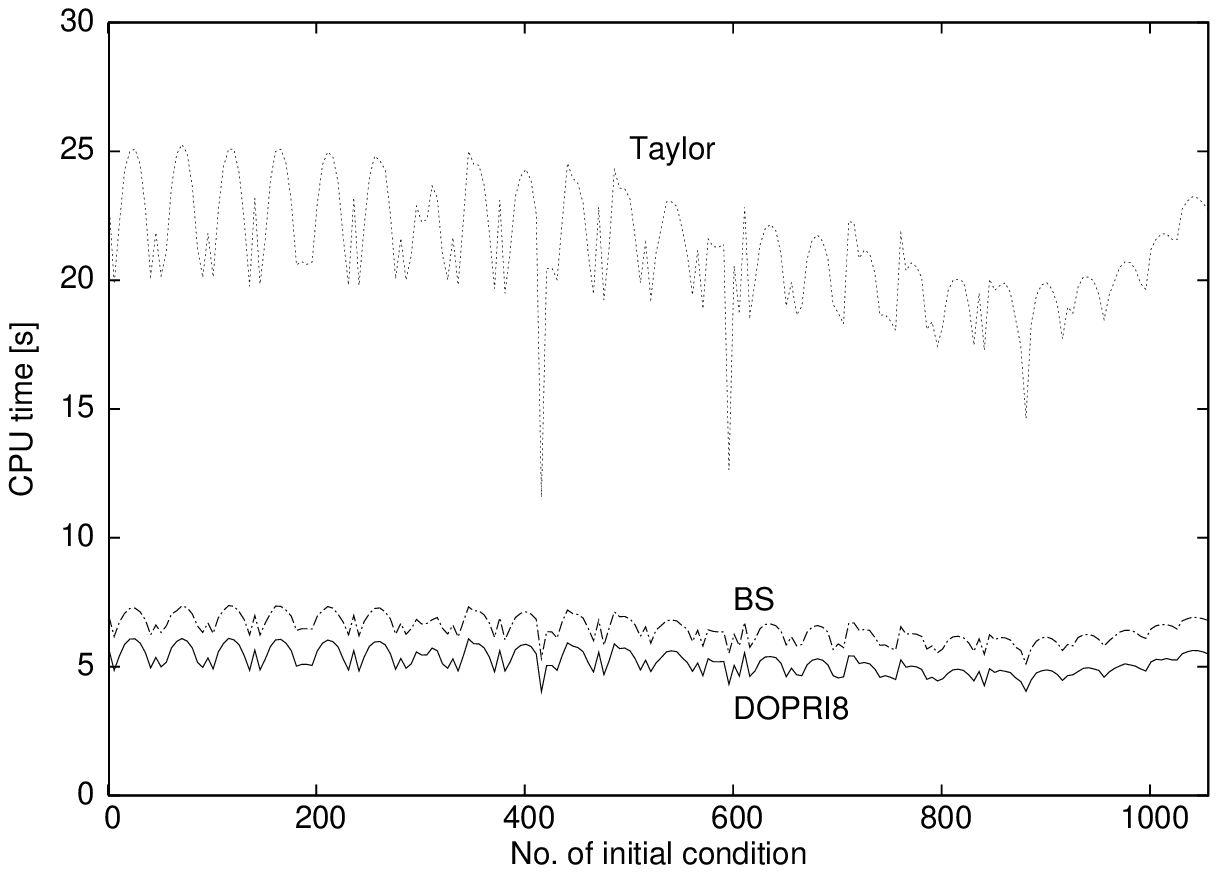, width=8cm}
    \epsfig{file=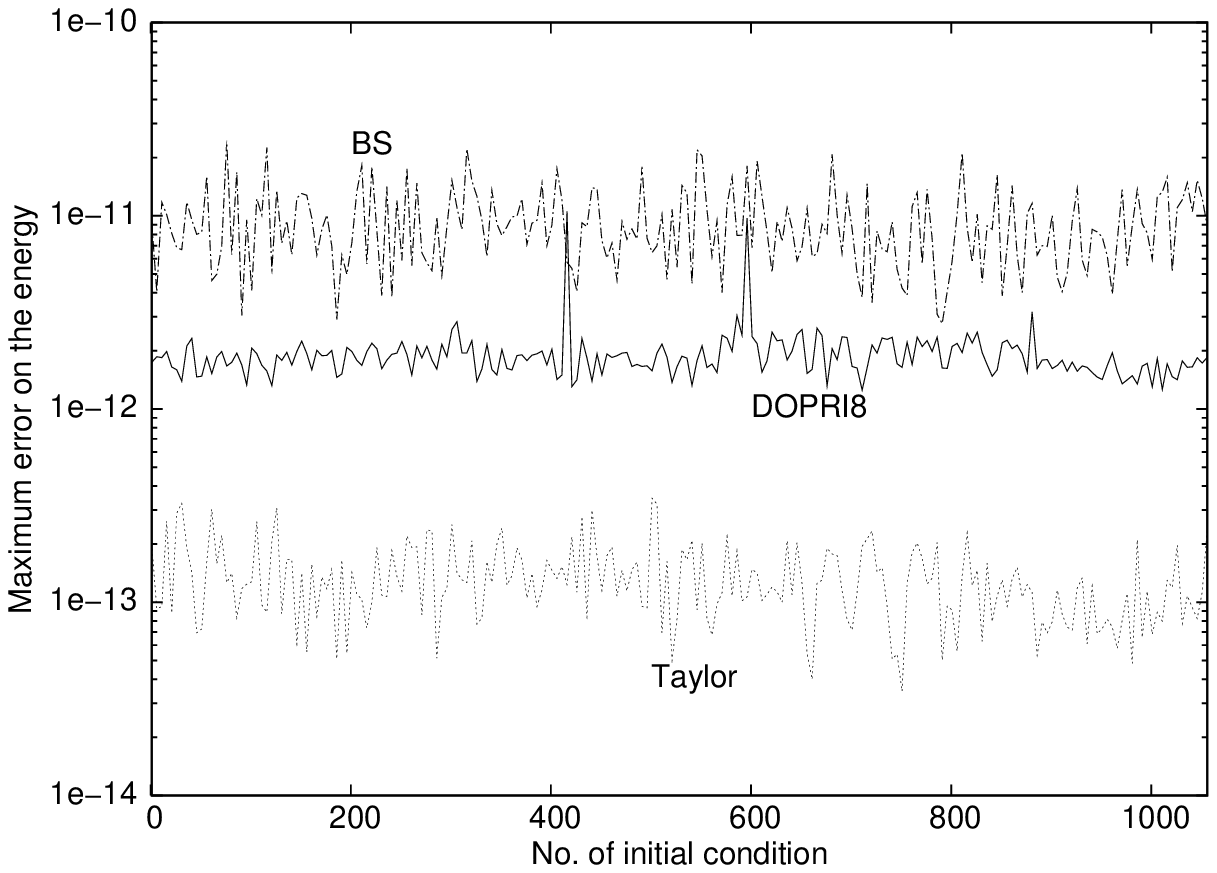, width=8cm}
  \end{center}
  \caption{CPU times (left) and maximum energy error (right) for several i.c. in the computation of the MEGNO for the quartic oscillator, using \texttt{taylor}, DOPRI8 and BS, for $t_{final}=10^5$ u.t.}
  \label{MEGNO_quartic_tcpu_maxerr}
\end{figure}

Figure \ref{MEGNO_quartic_tcpu_maxerr} (left) illustrates the
computational cost for evaluating the MEGNO in the current
application. It can clearly be seen that \texttt{taylor} required nearly
3 times the CPU time demanded by DOPRI8 and 2.3 times than that taken by
BS for the same task. In other words,  \texttt{taylor} turned out to be from about $2.3$ to $3$ times slower than the other two integrators, a factor of the same or even larger order than the one observed for the triaxial potential. 

The results in Figure \ref{MEGNO_quartic_tcpu_maxerr} (right) show that,
in this case, \texttt{taylor} is more accurate than DOPRI8 in a $\sim
15$ factor, that is a difference of approximately 1 order of magnitude
in the conservation of energy. On the other hand, the same difference as
in the case of the triaxial potential (see Figure
\ref{MEGNO_pot_tri_tcpu_maxerr}, right) is attained in the current
application when \texttt{taylor} is compared with BS performance, i.e. 2 orders of magnitude in favor of the former (more specifically a factor $\sim 70$).

Although \texttt{taylor} allows better energy conservation, the time
required to compute the MEGNO is too long to be a preferable alternative to the other two integrators, even when we are dealing with such a plain system as a perturbed quartic oscillator.

\subsection{The LI}
\label{section_LCN_quartic}
 
\begin{figure}[!htbp]
  \begin{center}
    \epsfig{file=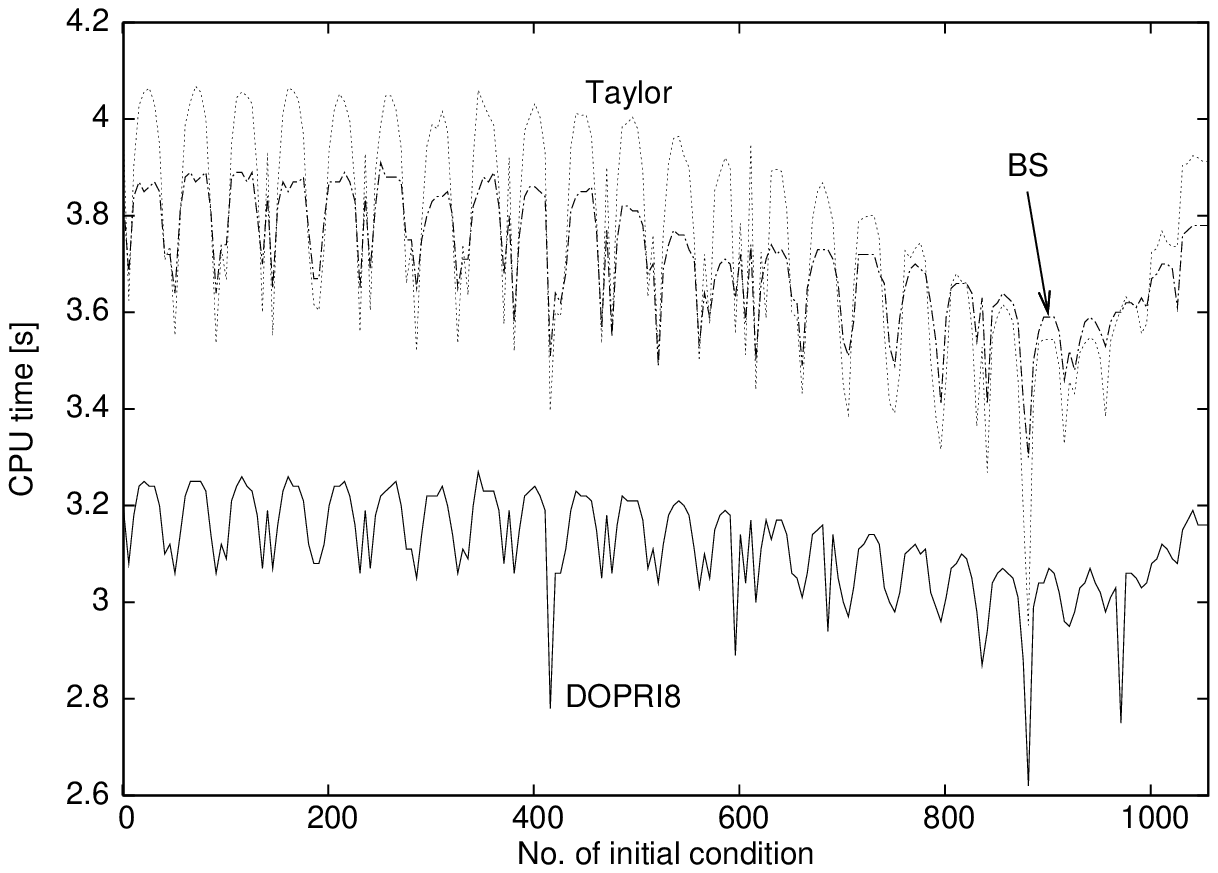, width=8cm}
    \epsfig{file=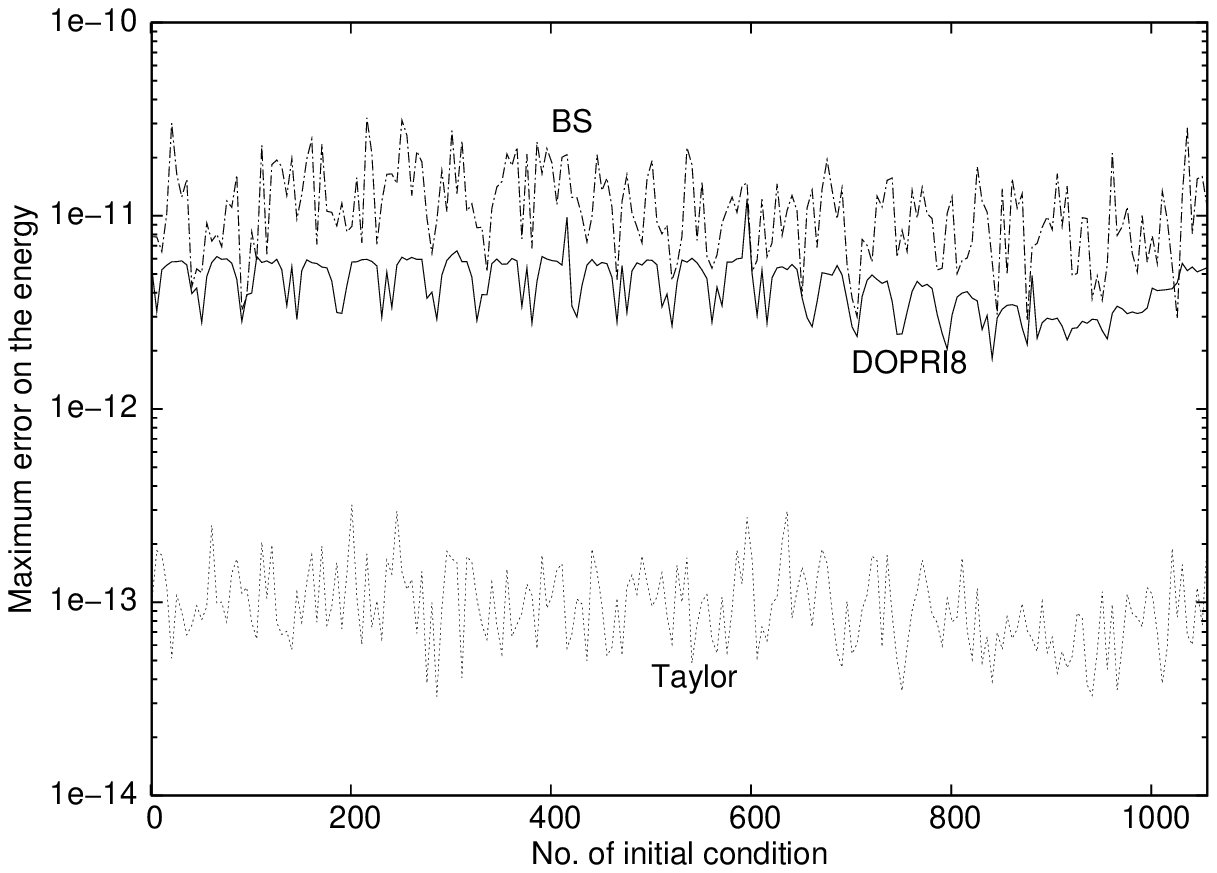, width=8cm}
  \end{center}
  \caption{CPU times (left) and maximum energy error (right) for several i.c. in the computation of the LI for the quartic oscillator, using \texttt{taylor}, DOPRI8 and BS, for $t_{final}=10^5$ u.t.}
  \label{LCN_quartic_tcpu_maxerr}
\end{figure}

The LI values were obtained for the same set of i.c. as those used in the previous section, imposing the same tolerance. The required computational effort and the achieved accuracy in the energy preservation are illustrated in Fig. \ref{LCN_quartic_tcpu_maxerr}.

From Figure \ref{LCN_quartic_tcpu_maxerr} (left) it can be deduced that
\texttt{taylor} was about $\sim 1.2$ times slower than DOPRI8 for the LI computation, which is a remarkably smaller difference than that observed for the CPU times for obtaining the MEGNO.

Regarding the resultant maximum energy error, Fig. \ref{LCN_quartic_tcpu_maxerr} (right) reveals a quite different behavior from that observed in the case of the MEGNO evaluation. Indeed, though there is still a difference of about 2 orders of magnitude between \texttt{taylor} and BS (a $\sim 100$ factor in favor of \texttt{taylor}), the difference between \texttt{taylor} and DOPRI8 grew considerably (to a $\sim 45$) compared to that shown on the right side of Figure \ref{LCN_pot_tri_tcpu_maxerr}.

Therefore, this result differs from the one obtained for the triaxial potential, where the differences in energy preservation turned out to be almost the same in the three problems analysed, namely, the computation of the MEGNO and  the LI values and the integration of the equations of motion. 

\subsection{The equations of motion}
\label{section_solomov_quartic}

This section is aimed at comparing both speed and precision when only
the equations of motion are computed for the quartic oscillator by testing again the three integrators under study.  

\begin{figure}[!htbp]
  \begin{center}
    \epsfig{file=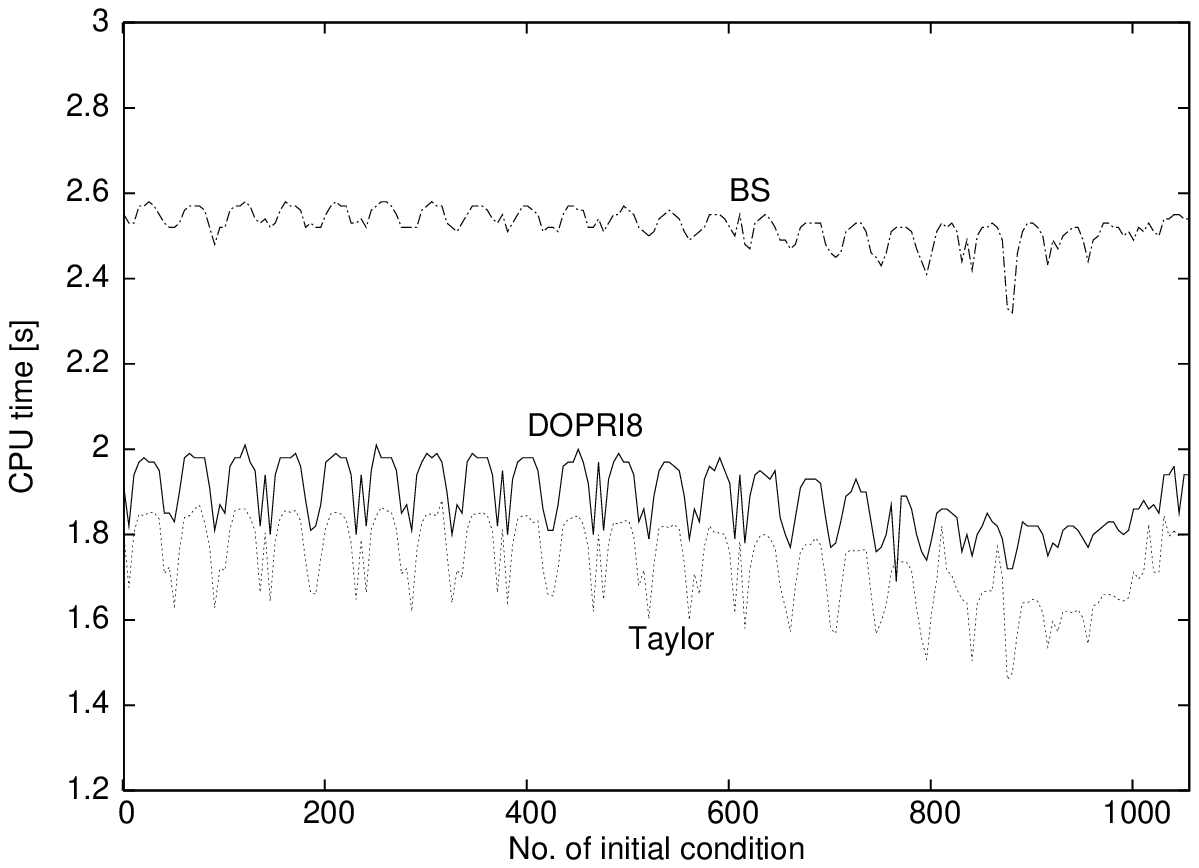, width=8cm}
    \epsfig{file=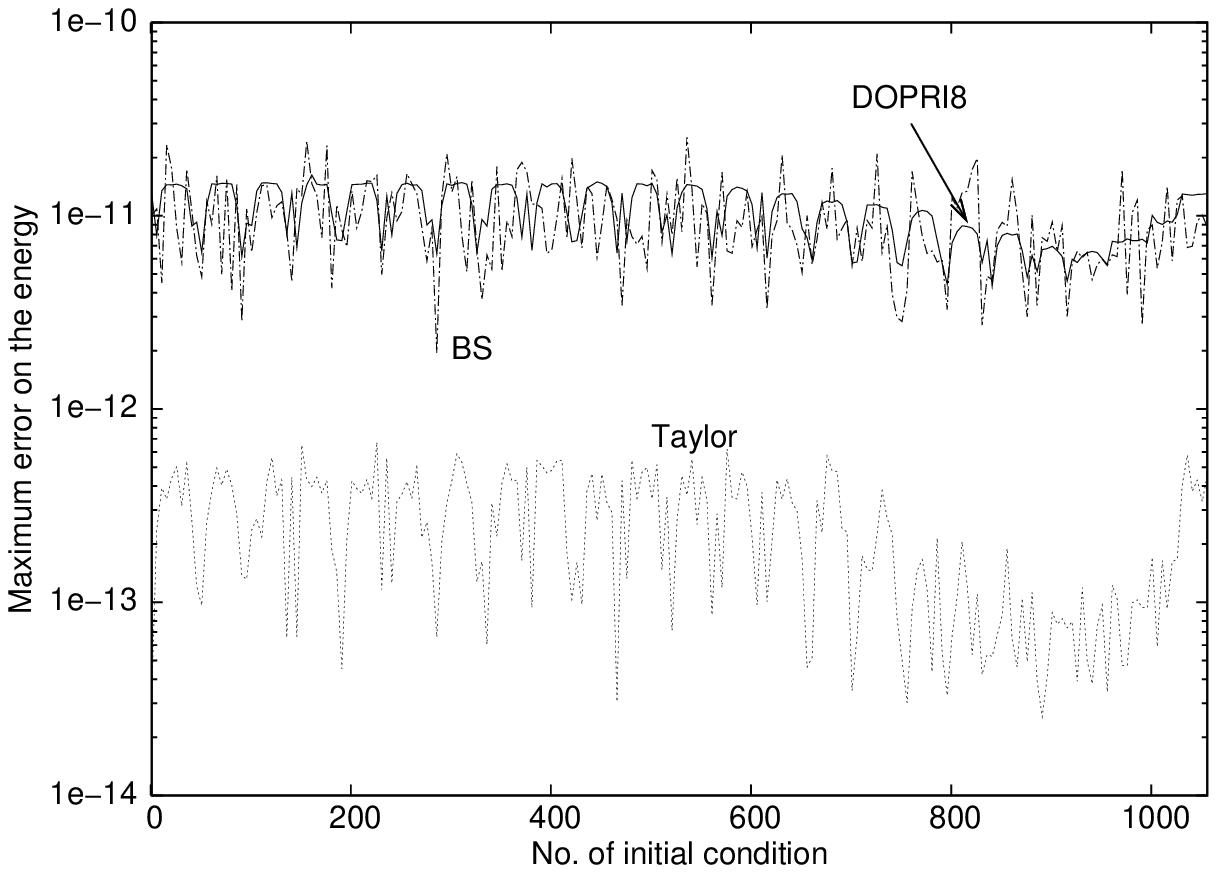, width=8cm}
  \end{center}
  \caption{CPU times (left) and maximum energy error (right) for several i.c. in the calculation of the equations of motion for the quartic oscillator, using \texttt{taylor}, DOPRI8 and BS, for $t_{final}=10^5$ u.t.}
  \label{solomov_quartic_tcpu_maxerr}
\end{figure}

As the left side of Figure \ref{solomov_quartic_tcpu_maxerr} shows,
\texttt{taylor} become the fastest one. In fact, a  similar behaviour
with respect to DOPRI8 is observed  as the one achieved for the galactic
model (i.e. \texttt{taylor} is just $\sim 1.08$ times faster than
DOPRI8), but the difference is greater when comparing \texttt{taylor}
with BS in a $\sim 1.3$ factor.

Concerning the maximum energy error presented on the right side of
Figure \ref{solomov_quartic_tcpu_maxerr}, the differences between
\texttt{taylor} and BS slightly decreased, falling below the 2 orders of
magnitude, i.e. nearly a $40$ factor. Moreover, this is the same difference between \texttt{taylor} and DOPRI8 results, the error curve for the latter lying behind the one corresponding to the BS method.

Thus, it should be highlighted that, when we solved the three different
sets of ODEs (those for the MEGNO, the LI and the equations of motion)
for the perturbed quartic oscillator, \texttt{taylor} was more accurate
than DOPRI8 and BS. In terms of CPU times, \texttt{taylor} was
comparable to DOPRI8 and faster than BS when integrating solely the equations of motion, and required the largest computational time when determining both the MEGNO and the LI values.

Again, the results obtained for the perturbed quartic oscillator
indicate that \texttt{taylor} is not well suited for systems involving
variational equations or ODEs like the ones defining the MEGNO. However,
it does offer some advantages when only the equations of motion are
integrated. The \texttt{taylor} performance slowdown is still not fully
understood and it is subject to further study (extracted from a private conversation with \`Angel Jorba).

As mentioned in Section \ref{section_introduction}, the study of
diffusive processes in phase space requires a fast and, above all, very
reliable method for integrating the equations of motion of such a
system. The experiments performed herein point out that \texttt{taylor}
is more efficient than the other two integrators considered, at least as
far as the integration of the equations of motion is concerned. Thus, we
are bound to advance  that it is the best candidate for studies on the
accurate determination of coordinates in phase space. This issue will be
discussed in Section \ref{section_trans_error}.

\subsubsection{Tolerance for a fixed $\Delta E$}
\label{Fidex_Delta_E}

In the previous section, we obtained the precision on the energy for a
fixed value of $\varepsilon$. The aim of this section is to obtain a
value of $\varepsilon$ for which we calculate a precision of $\Delta E$
in the preservation of the energy.

In order to do this we used the following steps: first, we fix the
precision on the energy $\Delta E_{0}$, then we set an initial value for
$\varepsilon_{0}=\Delta E_{0}$ and, by decreasing $\varepsilon$ by one
unit (i.e. $3 \times 10^{-11}$, $2 \times 10^{-11}$, $1 \times
10^{-11}$, $9 \times 10^{-12}$, etc.), we continue iterating until we
obtain a value of $\varepsilon$ for which $\Delta E \leq \Delta
E_{0}$. Table 3 shows the results obtained.

\begin{table}[!htbp]
  \begin{center}
    \tbl{Tolerance required to achieve the desired $\Delta E$ for $t_{final}=10^4$ (top), $t_{final}=10^5$ (bottom).}
    {\begin{tabular}{cccc}
      \toprule
      $t=10^4$ & \multicolumn{3}{c}{Tolerance ($\varepsilon$)}\\
      \hline
      $\Delta E$ & \texttt{taylor} & DOPRI8 & BS \\      
      \hline
      $10^{-11}$          & $3 \times 10^{-13}$ & $4 \times 10^{-14}$ &  $4 \times 10^{-13}$ \\
      $10^{-12}$          & $3 \times 10^{-13}$ & $4 \times 10^{-15}$ &  $2 \times 10^{-14}$ \\
      $7 \times 10^{-13}$ & $3 \times 10^{-13}$ & $3 \times 10^{-15}$ &  $2 \times 10^{-14}$ \\

      \hline \hline

      $t=10^5$ & \multicolumn{3}{c}{Tolerance ($\varepsilon$)}\\
      \hline
      $\Delta E$ & \texttt{taylor} & DOPRI8 & BS \\

      $10^{-9}$          & $1 \times 10^{-11}$ & $3 \times 10^{-13}$ & $5 \times 10^{-12}$ \\
      $10^{-10}$          & $3 \times 10^{-13}$ & $4 \times 10^{-14}$ & $4 \times 10^{-13}$ \\
      $10^{-11}$ & $3 \times 10^{-13}$ & $4 \times 10^{-15}$ & $2 \times 10^{-14}$ \\
      \botrule
    \end{tabular}}
  \end{center}
  \label{tol_for_different_integrators}
\end{table}

\begin{figure}[!htbp]
  \begin{center}
    \epsfig{file=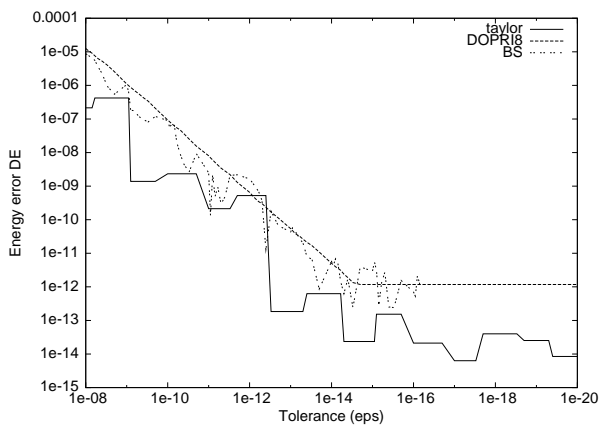, width=8cm}
  \end{center}
  \caption{Values of $\varepsilon$ for different values of $\Delta E$.}
  \label{taylor_step-like_behaviour}
\end{figure}

As indicated in Table 3, considerable less precision in $\varepsilon$ is
required by \texttt{taylor} to achieve the same error on the energy
preservation. Note that, in the case of $t_{final}=10^4$, the three
values of $\varepsilon$ given by \texttt{taylor} are exactly the
same. This is due to how \texttt{taylor} selects the order $p$ and the
stepsize $h$ (for further details, see \cite{JZ05}, Section 3.2). A
brief explanation is that the order $p$ depends on $\varepsilon$ and the
stepsize $h$ basically depends on $p$. Therefore, for all the values of
$\varepsilon$ which result in the same value of $p$, we will obtain the same $h$, in an identical integration. This step-like behaviour can be seen in Figure \ref{taylor_step-like_behaviour}.

Here we can see that, for $\varepsilon \approx 2 \times 10^{-15}$ in
DOPRI8 and $\varepsilon \approx 10^{-16}$ for BS, as both integrators
saturate, due to their intrinsic precision, it is useless to compare them beyond those values.

The step-like nature of \texttt{taylor}'s curve means that, if we take a
 given value of $\Delta E$, we can increase the tolerance up to the largest value in the same step. For instance, seeing figure \ref{taylor_step-like_behaviour}, if we ask $\Delta E \sim 3 \times 10^{-13}$, we can increase $\varepsilon$ up to $\sim 10^{-12}$. Or, from another perspective, if we change from $\varepsilon=10^{-12}$ to $3 \times 10^{-13}$, we win approximately 3 orders of magnitude, while DOPRI8 and BS will improve less than 1 order of magnitude.

\section{On the accuracy of the computation of coordinates in phase space}
\label{section_trans_error}

In this section we will try to elucidate which of the three integrators under study permits the most precise determination of the coordinates in phase space for a given initial value problem when investigating diffusive processes. 

To this end, the integration  using \texttt{taylor} with extended
precision arithmetic (i.e. GMP library with a mantissa of 256 bits, see
\texttt{http://gmplib.org/} for details) could be considered as the
``exact'' solution against which to contrast the result of integrating the equations of motion by means of the three methods. This choice is based on the results obtained in \cite{JZ05}, Section 5.1.4.

Of the $1056$ i.c. used in Section \ref{section_quartic}, in order to
perform the accuracy test, the perturbed quartic oscillator was
integrated with those which the MEGNO classified as regular (i.e. 1039
orbits). The reason for that selection is that it is useless to compute
the error on the coordinates of chaotic orbits, because of their
sensibility to the i.c., resulting in too different final coordinates
for each integrator. The results were then compared to the extended precision approximation. The integration was made for several tolerances, ranging from $10^{-10}$ to $10^{-16}$, and stopped at $10^5$ u.t.

In fact, the magnitude of the error vector in configuration space
provides an idea of the precision of the integrator
applied. Nonetheless, the magnitude of its component along the direction
normal to the trajectory would be more relevant in the present
application is, since the error in the tangential direction could be
assimilated as a mere shift of the integration time, having no impact on
a diffusion analysis. If we define $\theta$ as the angle between the position error vector and the tangent line of the actual orbit, given by the velocity vector of the ``exact'' trajectory, that is, 

\begin{equation}
  \mathbf{\Delta r} \cdot \mathbf{v} = \Delta{r} \,\rm{v} \,\rm{cos}(\theta), 
  \label{eq_theta}
\end{equation}  
the tangential and transverse components of the error can be recasted as $\Delta{r}\,\rm{cos}(\theta)$ and $\Delta{r}\,\rm{sin}(\theta)$, respectively.
 
Table 4 clearly shows the absolute value of the position error and its
component in the transverse direction as a percentage of the whole for
several tolerances. To get these results we considered all the i.c. mentioned above, and computed the errors in the following way:

\begin{equation}
\left\{\begin{array}{ll}
\overline{\Delta r} & = \frac{1}{N} \sum {\Delta r}_i \\
\\
  \overline{\Delta r}_\bot & = \frac{1}{N}\sum {\Delta r}_i *
  \sin(\theta_i) \\
\\
  R & = \frac{\overline{\Delta r}_\bot} { \overline{\Delta r}},
 \end{array}\right.
\end{equation}

\noindent
where $R$ clearly represents the fraction of the error on the transverse direction.

\begin{table}[!h]
\renewcommand\arraystretch{1.3}
  \begin{center}
  \tbl{Mean values of the magnitude of the position error for several tolerances and the transverse component of such a vector, for a  particular orbit of the perturbed quartic oscillator at $t_{final}=10^5$ u.t., written as a percentage.}
    {\begin{tabular}{ccccccc}
      \toprule
      &\multicolumn{2}{c}{\texttt{taylor}} & \multicolumn{2}{c}{DOPRI8} & \multicolumn{2}{c}{BS} \\
      \hline
      $\varepsilon$ & $\overline{\Delta{r}}$ & $R(\%)$ & $\overline{\Delta{r}}$ & $R(\%)$ & $\overline{\Delta{r}}$ & $R(\%)$ \\
      \hline
      1e-10 & 1.782e-4 & 31.8 & 8.132e-3 & 27.3 & 7.926e-3 & 34.2\\
      1e-11 & 1.423e-5 & 29.5 & 6.659e-4 & 27.6 & 1.388e-3 & 44.4\\
      1e-12 & 1.423e-5 & 29.5 & 5.420e-5 & 27.8 & 1.023e-3 & 44.3\\
      1e-13 & 3.661e-8 & 36.6 & 4.487e-6 & 27.5 & 9.352e-4 & 46.3\\
      1e-14 & 4.929e-8 & 31.0 & 1.068e-7 & 26.3 & 9.318e-4 & 46.4\\
      1e-15 & 3.197e-9 & 24.2 & 1.068e-7 & 26.3 & 9.317e-4 & 46.4\\
      1e-16 & 1.599e-9 & 22.2 & 1.067e-7 & 26.3 & 2.543e-2 & 50.5\\      
      \botrule
    \end{tabular}}
  \end{center}
  \label{table_trans_err_tol}
\end{table}

Table 4 reveals that the major component of the error lies along the
trajectory direction. The percentages for \texttt{taylor} and DOPRI8 can
be considered comparable ($\sim 25 - 30 \%$), although the latter
presents less dispersion, while BS gives higher values ($\sim 45 \%$).

\begin{figure}[!htbp]
  \begin{center}
    \epsfig{file=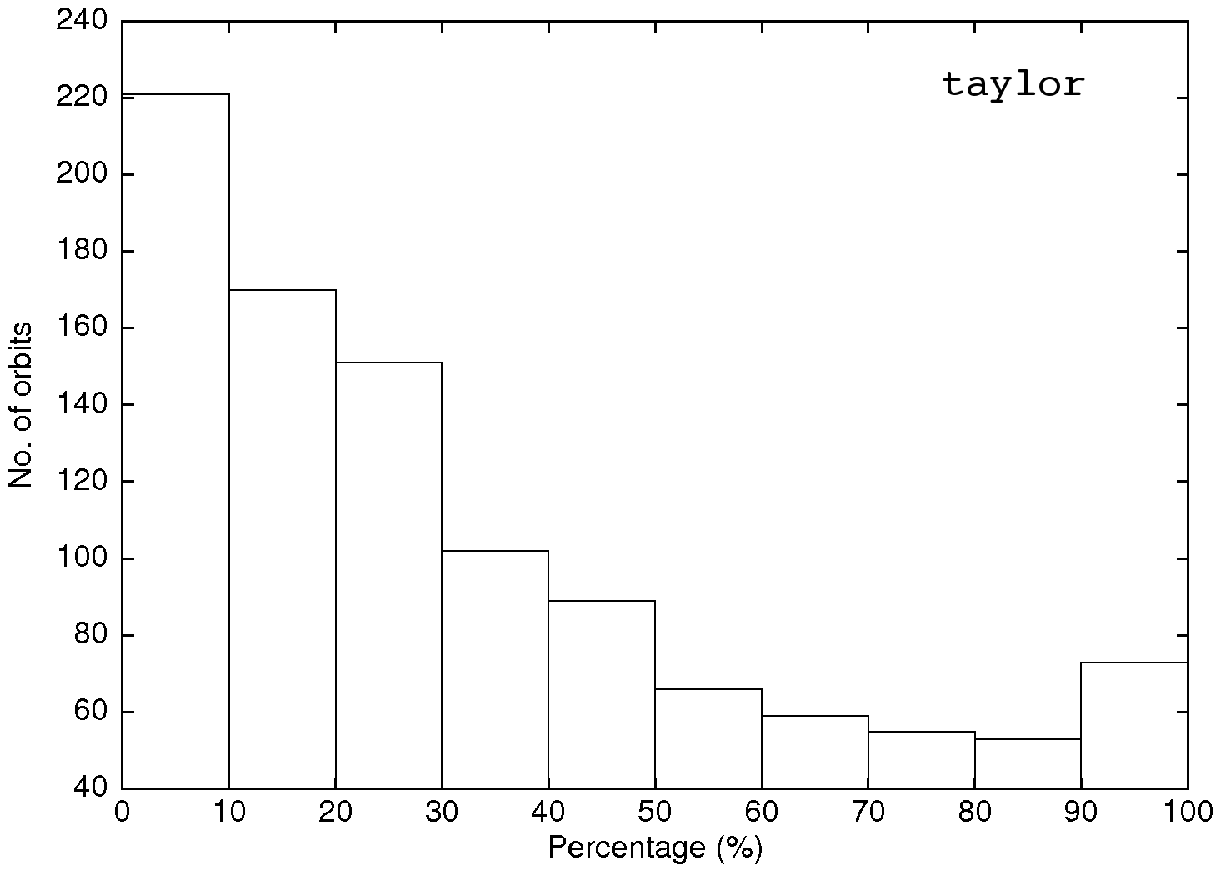, width=7.5cm}
    \epsfig{file=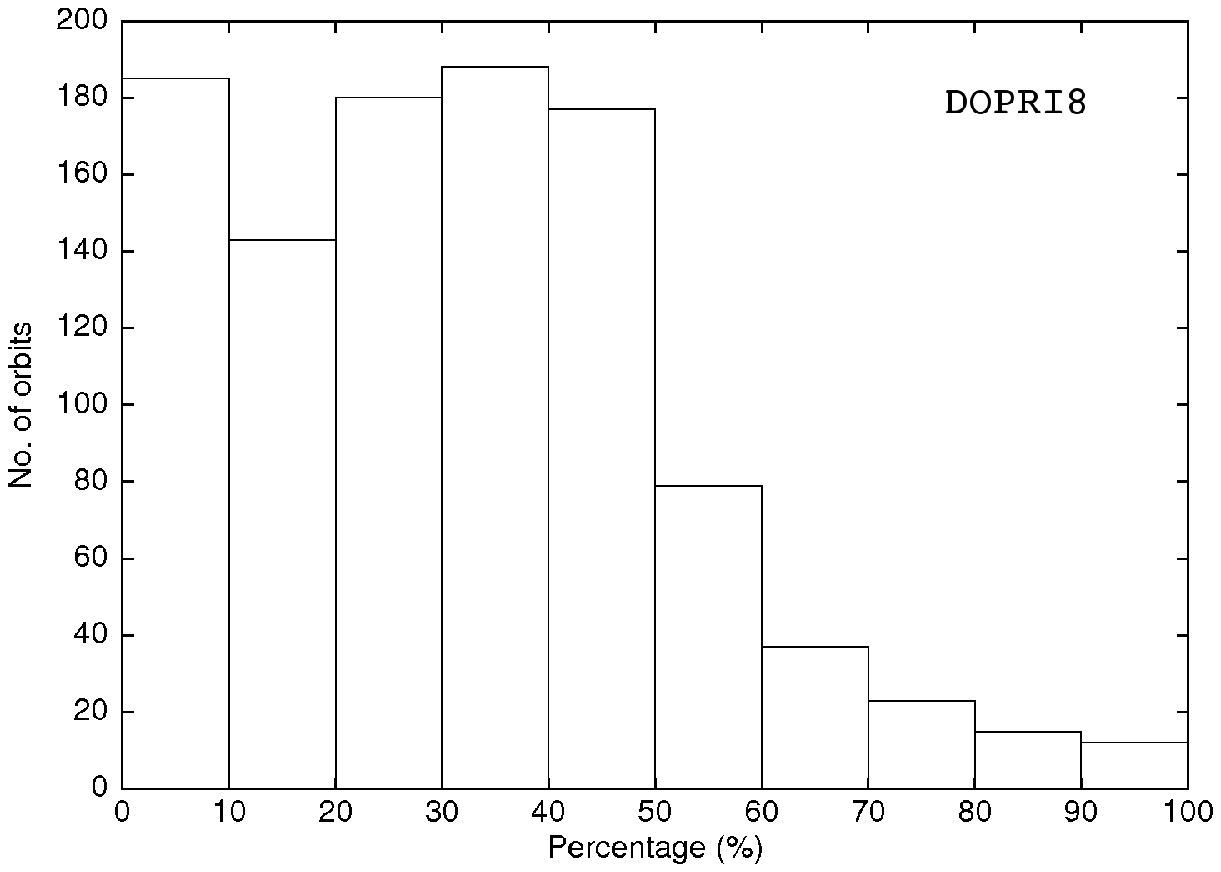, width=7.5cm}\\
    \epsfig{file=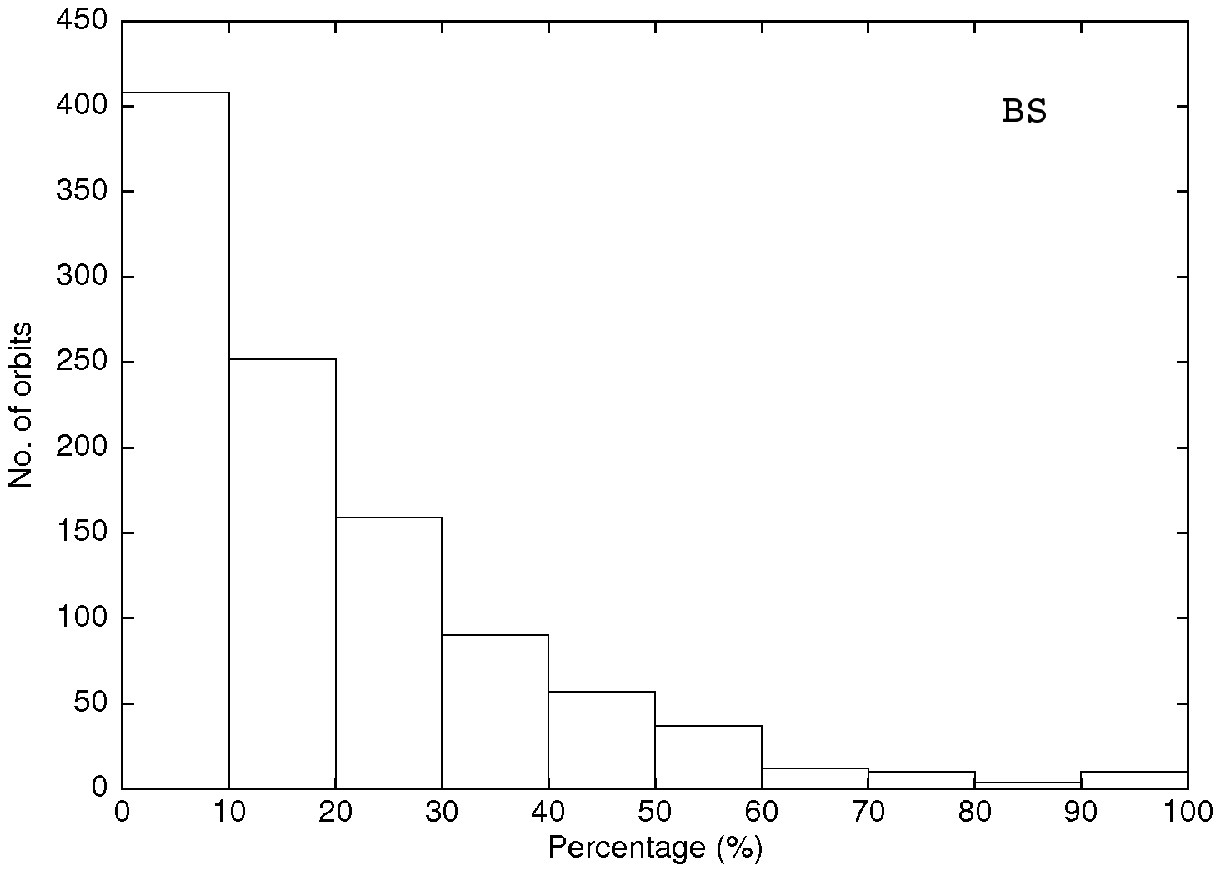, width=7.5cm}
  \end{center}
  \caption{Number of orbits whose fraction of the error in the perpendicular direction to the orbit falls in a given interval for \texttt{taylor}, DOPRI8 and BS respectively}
  \label{quartic_histograms}
\end{figure}

Taking into account the same orbits as in the previous table, an histogram for each integrator can be built, considering $\varepsilon=10^{-15}$, As shown in Figure \ref{quartic_histograms}.

The histograms show that \texttt{taylor} has the majority of its orbits below the mean value, as well as BS, while DOPRI8 has a near homogeneity in the interval $0 - 50 \%$.

In the attempt to determine which of the three integrators is the most precise one, the arithmetic mean (for $\varepsilon=10^{-15}$) of the absolute value of the total error is computed to yield,

\begin{equation}
  \left\{\begin{array}{ll}
      \rm{\texttt{taylor}:} & 3.1971267 \times 10^{-9}\\ 
      \rm{DOPRI8:} & 1.0677529 \times 10^{-7}\\
      \rm{BS:} & 9.3165436 \times 10^{-4}\\
    \end{array}\right.
\end{equation}

\noindent
and that of the component in the transverse direction

\begin{equation}
  \left\{\begin{array}{ll}
      \rm{\texttt{taylor}:} & 7.7557023 \times 10^{-10}\\  
      \rm{DOPRI8:} & 2.8064381 \times 10^{-8}\\
      \rm{BS:} & 4.3187811 \times 10^{-4}.\\
    \end{array}\right.
\end{equation}

Therefore, \texttt{taylor} results to be more accurate when computing
the coordinates of phase space than both DOPRI8 and BS, as already shown
in Table 4. In fact, the precision is really good, since it is about 2
orders of magnitude more precise than DOPRI8 and 6 orders of magnitude
more accurate than BS. These results support our previous assumption
that \texttt{taylor} is a very efficient alternative when dealing with
problems which involve the solution of the equations of motion and
requires the value of the coordinates to be quite precise.

\section{Discussion}
\label{section_discussion}
In this work, we reviewed the performances of some variational CIs as well as the numerical techniques for their computation over different known dynamical problems. 

As regards the comparison of the variational indicators studied here, we
can mention some particular points to take into consideration when
selecting appropriate tools to study a certain dynamical
problem. Notwithstanding the excellent performances shown by some
techniques, it is advisable to avoid using only one indicator of
chaos. On the other hand, it is imperative to reduce
the number of CIs in the package. Therefore, we aimed at selecting a
CIsF for a general Hamiltonian system.

In the previous experiments the RLI and the FLI/OFLI showed the robustest thresholds. The OFLI seems to be a reliable
suggestion if we can use its time evolution, i.e. to study small
samples of orbits or for single analysis. The OFLI and the MEGNO are the
recommended techniques to identify periodicity or stability levels,
respectively. The GALI$_2$ (or the SALI) --with a proper checking of the
i.d.v. because of their very low final values, \cite{Barrio05}-- is also appropriate for identifying periodicity, and the
GALI$_k$ for identifying regular motion on tori of lower dimensionality
(\cite{CB06}, \cite{Skokos07}, \cite{Skokos08}). Furthermore, to analyze
large samples of orbits by means of the final values, a combination of
the FLI/OFLI and the GALI$_4$ (or the GALI$_{2N}$ for a
$N$--d.o.f. Hamiltonian system) seems to be useful to describe the
divided phase space of the HH potential. The FLI/OFLI final
values did not show spurious structures in the regular component within the
experiments. The GALI$_4$ with the help of the corresponding saturation
times is useful both to rough out and describe thoroughly the chaotic
components. Finally, we ordered the CIs studied according to their
CPU--times with a fixed integration time. The LI is the most economical
in computational time followed closely by the MEGNO and the \textit{D} is the most CPU--time
consuming. However, the LI has a rather low speed of
convergence. We also showed that the FLI/OFLI is less
time--consuming due to an efficient implementation of the saturation values. 

The FLI/OFLI proved to be the most versatile technique tested. The FLI
showed a robust threshold and good performances using the final values to
study large samples of orbits and the OFLI showed good performances if we
use its time evolution for single analysis. Nevertheless, some techniques can be selected with the FLI/OFLI to give
richer descriptions of the system. 

The previous results showed that a reliable
suggestion for a CIsF for the HH potential and for a general Hamiltonian flow
is composed of the FLI/OFLI techniques, the MEGNO and the
GALI$_{2N}$. Furthermore, there is no reason to believe that such CIsF
can be only applied to Hamiltonian flows; it can also be used in mappings,
thus improving the efficiency of the CIsF selected in \cite{Maffione11a}. Therefore, the above--mentioned CIsF can be applied to a general Hamiltonian system. 

We also present the results of a comparative study of the \texttt{taylor} package developed by \cite{JZ05} against two other well--known and widely used integrators, namely, DOPRI8 and BS. The three methods were applied to diverse dynamical problems and different vector fields as well,  with the aim of determining their relative efficiency. Both speed and accuracy were tested for three dynamical problems: the computation of a fast indicator as the MEGNO, that of the largest LCN (i.e. LI), and the integration of the equations of motion for a model of triaxial galaxy and a perturbed quartic oscillator. 
The numerical experiments performed, showed that one of
\texttt{taylor}'s strongest features is its capacity to preserve the
energy integral. In this regard, in the more complex system (the
triaxial potential), DOPRI8 showed to be nearly as accurate as
\texttt{taylor} and both turned out to be about two orders of magnitude more accurate than BS. In the case of the perturbed quartic oscillator, it resulted about two orders of magnitude more precise than both DOPRI8 and BS.

A characteristic found in this experiments is that the energy
preservation yielded by the use of \texttt{taylor} turned out to be
problem-independent for a given vector field, in contrast to what occured when DOPRI8 was applied.

Moreover, the computational effort demanded by the three integrators was subjected to comparison to conclude that \texttt{taylor} was considerably slower than DOPRI8 and BS while computing the MEGNO. The differences in the required CPU time decreased significantly for the computation of the LI, but even in this case \texttt{taylor} remained the slowest one. When dealing with the integration of the equations of motion, a change in the behavior was observed, \texttt{taylor} advantaging in speed  both DOPRI8 and BS (Figs. \ref{solomov_pot_tri_tcpu_maxerr} and \ref{solomov_quartic_tcpu_maxerr}).

Although this is not a thorough test, in the light of the results obtained we can say that \texttt{taylor} is rather slow when the variational equations or the ones defining the MEGNO are integrated. In general, its efficiency seems to diminish  as the complexity of the system of ODEs increases.

However, not only the dynamical problem being attacked, but also the vector field under study impact on the differences in the CPU times required by the different integrators. Therefore, for instance, while computing the MEGNO the time differences between \texttt{taylor} and the other two integrators are greater for the perturbed quartic oscillator than those for the triaxial potential (as seen in  Figs. \ref{MEGNO_pot_tri_tcpu_maxerr} and \ref{MEGNO_quartic_tcpu_maxerr}).

In an attempt to unravel the observed behavior we adventure to say that whenever dealing with a plain formulae for the vector field, such as in the case of the perturbed quartic oscillator, \texttt{taylor} takes too much time for computing the recurrences for the automatic differentiation, determining each coefficient and building the power series, while the other integrators  only evaluate the quite simple functions defining the system of ODEs in some points conveniently chosen, thus resulting more efficient than   \texttt{taylor}. For  a more complex system of ODEs, like the ones involved in the model of a triaxial potential, the mere evaluation of the field in DOPRI8 or BS would be not as fast as for the quartic oscillator which only encompasses polynomic expressions, causing  their time differences with \texttt{taylor} to decrease.

Conversely, regarding the energy preservation the opposite behavior is
observed, \texttt{taylor} advantaging DOPRI8 for the simpler vector
field more than for the one defined by rather cumbersome expressions. In
fact, in the case of the triaxial potential the obtained differences in $\Delta E_{max}$  between \texttt{taylor} and DOPRI8 are quite small (barely reaching a factor of $\sim 2$), while for the perturbed quartic oscillator the errors in the energy differ in a factor ranging from $ \sim 14$  to $\sim 100$ (i.e. in 2 orders of magnitude). Let us say that the comparison against BS has been left behind since, even when BS demanded CPU times comparable to those required by DOPRI8, the resulting energy preservation in all cases was not as good as that obtained by any of the other two integrators.

In the light of our experiments, a crosswise situation was observed when
comparing both the computational effort and the energy preservation for
the integrators being applied to vector fields of different
complexity. Consequently, generally speaking the supremacy of one
integrator over the other could not be claimed out. However, what remains clear is that the \texttt{taylor} package results unsuitable for computing the MEGNO or integrating the variational equations (this could also be seen on Gerlach \& Skokos [2010]), where DOPRI8 shows up as a good choice to perform the required integrations (let us remark that on the previous papers \cite{Cincotta03} and \cite{Cincotta00}, for instance, the DOPRI8 subroutine was used for this purpose).

On integrating only the equations of motion, the analysis of the
accuracy reached by means of the three integrators exposed that the
\texttt{taylor} package turned out to be the most efficient of the
three. In this application, not only the final errors in the position
vector but also their component in the perpendicular direction to the
orbit were compared in order to estimate how much the trajectory
obtained departed from the current one. There, \texttt{taylor} was found to be considerably more precise than both DOPRI8 and BS (about 2 and 6 orders of magnitude, respectively).

Thus, considering that the \texttt{taylor} package also revealed itself as advantageous regarding the required CPU time and the preservation of the energy integral when integrating  the equations of motion of the 3D perturbed quartic oscillator, we state that \texttt{taylor} is the most convinient tool to be used in those problems that only involve the integration of the equations of motion, such as the study of chaotic diffusion in phase space.

Thanks to investigations like the presented above, we can make smart decisions about the most efficient group of CIs and the appropriate numerical integrator to study a given system. Nevertheless, a suitable tool should attend the preceeding studies. Thus, in Section \ref{section_Flows} we mentioned that the computing of the CIs were accomplished with a code we are developing, the LP-VIcode. The LI, the RLI, the MEGNO, the D and the SSNs, the SALI and the GALIs and the FLI and the OFLI are already programmed for the systems tested, being such systems either a mapping or a flow. They are grouped in several modules, so the user can select the desired module to compute the preferred CIs. Moreover, the integration routine used is the DOPRI8. However, our final goal is to introduce a code with a growing CIs' database (variational CIs at first), where the user can easily program the model to study, and use the desired integrator.  

\section*{Acknowledgments}
The authors want to thank both anonymous referees for their suggestions as well as the English department of FCAG--UNLP for improving the English of the reported research. This work was supported with grants from the Consejo Nacional de Investigaciones Cient\'{\i}ficas y T\'ecnicas de la Rep\'ublica Argentina (CCT--La Plata) and the Universidad Nacional de La Plata.

\end{document}